\pdfoutput=1
\documentclass[a4paper,11pt]{extarticle}  

\usepackage{stylesheet}

\usepackage[left=1in,top=1in,bottom=.65in,right=1in]{geometry}

\usepackage{marvosym}

\usepackage{nomencl}
\makenomenclature

\title{Evaluation of 'Dunkelflaute' event detection methods considering grid operators' needs}

\author{%
\textbf{
Benjamin Biewald\textcolor{Accent}{\textsuperscript{1,\Writinghand}}\orcidlink{0009-0001-0721-5878}, %
Bastien Cozian\textcolor{Accent}{\textsuperscript{2,\Writinghand,\Letter}}\orcidlink{0000-0003-0572-0907}, %
Laurent Dubus\textcolor{Accent}{\textsuperscript{2,3}}\orcidlink{0000-0002-3987-646X}, %
William Zappa\textcolor{Accent}{\textsuperscript{4}}\orcidlink{0000-0001-6810-7224}, %
Laurens Stoop\textcolor{Accent}{\textsuperscript{4}}\orcidlink{0000-0003-2756-5653}, %
}\\[0.5em]
\begin{small}%
\textcolor{Accent}{\textsuperscript{1}}TenneT TSO GmbH, Bayreuth, Germany\\[0.5em] 
\textcolor{Accent}{\textsuperscript{2}}Reseau de Transport d'Electricite, Paris, France\\[0.5em]
\textcolor{Accent}{\textsuperscript{3}}World Energy \& Meteorology Council, Norwich, UK\\[0.5em] 
\textcolor{Accent}{\textsuperscript{4}}TenneT TSO B.V., Arnhem, the Netherlands\\[0.5em] 
\textcolor{Accent}{\textsuperscript{\Writinghand}}These authors contributed equally to this work. \\[0.5em] 
\textcolor{Accent}{\textsuperscript{\Letter}}Corresponding Author: \textcolor{Accent}{bastien.cozian@rte-france.com} \\ \end{small}
}

\date{\today}

\makeatletter
\let\newtitle\@title
\let\newauthor\@author
\let\newdate\@date
\makeatother

\begin{document}

\thispagestyle{empty}
\begin{center}

\parbox[c]{350pt}{This is the author preprint of the following submitted article:}  \vspace{20pt}

\begin{tabular}{|r|l|}\hline
 & \\ \hline
\textbf{author(s)} & \parbox{220pt}{Benjamin Biewald, Bastien Cozian, Laurent Dubus, William Zappa, Laurens Stoop} \\ \hline 
\textbf{title} & \parbox{220pt}{Evaluation of 'Dunkelflaute' event detection methods considering grid operators' needs} \\ \hline
\textbf{journal} & Renewable and Sustainable Energy Reviews \\ \hline
\end{tabular}

\vspace{30pt}
\parbox[c]{350pt}{This manuscript version corresponds to the author's submission of the article. It has not been copy-edited or formatted by the journal.     
This preprint is deposited under a Creative Commons Attribution (CC-BY) license.}    
\end{center}

\newpage

\pagenumbering{arabic} 

\maketitle

\begin{abstract}
Weather conditions associated with low electricity production from renewable energy sources (RES) can result in challenging 'dunkelflaute' events, where 'dunkel' means dark and 'flaute' refers to low windspeeds. 
In a power system relying significantly on RES, such events can pose a risk for maintaining resource adequacy, i.e. the balance between generation and demand, particularly if they occur over a large geographical area and for an extended period of time.
This risk is further emphasized in periods of cold ('kalte') temperature, known as 'kalte dunkelflaute'. 

In this paper, we perform a literature review of different methods to identify dunkelflaute events from hourly RES production and load data alone. 
We then validate three of these methods by comparing their results with periods of shortage identified from a detailed power system simulation model used by grid operators (ERAA2023). 
Strengths and weaknesses of these methods are discussed in terms of their data requirements, ease of application, and skill in detecting dunkelflaute events.

We find that all three 'dunkelflaute' event detection methods have some ability to identify potential energy shortages, but none are able to detect all events. 
Most likely other factors such as the presence of energy storage capacity, non-weather-dependent outages, and model-related factors limit the skill of these methods. 
We find that all three methods perform best if the residual load is used as input, rather than hourly RES production or load alone. 
Overall, we find that Otero'22 is the method that yields the best results while being straightforward to implement and requiring only data with daily resolution.
The results hold for countries relying on a small or a large share of RES production in their electricity mix.
\end{abstract}

\noindent{\it \color{Highlight} Highlights}:  
\begin{itemize}
    \item Residual load is the energy variable that best detects energy shortages
    \item Dunkelflaute detection methods are validated with a reference energy shortage dataset
    \item These methods improve identification of challenging years for adequacy studies
\end{itemize}

\noindent{\it \color{Highlight} Keywords}: dunkelflaute, resource adequacy, power system, climate data, energy-meteorological variability, energy shortage events

\section{Introduction}
Earth's changing climate is a threat to human well-being~\autocite{ipcc6_wg2}.
A significant contributor to climate change are the greenhouse gas emissions resulting from combustion of fossil fuels for electricity generation~\autocite{ipcc6_wg1_synth}. 
An energy transition is ongoing to provide societies with their required energy services to meet basic human needs~\autocite{ipcc6_wg1_synth}.

The monumental task of guiding the energy transition and providing a resilient electricity grid with reliable and affordable energy in the future lies with energy system planners and operators around the world~\autocite{gernaat2021climate,craig2022disconnect,Bloomfield2021nextgen}. 
For this they need, among others, insightful knowledge on the impact of the energy-meteorological variability of renewable energy sources (RES) and specifically the weather-driven threats that might pose a risk to adequacy in the future~\autocite{Stoop2024phd}. 

\begin{figure}
    \centering
    \includegraphics[trim={5mm 48mm 37mm 3mm},clip,width=\textwidth]{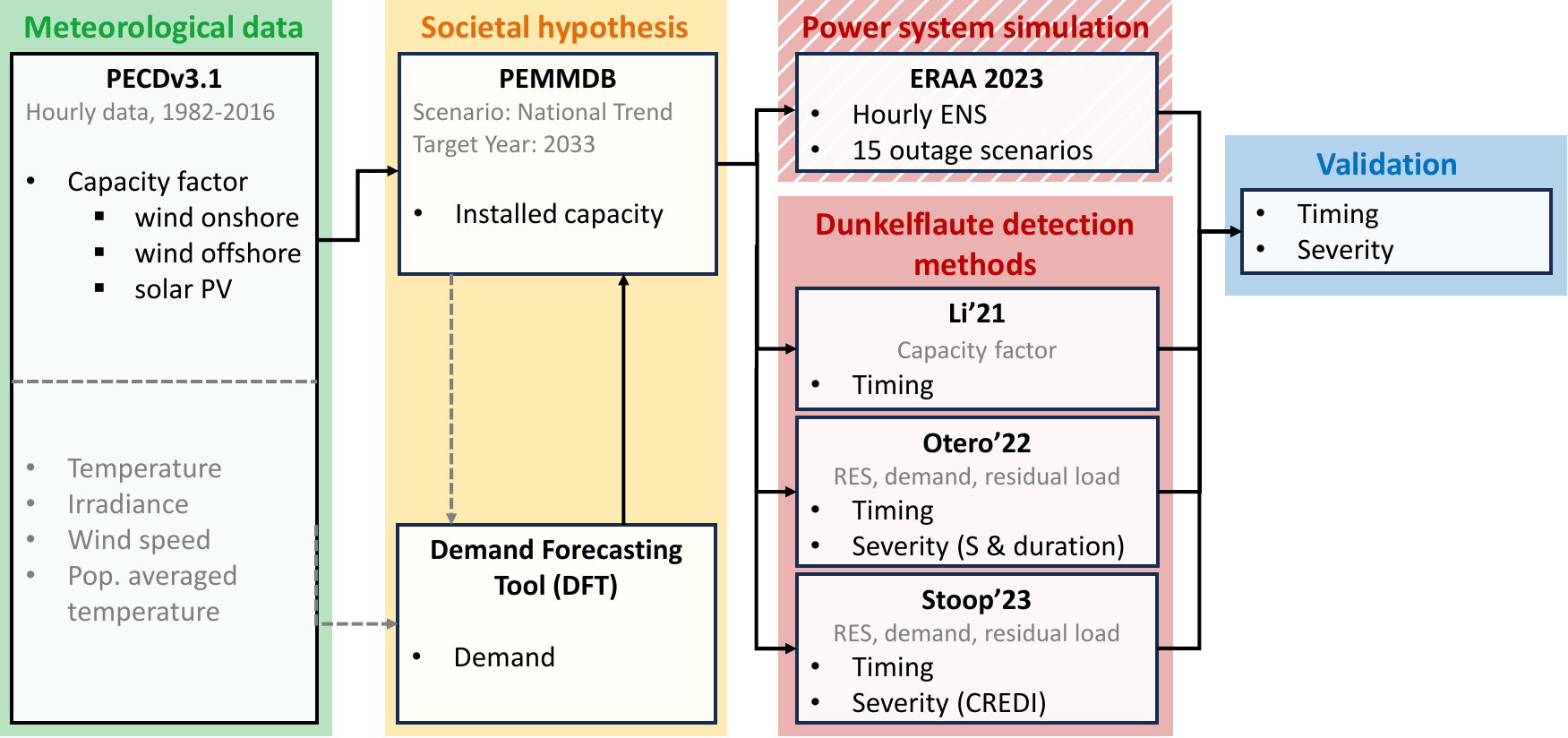}
    \caption{
    Flowchart of data sources used for the validation and comparison of various dunkelflaute identification methods.
    The same set of consistent meteorological data (PECDv3.1) and societal hypothesis (National Trends 2033 scenario) is used for the dunkelflaute detection methods investigated in this paper and the operational power system simulation (data downloaded from~\textcite{ERAA2023}).
    Further information on the specific methods used can be found in Section~\ref{sec:methods}. 
    For the data sources used, see Section~\ref{sec:data}.
    }
    \label{fig:flow_chart}
\end{figure}

Weather-driven periods of low electricity production from RES can result in so-called 'dunkelflauten'\footnote{
For clarity and consistency, and as the terminology changes within literature, we will use the generic term `dunkelflauten' (plural form of `dunkelflaute') to refer to all challenging events identified from weather data alone (even if defined with demand only), and to all events identified by power system simulations as `shortages'.
From a power system simulation perspective, dunkelflauten are based on input data (Fig.~\ref{fig:flow_chart}), while shortages are output-based.}, 
also referred to as 'energy droughts'. 
When these weather phenomena occur over a large geographical area for extended periods of time in a highly renewable (future) power system, they pose a risk of energy `shortage', i.e. electricity demand not met by supply. 
Coincidence of these periods with widespread cold temperatures (e.g. a 'kalte dunkelflaute'), and the associated increased electricity demand, might exacerbate this risk.
Being able to identify and assess both the severity and frequency of dunkelflaute events is crucial for monitoring resource adequacy, and planning a reliable future power system.

The most robust way of identifying potential energy shortages is to use detailed hourly power system simulations to assess resource adequacy\footnote{In Europe, the reference study is the European Resource Adequacy Assessment (ERAA)~\autocite{ERAA2023}, performed by the European Network of Transmission System Operators for Electricity (ENTSO-E).} -- so-called unit commitment and economic dispatch models.
These models capture the effect of weather variability on both the demand and supply by integrating different climate years.
Traditionally, only historical climate datasets such as reanalyses were available for the energy sector. 
In recent years new datasets have become available based on models of the future climate~\autocite{Dubus2022PECD,Dubus2023C3S}, thus including many more potential climate years.

However, power system simulations are computationally intensive and most Transmission System Operators (TSOs) can only consider a limited number of climate years (typically 30) in a given study. 
It is therefore not feasible to directly assess the impact of climate change on these shortages, or the climate change risk on a power system, as it would require much more climate years~\autocite{craig2022disconnect}.
Therefore, a robust way to identify (kalte) dunkelflauten from climate data alone is needed in order to identify challenging years or periods 
and correctly sample both typical and extreme weather situation, while keeping the simulations computationally tractable. 

In the literature, different methods are used to define dunkelflauten such as statistical methods which detect singular events by defining a threshold for renewable energy production (e.g.~\textcite{Kaspar2019}) or how much (net) load is covered by RES (e.g.~\textcite{Raynaud2018}), and those which use contextual information to identify a dunkelflaute (e.g.~\textcite{Ruhnau2022}).
While a comparison of the properties of various methods exist~\autocite{Kittel2024}, there is no consistent validation of the identified events with likely shortages as derived from power system simulations. 

In this paper we provide a short review of dunkelflaute detection methods, from which we select three methods of different complexity that align with operational requirements of grid operators.
Based on a dataset of potential shortages derived from the detailed hourly simulation used within the 2023 European Resource Adequacy Assessment (ERAA)~\autocite{ERAA2023}, the reference adequacy study within Europe (Figure~\ref{fig:flow_chart}), we address the following research questions:
do the dunkelflaute events detected by these methods correctly identify the potential energy shortages?
What are the respective strengths and weaknesses of these methods?
How do the energy variables (capacity factor, total RES production, demand) used in the definition of dunkelflaute help to detect the energy shortage?

The paper is structured as follows. 
In Section~\ref{sec:review}, we analyze the various detection methods and discuss their use.
In the Section~\ref{sec:methods} we describe the three selected methods and the procedure to evaluate them in terms of timing and severity of the dunkelflaute events.
In Section~\ref{sec:data}, we indicate the data used.
In Section~\ref{sec:results}, we present the results of our findings and compare the three methods. 
In Section~\ref{sec:disc} we discuss our methods and findings. 
Finally, in Section~\ref{sec:conc}, a synthesis of our findings is presented and potential use cases in research and/or operational applications are outlined. 
Supporting Information (SI) with additional figures and analysis is available online.

\section{Literature review}
\label{sec:review}

The literature identifies different (kalte) dunkelflauten, namely low RES production and low RES combined with high demand.
Low RES production are characterized by a fall in (renewable) energy production alone, but do not consider load.
These are mainly due to low wind speed and insolation and therefore often named 'dunkelflauten'.
Consequently, they may neglect periods with exceptionally high energy demand and vice versa exaggerate periods with very low demand.
Some works focus only on the capacity factor (CF, in \%), which is the ratio of actual production to nominal production, for one~\autocite{Ohlendorf2020} or multiple RES technologies~\autocite{Hu2023,Li2021,Kaspar2019,Kapica2024} (e.g. wind onshore, wind offshore, solar PV).
This allows to investigate the relationship between technologies, such as correlation of co-occurring low production events. 
If the installed capacities (in MW) are available, the production of each RES technology can be computed by multiplying the capacity factor by installed capacity, allowing to calculate the total RES production (in MW) or a composite capacity factor index~\autocite{Mockert2023}, which accounts for the relative contribution of each technology given by their respective installed capacity. 

Some methods also take load into account~\autocite{Mockert2023,Raynaud2018}.
In this case, the residual load (demand minus RES production) is generally used, as it can be interpreted as the required energy supply from other resources than wind and solar (e.g.~\textcite{vanderWiel2019,Otero2022,Allen2023}). 
Since the increased energy demand is often correlated with low temperatures, these events are sometimes also referred to as 'kalte dunkelflauten'. 
However, this term may be misleading, as in winter-time they often co-occur with an increased solar PV production~\autocite{vanderWiel2019}.

Regardless what type of dunkelflaute is considered, most methods can be divided either into having a \emph{statistics inspired} approach, that analyses the statistical nature of dunkelflaute events, or a \emph{context-informed} approach, that focuses on physical consistency and incorporates contextual information in the identification of a dunkelflaute.
Table~\ref{tab:literature} provides an overview of selected papers and the methods they present into these categories. 

\begin{table}[h]
    \centering
    \caption[Table]{Overview of selected methodologies for defining and analyzing dunkelflauten.}
    \label{tab:literature}
    \begin{tabular}{|c|c|c|}
    \hline
    \multicolumn{2}{|c|}{Statistical approach} &  Context-informed \\
    \cline{1-2}
    Low RES production & Low RES \& high demand &  approach \\
    \hline
    \textcite{Hu2023} & \textcite{Raynaud2018} & \textcite{Ruhnau2022} \\
    \textcite{Li2021} & \textcite{Allen2023} & \textcite{Scholz2023} \\
    \textcite{Kapica2024} & \textcite{Otero2022} & \textcite{Stoop2024} \\
    \textcite{Kaspar2019} & \textcite{vanderWiel2019} & \textcite{Antonini2024}  \\
    \textcite{Ohlendorf2020} & \textcite{Mockert2023} &  \\
    \textcite{Li2022} & \textcite{Kittel2024} & \\
    \hline
    \end{tabular}
\end{table}

\subsection{Statistical Approach}
The ``statistical'' group of approaches identifies distinct dunkelflaute events and assesses their properties by statistical methods. 

Most commonly a dunkelflaute is defined as a period of time where either RES generation~\autocite{Kapica2024,Kaspar2019,Ohlendorf2020,Mockert2023} or residual load~\autocite{Allen2023,Raynaud2018,Otero2022,vanderWiel2019} fall below or above a predefined threshold. 
The threshold can be either defined as an absolute value (in \% for capacity factor~\autocite{Kaspar2019,Li2021,Mockert2023,Ohlendorf2020}, or in MW for RES production, demand, or residual load) or from a statistical quantity derived from its underlying distribution such as quantiles~\autocite{Hu2023,Otero2022,Kapica2024}, daily mean production~\autocite{Raynaud2018}, or a return period (1-in-10 years occurrence probability, e.g. \textcite{vanderWiel2019}).
\textcite{Allen2023} define standardized indices facilitating comparison between regions and applicable for detecting both low RES production and high residual loads. 
These indices are calculated from a probability integral transform of the cumulative distribution function of either generation or residual load, which enables to use certain percentiles of a standard distribution as thresholds. 

Using thresholds to detect dunkelflaute events may result in the identification of two separate events near the threshold instead of a single prolonged event.
To prevent an over-detection of short events, a clustering of consecutive events might be done whenever the time in between is too short~\autocite{Otero2022}.
An alternative is to consider only events of at least a predefined minimum duration~\autocite{Mockert2023}. 
\textcite{Kittel2024} advocate for two methods which ensure that each identified event is unique, can be of any duration and pool together nearly consecutive events, namely the Variable-duration Mean Below Threshold (VMBT) introduced by the authors and the Sequent Peak Algorithm (SPA) originating from literature in hydrology.
The VMBT method identifies events whose average moving windows value exceed a threshold.
It then iteratively reduces the size of the moving window and eliminates events that exceed the threshold in the previous iteration to prevent overlap of events. 
The SPA algorithm calculates the cumulative value that surpasses a specified threshold until the cumulative value returns to zero. 
Subsequently, it defines the event as the interval between the initial point and the maximum cumulative value attained.
All these methods require to define a threshold beforehand.

Having identified distinct dunkelflaute events, main properties can be assessed, namely duration and frequency. 
Often duration is used as a measure of severity, but some methods~\autocite{Allen2023,Otero2022} calculate severity separately, e.g. as cumulative energy shortfall. 
If the size of the dataset permits, the distribution of these properties is assessed empirically. 
However, smaller datasets may lack a representation of High Impact, Low Probability events (``HILP'') and hence when using limited data, often more sophisticated methods (e.g. best fitting parametric distribution function by maximum likelihood estimation) are used to estimate the underlying distributions of duration (and severity). 
Copula-based methods are common to compute joint probabilities of dunkelflaute properties (e.g. duration and severity~\autocite{Otero2022}) or investigate the risk of compound events (e.g. co-occurring weather conditions related to low RES generation and high demand~\autocite{Tedesco2023}). 
Some studies also connect observed dunkelflaute events with weather patterns and identify underlying meteorological correlations~\autocite{Li2021,Li2022,Mockert2023,vanderWiel2019}.

A less prominent statistical approach is the use of Machine-Learning methods to identify dunkelflaute.
\textcite{Li2022} uses a convolutional neural network (WISRnet) to classify periods as dunkelflauten based only on spatially resolved meteorological data.

In conclusion, due to their simplicity, approaches based on thresholds are commonly used to investigate the statistical nature of energy dunkelflauten. 
As there is no need to pre-define a time-window of interest, properties as duration, frequency and severity can be easily obtained and compared between regions or in time. 
However, the simplicity comes with a lack of accuracy due to the harsh thresholding and assumptions must be made to define a threshold in the first place.

\subsection{Context-informed approach}
Besides assessing the statistics of dunkelflaute events, there is a smaller, second group of analysis methodologies, that is inspired by the physical nature of dunkelflaute events and focuses on their effect on power supply. 
These methods do not necessarily identify distinct events, but rather define a measure to indicate the impact of dunkelflaute events on the energy system.

One way to asses the impact of a dunkelflaute is by using an index, e.g. the Climatological renewable energy deviation index (CREDI), which is the cumulative anomaly of renewable energy potential (capacity factors) over a given time frame~\autocite{Stoop2024}. 
Hence, instead of defining events by thresholds and analyzing their duration consequently, here a time frame is chosen beforehand. 
When applying this index to time windows of a few days, it can also be used to identify distinct events.

Other methods aim directly to asses the additional energy demand caused by a dunkelflaute. 
The maximum energy deficit, introduced in \textcite{Ruhnau2022}, calculates the maximum of the integrated residual load of all possible time frames for a given duration.
Similarly, the method by \textcite{Scholz2023} uses rolling averages of residual load anomalies to evaluate the maximum additional energy demand over a given period of time.
However, methods based on annual maximum values~\autocite{Ruhnau2022,Antonini2024} prevent the detection of multiple potentially relevant events within the same year.

All of these methods can be used to identify and compare particularly challenging periods, although a duration of interest must be chosen beforehand. 
However, for further adequacy analysis the advantage of this group of methods is that they measure the impact of dunkelflaute events in the long-term and consequently can help assess the need for additional energy storage in a future renewable energy system.

\section{Selection of dunkelflaute detection approaches}
\label{sec:methods}
Based on the literature review, three dunkelflaute detection methods are selected. 
First, a brief description of each method is provided. 
The procedure for validating these methods is then presented.

\subsection{Selection of dunkelflaute detection approaches}
We select the dunkelflaute detection methods based on the type of approach, the increasing levels of data requirements (see Tab.~\ref{tab:required_data}), and the degree of complexity of their implementation.
The methods are described below, and we refer to the main papers for additional information.

\begin{table}[h]
    \centering
    \caption[Table]{Overview of the type of approach, dataset and time resolution required for each selected method.}
    \begin{tabular}{llccc}
    & & Li'21~\autocite{Li2021} & Otero'22~\autocite{Otero2022} & Stoop'23~\autocite{Stoop2024} \\
    \hline
    Approach & & Statistical & Statistical & Context-informed \\
    \hline
    & Capacity factor (CF) & Hourly & & \\
    Required data & CF + Installed capacity & & Daily & Hourly \\
    & Demand & & Daily & Hourly \\
    \hline
    \end{tabular}
    \label{tab:required_data}
\end{table}

\subsubsection{Li'21: Low capacity factor events}\label{sec:Li21}
The first method is based on~\textcite{Li2021} and defines a dunkelflaute using only capacity factors of three technologies: wind onshore, wind offshore, and solar photovoltaic.
A dunkelflaute is defined as a period where the capacity factor of each technology falls below the same pre-defined threshold value (e.g. 20\%) for more than 24 consecutive hours.

This method is rather simple and easy to implement when little information is available: no assumption on installed capacity of each technology is required, and the demand is not considered. 
The potential drawback lies in the detection of the shortages, as discussed in section~\ref{sec:results}.

\subsubsection{Otero'22: low RES, high demand and high residual load}
The second method from~\textcite{Otero2022} considers different types of dunkelflaute events: low RES production (the sum of wind onshore, offshore, and solar PV, in MW), high demand, and high residual load.
A comparison of these different energy variables makes it possible to assess how they each perform in the detection of energy shortages.
In contrast to the method of \textcite{Li2021}, here daily data is used instead of hourly data. 
All of these dunkelflaute events are defined by values exceeding a certain percentile of the distribution (e.g. the lowest 10th percentile of total RES daily generation, or the highest 10th percentile of demand or residual load). 
Consecutive days classified as dunkelflaute are treated as a single event but become distinct events if they are separated by at least two days. 
This ensures that a longer event caused by certain long-lasting meteorological conditions is not split into several smaller events, due to single days in-between crossing the threshold.

This definition provides timing properties for each dunkelflaute: when it starts and its durations. 
In addition, \textcite{Otero2022} define a measure of severity as the difference between the daily energy value $E_{i}$ (residual load or demand) and its corresponding threshold $E_{th}$, summed over all days $D$ of the event\footnote{
We use a slightly modified version of the original by \textcite{Otero2022}: the absolute value is removed so that days that temporarily fall below the threshold do not increase the severity value.
}:
\begin{align}
    S = \sum^D_{i=1} \frac{E_{i}-E_{th}}{\sigma}
    \label{eq:OteroSeverity}
\end{align}
For RES production, the opposite sign is used in Eq.~\eqref{eq:OteroSeverity} as for adequacy purpose we are also interested in events of low production.
To make the severity of different event types comparable, it is normalized by the standard deviation $\sigma$ of the daily energy value distribution.

\subsubsection{Stoop'23: Climatological renewable energy deviation index (CREDI)}
\label{sec:CREDI}
The third method follows a context-informed approach.
It is based on the Climatological renewable energy deviation index (CREDI) introduced by~\textcite{Stoop2024}.
CREDI is originally defined as the cumulative anomaly of capacity factor time series over a duration $T$.
This allows to study different events with a specific duration $T$ fixed beforehand.
In the present paper, we adapt the CREDI method for other energy variables: demand, RES production, and residual load.

As the goal is to compute how far the demand (resp. RES generation and residual load) is from its expected value, the latter must be correctly computed.
To this end, the expected value -- or climatology -- must be physical and smooth, as explained hereafter.
Indeed, the climatology should take into account all relevant timescales. 
For instance, we expect solar PV generation to be highest at solar noon and null during the night, and higher in July than in December (in the Northern Hemisphere).
Additionally, we expect the demand profile to be different in working days and week-ends.
Therefore, care must be taken to compute a climatology that is physical, i.e. that accounts for all these timescales (daily, weekly, and annual) to then correctly estimate the anomaly, i.e. the deviation of the signal from this climatology. 
However, a simple average at each ordinal hour over a few decades leads to random fluctuations due to the small sample size, as shown in the Supporting Information of~\textcite{Stoop2024}.

To smooth out these fluctuations in the climatology, we define an \emph{hourly and weekly rolling window} (HWRW).
This rolling window is adapted from the \emph{hourly rolling window} introduced in the original paper which accounts for the daily and annual cycles of capacity factor time series, but additionally accounts for the weekly cycle that is particularly relevant for demand and residual load.

The hourly and weekly rolling window climatology $C$ of an energy variable $P$ for hour-of-the-year $h$ is computed as:
\begin{align}
    C_P(h) &= \frac{1}{n} \sum_{y=1}^{n} \sum_{h'\in\{h+7\times 24d\}_{d=-\Delta}^{d=+\Delta}} \frac{P(y,h')}{2\Delta+1}, \label{eq:HWRW}\\
    h &= 1,\ldots,8760 \nonumber
\end{align}
where $h$ is the hour of the year, $n$ is the number of years, $\Delta$ is the half of the averaging window (in weeks), and $P(y,h')$ is the value of the energy variable for hour $h'$ of year $y$.

\begin{figure}
    \centering
    \includegraphics[width=\textwidth]{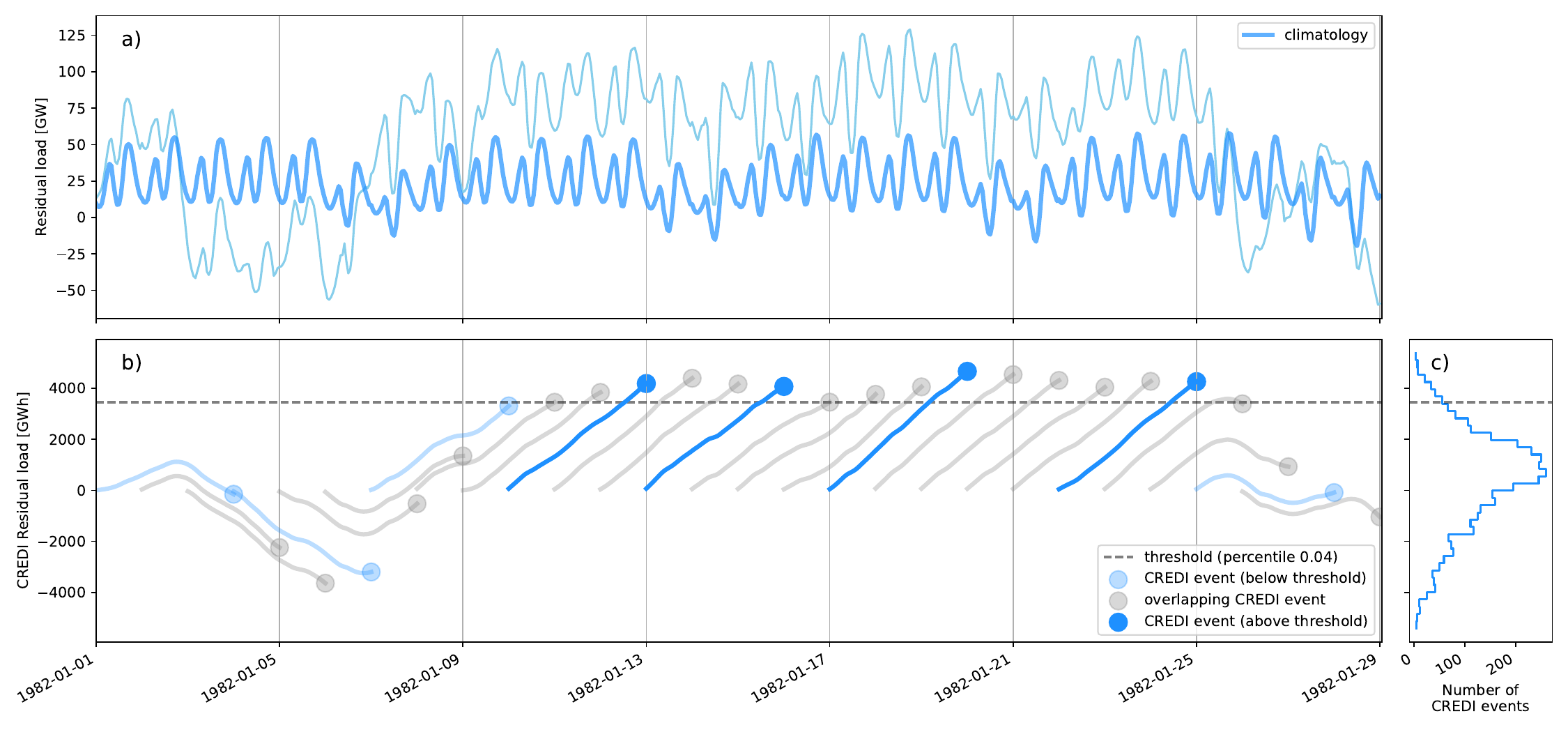}
    \caption{
    Panel \textbf{(a)} illustrates the residual load and the climatology for the first 4 weeks of 1982, in Germany.
    The climatology is computed with an hourly and weekly rolling window size of 9 weeks.
    The average daily and weekly cycles of residual load due to the respective demand cycles are visible.
    Panel \textbf{(b)} shows $T=3$-day CREDI events.
    CREDI is computed as the cumulative anomaly (difference between the residual load and its climatology) over $T=3$~days, at 24h interval.
    Overlapping CREDI events (in gray) are discarded by ranking the CREDI end value (circles) from highest to lowest.
    CREDI events with end value above a threshold are labeled as ``dunkelflaute'' events (dark blue).
    Histogram of residual load CREDI events for the 1982-2016 period is also shown \textbf{(c)}.
    }
    \label{fig:Illustration_CREDI}
\end{figure}

For example, by calculating the climatology for an hourly and weekly rolling window size of $2\Delta+1=3$ weeks, the climatology for Monday 8 January at 9:00 is the average of Mondays 1, 8 and 15 January at 9:00, for all $n$ years considered.
Figure~\ref{fig:Illustration_CREDI}a illustrates the climatology in the first 4 weeks of January.
In the following, an hourly and weekly rolling window of 9 weeks is selected as a trade-off to smooth out the random fluctuations caused by the limited sample size while preserving the seasonal cycle (see Section~\ref{sec:Stoop_HWRW} in Supporting Information).

CREDI is then computed as the cumulative anomaly, starting at ordinal hours $t_i$ of year $y$ over a time period $T$:
\begin{equation}
    \mathrm{CREDI}(y, t_i, t) = \sum_{h=t_i}^{t} \left( P(y,h) - C_P(h) \right), \quad t\in[t_i, t_i+T].
    \label{eq:CREDI}
\end{equation}
We define $T$-day CREDI events of an energy variable by computing the CREDI index starting at 00:00 am and lasting for $T$ days, as illustrated in Figure~\ref{fig:Illustration_CREDI}b.
The CREDI index increases when the energy variable is above its climatology and decreases when it is below.
The overlapping events with the lowest CREDI end value are discarded.
An overlap of up to $25\%$ is accepted (e.g. 1 day overlap for $T=4$ day events).
CREDI events are labeled as a dunkelflaute events if their CREDI end values are above a threshold, defined as a percentile of the distribution of CREDI values (Fig.~\ref{fig:Illustration_CREDI}c).

\subsection{Validation process}
\label{sec:analysis}

Dunkelflaute detection methods are validated against a dataset of energy shortage derived from detailed hourly power system simulation.
Specifically, we use an hourly dataset of Energy Not Served (ENS), which is the amount of electricity demand (in MWh) that is not met by all production means, including RES production.
The ENS dataset is further detailed in Sec.~\ref{sec:data}.

Two aspects of the dunkelflaute detection methods are validated, namely (i) \emph{time accuracy} and (ii) the assessment of \emph{severity}.
To measure the predictive performance of time accuracy, we use the F-score~\autocite{Christen2023}:
\begin{align}
    F = \frac{ 2 \mathrm{TP}}{2\mathrm{TP} + \mathrm{FN} + \mathrm{FP}}
    \label{eq:F-score}
\end{align}
where $F=1$ is the best prediction and $F=0$ the worst.
Every day is marked either as a true positive (``TP'', i.e. dunkelflaute detected and positive ENS), false positive (``FP'', i.e. dunkelflaute detected, but no ENS), false negative (``FN'', i.e. no dunkelflaute detected, but positive ENS) or a true negative (``TN'', i.e. neither dunkelflaute, nor ENS).

Each day is labeled as a dunkelflaute or shortage as follow.
A day is marked as a shortage if it experiences ENS for at least one hour, i.e. if the sum of the hourly ENS is positive. 
As the three detection methods use different time resolution (hourly data for Li'21, daily data for Otero'22, hourly data to compute CREDI events at daily interval), the following choice is made to mark each day as a detected dunkelflaute or not.
For Li'21, a day is labeled 'dunkelflaute' if at least 12h of that day overlap belong to a dunkelflaute.
For Otero'22, a day is simply labeled as dunkelflaute if the daily RES production (resp. demand or residual load) is above (resp. below) a threshold.
For Stoop'23, a day is labeled as dunkelflaute if it belongs to a CREDI event with a value above a threshold.

For all three methods, dunkelflaute events are defined based on a threshold value: a capacity factor threshold (the same for all three technologies) for Li'21; a threshold based on the percentile of the daily distribution of demand, renewable generation, or residual load for Otero'22; and a threshold based on the percentile of the distribution of CREDI events for Stoop'23. 
As the choice of threshold is arbitrary, we compute the F-score on a range of thresholds.
This allows to obtain the best-fitting threshold that maximize the F-score, or in other words, the threshold value for which the detection methods reproduce the timing of ENS events the best. 

The severity assessment evaluates how close the method is in assessing the energy shortfall caused by an ENS event.
The best-fitting threshold value is selected to compare the severity of the detected events ('Severity' defined in Eq.~\eqref{eq:OteroSeverity} for Otero'22, CREDI value for Stoop'23, no severity index is defined for Li'21) with the total ENS, i.e. the sum of hourly ENS values during these events. 
Two metrics are computed : the Pearson correlation coefficient $r$, which measures the linear relationship between the severity index and the total ENS; and the Spearman's rank correlation coefficient $\rho$, which assesses the monotonic relationship between two variables.
The latter metric adds additional relevant information as it measures whether the severity index correctly ranks the dunkelflaute events from most extreme to least extreme, even if the relationship is not linear.
Spearman's coefficient ranges from $\rho=-1$ for a decreasing monotonic relationship to $\rho=1$ for an increasing monotonic relationship.

\section{Data}
\label{sec:data}

In this section, we describe the different datasets used in this study. 
For reference, a flowchart of the data is provided in Fig.~\ref{fig:flow_chart}. 
The Pan-European Climate Database (PECD) provides the wind and solar capacity factors, which are multiplied by scenario-specific installed capacities to get the generation time series. 
The demand time series also include both meteorological components of the PECD and the scenario-specific inputs.
The European Resource Adequacy Assessment (ERAA) power system simulation uses the generation and demand time series and other scenario specifics to simulate an optimal power system under 15 forced outage scenarios.

\subsection{PECD}
The PECD version 3.1 provides the meteorological and potential renewable generation data.
The PECD is the standard climate dataset used for studies of the European Network of Transmission System Operators for Electricity (ENTSO-E), such as the European Resource Adequacy Assessment (ERAA).
While a newer version of the PECD is available (PECDv4.1)~\autocite{Dubus2023C3S}, the PECDv3.1 is used for consistency as that version was used within the ERAA 2023 study which provides the shortage data used for validation~\autocite{ERAA2023}.

The PECDv3.1 is derived from the ERA5 reanalysis product~\autocite{Hersbach2020}, which is a state-of-the-art reconstruction of the past state of the atmosphere combining observation and weather model simulations.
The PECD provides capacity factor for renewable energy sources (RES), along with climate variables such as temperature, wind speed and solar radiation.
The capacity factors are calculated based on existing wind and solar technologies~\autocite{Dubus2022PECD,ERAA2023}.
The climate and energy variables are aggregated at sub-national levels, and available at hourly resolution over the 1982-2016 period.

\subsection{PEMMDB capacities and demand}
The installed capacities (in MW) of wind onshore, wind offshore, and solar PV, in all sub-national zones are provided by the Transmission System Operators (TSOs) via the Pan-European Market Modelling Database (PEMMDB)~\autocite{ERAA2023}.
These capacities correspond to the TSOs' estimations based on current national trends and future targets (2025, 2028, 2030, 2033).
In this paper we use the installed capacity of the target year 2033, which is the year with the highest installed renewable capacity. 

The solar PV, wind onshore, and wind offshore generation are then computed by multiplying the capacity factor by their respective installed capacity.
For each technology, the national generation is the sum of the generation of all sub-national regions.
We compute the total RES production as the sum of all three technologies.
For Li'21 (see Section~\ref{sec:Li21}), the capacity factor at national scale for each technology is the average of sub-national regions weighted by their respective installed capacity.

Hourly demand time series for each country are calculated with a Demand Forecasting Tool (DFT)~\autocite{ERAA2023}.
The demand depends on consumption assumptions based on the same National Trend scenario at the 2033 horizon, as well as meteorological variables (temperature, wind speed and solar radiation).

\subsection{Energy Not Served (ENS) dataset}
The dunkelflaute metrics are tested against a reference dataset of Energy Not Served (ENS), which is the amount of demand that is not met by supply, expressed in MWh.

We use the ENS derived from the European Resource Adequacy Assessment (ERAA) 2023~\autocite{ERAA2023}, again for the target year 2033 and the national trends scenario.
ERAA studies assess the security of electricity supply for European consumers.
They consist of a probabilistic assessment of whether the power system capacities planned in the next decade (thermal power plants, wind and solar capacities, transmission line capacities, etc.) will be able to meet electricity demand at any hour, even in potentially adverse situations. 

A set of Monte Carlo scenarios are used to test the adequacy of the power system.
In the ERAA 2023, these Monte Carlo scenarios are a combination of weather scenarios (35 historical climate years from the PECDv3.1 dataset) and 15 outages scenarios consisting of random outages of thermal plants and power lines.

In this paper, we consider the days with ENS as the ``true'' dunkelflaute events, which we try to detect with dunkelflaute metrics derived from weather-driven energy variables (wind and solar capacity factor, renewable energy production, demand and residual load).
We emphasize that we use the same energy variable dataset (PECDv3.1) as that used to calculate the ENS dataset in the ERAA 2023.
The 15 outages scenarios are used to assess the robustness of the detection of weather-related energy shortages despite varying random outage situations. 

Additionally, 2 Economic Viability Assessment (EVA) scenarios (scenarios A and B) are available in the ENS dataset.
While this paper only considers the EVA scenario B, the results for scenario A are also discussed in the SI (Fig.~\ref{fig:Otero_EVA}).

\section{Results}
\label{sec:results}
In this section, we present the results of the dunkelflaute detection method validation.
We first validate each method individually by evaluating its ability to assess the timing and severity of potential shortages. 
Then, we compare the three methods and draw conclusions on their potential strengths and weaknesses.
For all methods, we focus on Germany\footnote{In the PECD, the regions correspond to areas relevant for the electricity market, called bidding zone. Here, Germany corresponds to the bidding zone `DE00'.} as an illustrative case.
We discuss the results for France and a larger European region in section~\ref{sec:Comparison_region} and refer to the Supporting Information~\ref{sec:Otero_other_countries} \& \ref{sec:Stoop_other_countries} for the results on other countries.

\subsection{Validation of dunkelflaute detection methods}

\subsubsection{Li'21}
\begin{figure}
    \centering
    \includegraphics[width=0.55\textwidth]{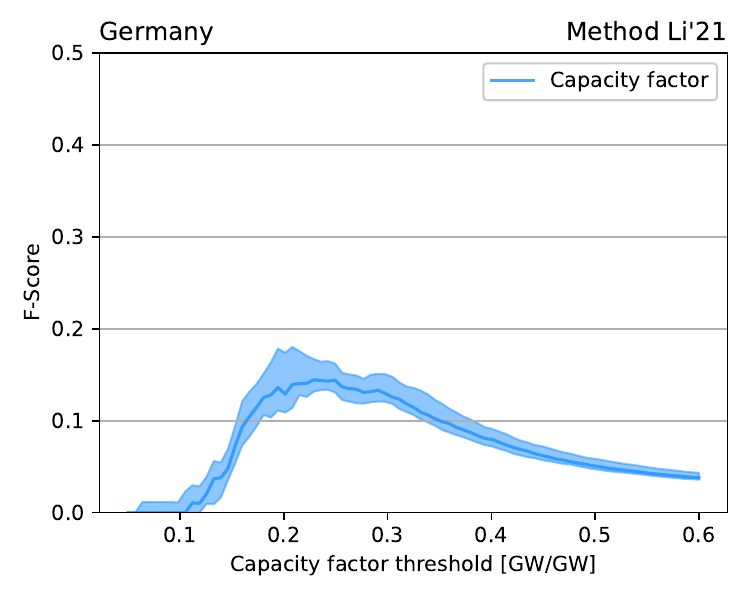}
    \caption{
    F-score for low capacity factor events (based on~\textcite{Li2021}) for Germany.
    A low capacity factor event occur when the capacity factor of the all three technologies (solar PV, onshore and offshore wind) fall below the same threshold value for at least 24 consecutive hours. 
    Uncertainty is evaluated with the 15 outage scenarios.
    The solid line is the median F-score value, while filled areas shows the minimum and maximum F-score values.
    }
    \label{fig:Li_DE00}
\end{figure}

Figure~\ref{fig:Li_DE00} shows the F-score as a function of the capacity factor threshold value for Li'21.
The F-score peaks at a threshold of 23\%, meaning that daily shortages are best detected when the capacity factor of all three RES technologies (solar PV, onshore and offshore wind) falls below 23\% for 24 consecutive hours.
The maximum F-score value of 0.14 shows the detection is relatively poor\footnote{$F=0.14$ implies that $12 \mathrm{TP} \approx \mathrm{FP} + \mathrm{FN}$ according to Eq.~\eqref{eq:F-score}, which means that for one correct prediction the method makes about 12 mistakes, either FP or FN.}.
The result is robust to the scenarios of random forced outages of thermal plants and power lines, with a relative change of about 10\%.
Although we focus here on Germany, the same conclusions can be drawn looking at other countries where relatively low F-score values are also found.
We also performed a filtering of non-weather related ENS events by only considering positive ENS that occur in a given number of outage scenarios (see Section~\ref{sec:SI_filter}).
However, it only slightly improves the F-score. 
This illustrates that only considering capacity factors of RES is likely insufficient to detect the occurrence of shortages.

\subsubsection{Otero'22}
\begin{figure}
    \centering
    \includegraphics[width=0.55\textwidth]{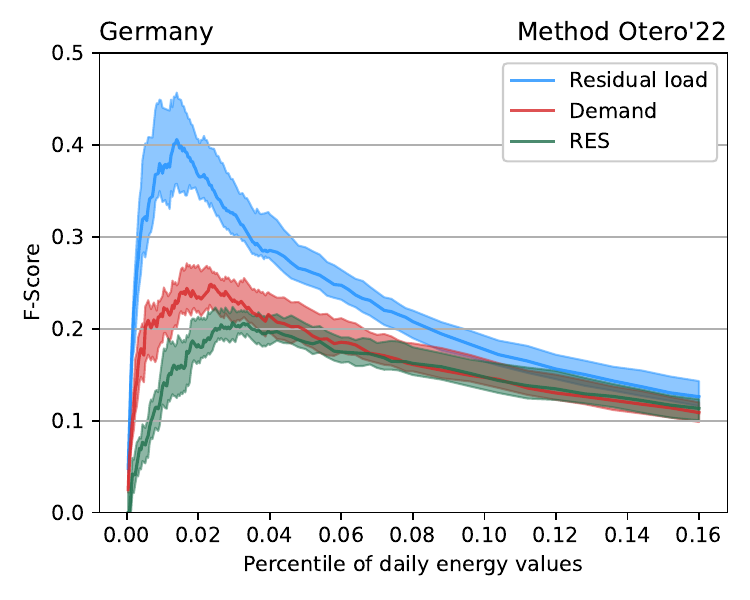}
    \includegraphics[width=0.45\textwidth]{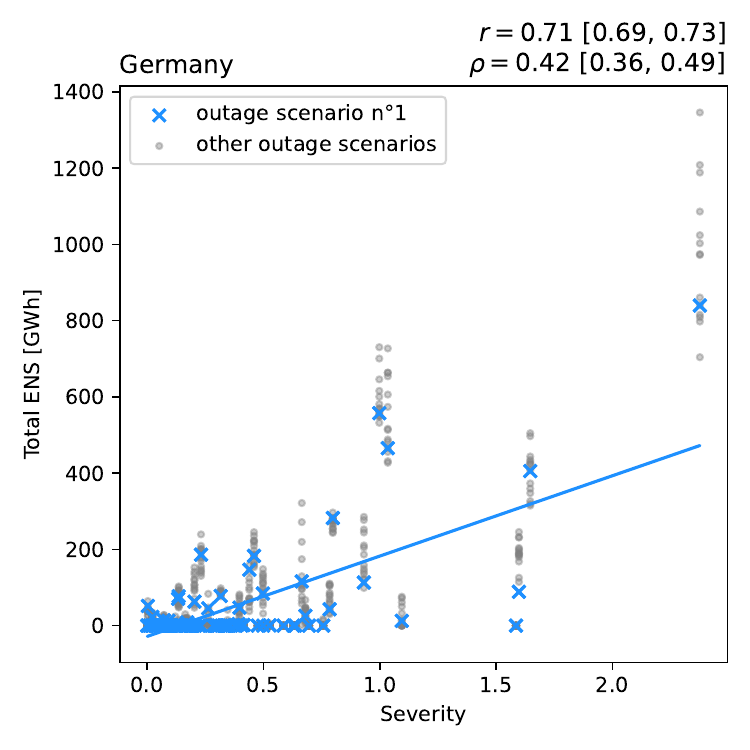}
    \includegraphics[width=0.45\textwidth]{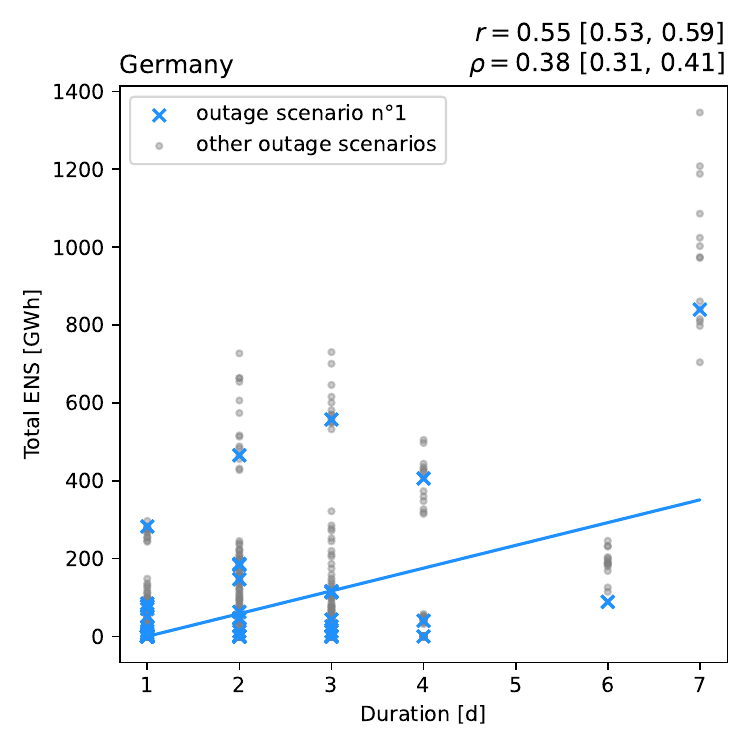}
    \caption{
    Top panel shows the F-score for dunkelfauten (based on~\textcite{Otero2022}) for Germany.
    Dunkelflauten are calculated based on residual load (blue), demand (red) and RES generation (green).
    Uncertainty is evaluated with the 15 outage scenarios as in Fig.~\ref{fig:Li_DE00}.
    Bottom left and bottom right panels show, respectively, the correlation between the total ENS during a dunkelflaute with the severity (see Eq.~\eqref{eq:OteroSeverity}) and the duration of the dunkelflaute.
    The total ENS is shown for all outage scenarios (the first scenario is shown with blue crosses, while the others are shown with gray dots). 
    The blue line shows the linear regression for the first outage scenario.
    The median value of the Pearson correlation coefficient $r$ and Spearman's rank correlation coefficient $\rho$ are indicated on top of each panel, with the minimum and maximum values from all of the outage scenarios shown in brackets.
    }
    \label{fig:Otero_DE00}
\end{figure}

Figure~\ref{fig:Otero_DE00}a shows the F-score for dunkelflaute events computed based on three energy variables: RES production, demand, and residual load.
The thresholds used to define each type of dunkelfaute event correspond to the percentile of their underlying distribution.

All types of dunkelflaute events show some shortage detection skill.
Both low RES and high demand have modest F-score values, with a maximum of about 0.21 and 0.25, respectively.
The best performing energy variable to detect the occurrence of ENS is residual load, which combines the information on demand and RES production.
The F-score peaks at 0.41, when the top 1.4\% events are considered as dunkelflaute events.
This illustrates that ENS occur more frequently when both demand is high and RES production low, rather than just one or the other.
Accounting for both RES generation and demand is therefore important for dunkelflaute detection.

Nonetheless, even based on the residual load, the timing detection skill remains limited\footnote{$F=0.4$ implies that $3 \mathrm{TP} = \mathrm{FP} + \mathrm{FN}$, meaning that for one correct prediction the method makes 3 mistakes.}.
The majority of shortages does not correspond to the detected dunkelflaute events. 
We will discuss this point in more detail in the Sec.~\ref{sec:disc}.

We now consider high residual load events with the 1.4\% threshold maximizing the F-score.
Figure~\ref{fig:Otero_DE00}b compares the value of the total ENS of these events with their severity $S$ (see Eq.~\eqref{eq:OteroSeverity}). 
Looking first at a single outage scenario (blue crosses), the dunkelflaute events with high severity tend to be associated with high total ENS, while low severity is generally associated with low total ENS.
However, the relationship is not linear (Pearson coefficient has a median value of $r=0.71$) and even less so monotonic (the median Spearman coefficient is $\rho=0.42$).
This is illustrated by the fact that some dunkelflaute events with a relatively high severity have a low (if not null) total ENS. 
The dispersion of total ENS values from one outage scenario to another illustrates that the depth of the shortage strongly depends on the outages experienced by the power system, with a total ENS that almost doubles for the strongest event from about 700~GWh to nearly 1400~GWh.
Still, as the severity increases, the total ENS tends to increase, as indicated by the positive signs of Pearson and Spearman coefficients.

Similar conclusions can be drawn when comparing the total ENS with the duration of dunkelflaute events (Fig.~\ref{fig:Otero_DE00}c).
Long events tend to be associated with high total ENS, but the relationship is less linear or monotonic than for the severity metric (with median coefficient values of $r=0.55$ and $\rho=0.38$).

\subsubsection{Stoop'23}
\begin{figure}
    \centering
    \includegraphics[width=0.55\textwidth]{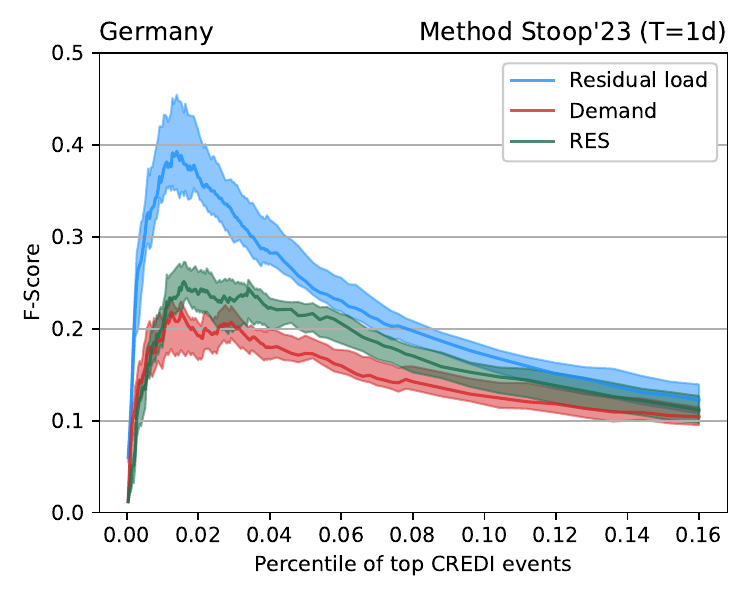}
    \includegraphics[width=0.44\textwidth]{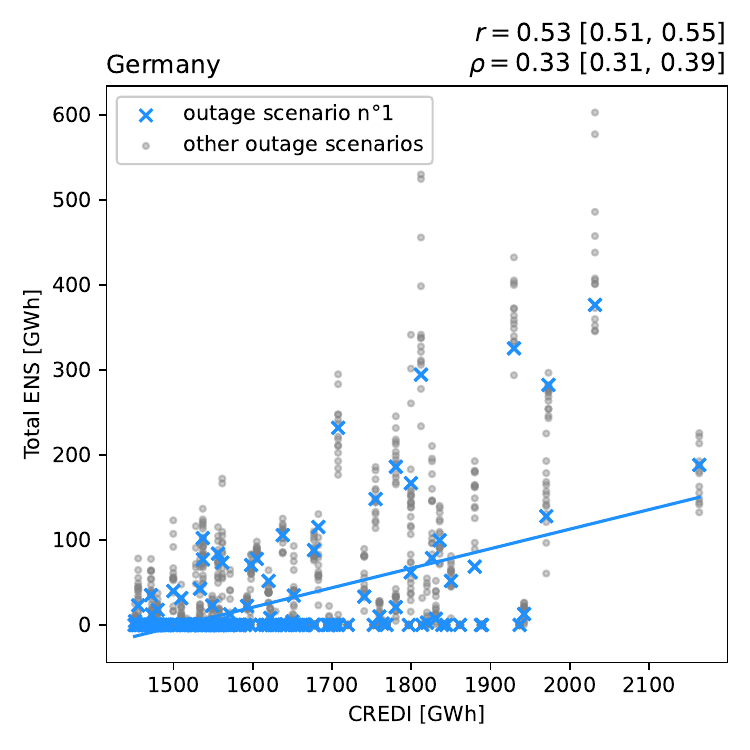}
    \caption{
    Left panel shows the F-score for $T=1$~day CREDI events (based on~\textcite{Stoop2024}) for Germany.
    CREDI events are calculated based on residual load (blue), demand (red) and renewable generation (green).
    Uncertainty is evaluated as in Fig.~\ref{fig:Li_DE00}.
    Right panel shows the correlation between the total ENS value of a CREDI event and its CREDI value, similarly to Fig.~\ref{fig:Otero_DE00}.
    }
    \label{fig:Stoop_DE00_T1}
\end{figure}

The timing detection skill for dunkelflaute events computed for daily CREDI method ($T=1$~day, see Eq.~\eqref{eq:CREDI}) is shown in Fig~\ref{fig:Stoop_DE00_T1}.

While both RES and demand show some detection skill (with a maximum F-score of about 0.25 and 0.22, respectively), the residual load is again the best energy variable for dunkelflaute detection, with $F=0.39$ for the top 1.4\% CREDI events.

Looking at the severity of dunkelflaute events, we compare the total ENS of $T=1$~day events with their end CREDI value.
Similarly to Otero'21, high CREDI values tend to be associated with high summed-up ENS. 
However, the linearity and monotonicity is relatively weak ($r=0.53$ and $\rho=0.33$).

Here, we have fixed the duration $T=1$~day.
We could expect adverse weather situations lasting for a long time to be better associated with the outage intensity.
However, this is not the case: correlation plots for $T=3$ days and $T=5$ days are shown in Figure~\ref{fig:Stoop_corr_T} for several regions. 
The correlations between CREDI values and total ENS are lower (for both $r$ and $\rho$) for longer events than for daily events.

\subsection{Comparison of KDF methods}

\subsubsection{Timing and severity of events}
The methods Otero'22 and Stoop'23, with a $F$-score of about $0.4$ for dunkelflaute events computed based on residual load, clearly outperform Li'21 ($F=0.14$).
The fact that Otero'22 and Stoop'23 yield similar results despite the different approaches (absolute residual load value vs. difference with a climate normal, free vs. fixed event duration, etc.) shows that the specific details of the definition are not so important compared to taking the best available energy variable, here the residual load.

Otero'22 demonstrates a slightly better performance at predicting the severity of dunkelflaute events: its metrics (duration and particularly severity, see Fig.~\ref{fig:Otero_DE00}b-c) correlate better with the total ENS than the CREDI value of Stoop'23.

In case only data of the RES production is available, considering not only the capacity factor but also the installed capacity of each technology also yields a better detection skill. 
This is shown by the higher F-score values for low RES in both Otero'22 and Stoop'23 (green curves in Fig.~\ref{fig:Otero_DE00}a and Fig.~\ref{fig:Stoop_DE00_T1}a) compared to Li'21 (Fig.~\ref{fig:Li_DE00}).
Although taking the same threshold for all three technologies, as done in Li'21, allows for a more general analysis without assumptions on installed capacity, this approach implicitly assumes that their relative share of capacity is fixed, thus assuming a different RES production profile than the actual one, which leads to lower detection skill.

\subsection{Comparison of different regions}
\label{sec:Comparison_region}
\begin{figure}
    \centering
    \includegraphics[width=0.49\textwidth]{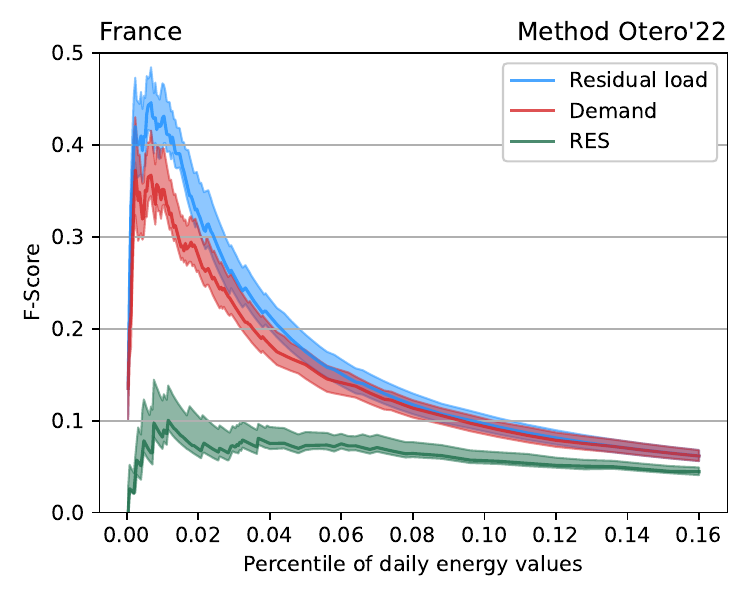}
    \includegraphics[width=0.49\textwidth]{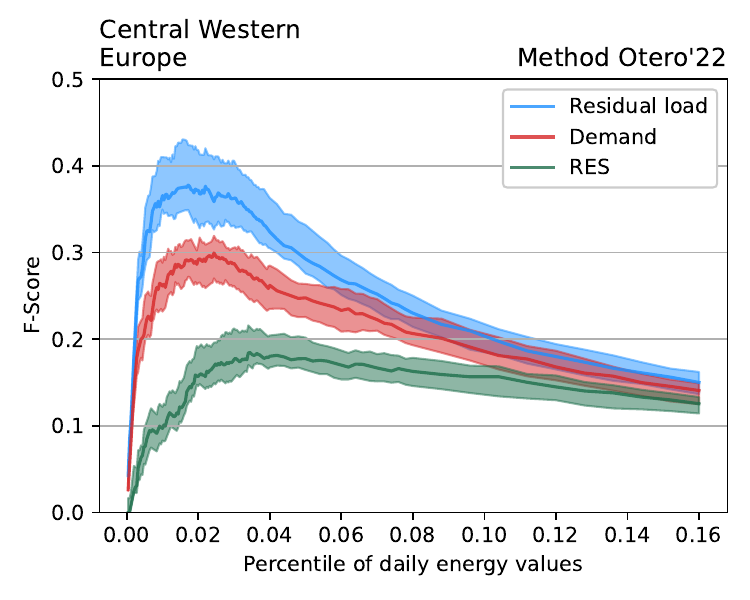}
    \caption{
    Same as Fig.~\ref{fig:Otero_DE00}, for France (left panel) and Central Western Europe (right panel).
    }
    \label{fig:Otero_FR_CWE}
\end{figure}

Although we focus on Germany, the same analysis was carried out on other countries.
Figure.~\ref{fig:Otero_FR_CWE} shows the F-score computed with Otero'22 for France and for the wider Central Western Europe\footnote{Here, 'France' region means continental France without Corsica. The Central Western Europe region includes Austria, Belgium, Switzerland, Germany, France, and the Netherlands.}.  
For analysis at this larger spatial scale, the demand, RES production, and ENS of all countries are summed-up.
The results for other countries are shown in the Supporting Information (Fig.~\ref{fig:Otero_other_countries}).

Interestingly, the F-score curves for France and Germany (Fig.~\ref{fig:Otero_DE00}a) highlight the different structures of the power system.
In France, the electricity demand is historically high compared to other European countries, with a strong dependence on outdoor temperature due to the high share of electric heating.
The F-score for high demand events is therefore significantly higher ($F=0.38$) in France than in Germany ($0.22$).
On the contrary, in Germany, the shortages are more induced by RES production than in France due to the high share of RES production (88\% of the annual demand is met by RES in this 2033 power system, vs. 27\% in France).
At a larger European scale, shortages are better detected using demand than RES production. 

In all cases, the residual load remains the best energy variable for ENS detection, even when the share of RES production is relatively low.
The same result holds for Stoop'23 (Fig.~\ref{fig:Stoop_other_countries}).
This further stresses the need to consider both RES generation and demand when studying shortages events.

\subsection{Climate years selection}
\label{sec:year_selection}

Detecting dunkelflaute events based on meteorological data allows to identify \emph{a priori} challenging periods for the power system, prior to perform power system simulations.
Given the fact that power system simulations are very computationally expensive, one possible use case is to identify which years are the most challenging for the power system, and focus the simulations and analyses on these years.

\begin{figure}
    \centering
    \includegraphics[width=0.49\textwidth]{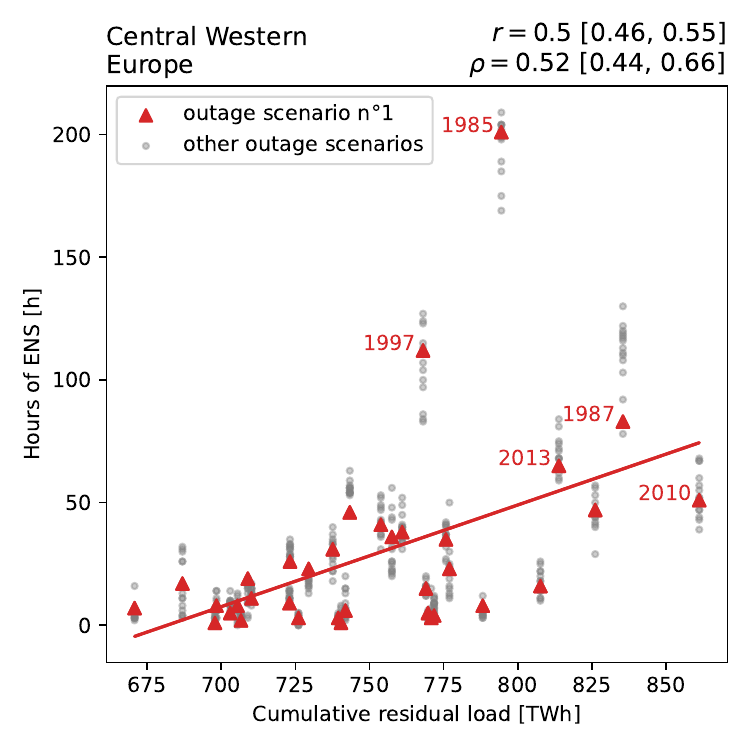}
    \includegraphics[width=0.49\textwidth]{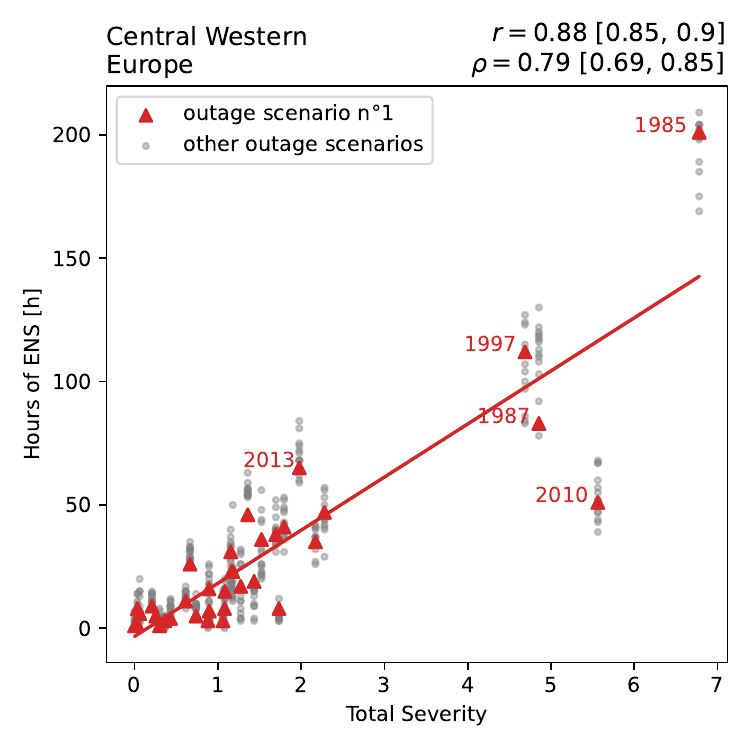}
    \caption{
    Correlation between the number of hours of ENS per year and the cumulative residual load of each year (left panel), or the total severity of all dunkelflaute occurring each year using Otero'22 (right panel).
    }
    \label{fig:yeark_ranking}
\end{figure}

To allow such a selection at a large European scale, a simple approach is to compute, for each year, the total cumulative residual load of several countries.
Figure~\ref{fig:yeark_ranking}a compares the hours of ENS per year with the cumulative residual load for all years, for the Central Western Europe region.
Years with a large number of ENS hours are considered to be challenging from an adequacy perspective, while
years with high cumulative residual load are expected to be a good proxy to identify challenging years.
Interestingly, over the wide region Central Western Europe, the years 1985, 1987, and 1997 stand out as the most extreme in terms of annual hours of ENS, despite the variability between outage scenarios.

However, the cumulative residual load is relatively poorly correlated with the hours of ENS, with a median correlation coefficient $r=0.5$.
It also has a poor skill in ranking the climate years from the most extreme to the least extreme ($\rho=0.52$).
For instance, while 1985 stands out as challenging year for shortage events, it is only ranked as the 6th year out of 35 based on cumulative residual load.

An alternative is to use a metric based on a dunkelflaute detection method.
Figure~\ref{fig:yeark_ranking}b uses the total severity, which is calculated as the sum of the severity (see Eq.~\eqref{eq:OteroSeverity}) of all dunkelflaute events detected per year with the method Otero'22.
This approach improves the correlation ($r=0.88$) and ranking skill ($\rho=0.79$) of the challenging years, and correctly identifies 1985, 1987 and 1997 as the most challenging years.
Similar results are also found for Stoop'23 (see Fig.~\ref{fig:year_rank_Stoop}).

Thus, dunkelflaute detection methods such as Otero'22 and Stoop'23 can be used to identify \emph{a priori} challenging years more robustly than a simpler approach that calculates the cumulative residual load.

\section{Discussion}
\label{sec:disc}

The validation of different dunkelflaute detection methods against a reference ENS dataset was a first and necessary step to understand the relationship between dunkelflaute events, defined solely based on time series of RES production and demand, and shortage events computed with complex power system models. 
A potential use case for these methods is to identify the most extreme dunkelflaute events, or most challenging climate years, which could then be used to stress test the resilience of future power systems. The methods could also be used for selecting representative climate years for performing more detailed simulations
Even though the relationship between ENS and severity is not perfectly monotonic, we find that the highest total ENS events are associated with high dunkelflaute severity values, whether for Otero'22 (Fig.~\ref{fig:Otero_DE00}) or Stoop'23 (Fig.~\ref{fig:Stoop_DE00_T1}), thus providing some skill for detection of the most extreme events.

Nevertheless, as shown by the relatively low F-score values (e.g. 0.41 at best in Germany), we find a significant mismatch in timing between identified dunkelflaute and simulated shortages, and that the dunkelflaute detection methodologies are not able to identify all the ENS events from the detailed simulations. 
There are several possible reasons for these relatively low F-scores.
Firstly, generation from run-of-river hydroelectric power plants is another weather driven parameter which was not included in the RES production  time series, and including this may lead to better F-Scores.
Secondly, the power system scenarios also include significant non-RES dispatchable capacity such as thermal power plants, transmission and storage, which can supply demand even during dunkelflaute events, and impact the timing and severity of shortages. For example, storage technologies such hydropower reservoirs can store energy from RES in periods of surplus, and use this to meet demand at a later point in time. 
Especially the long-term annual optimisation of hydro storage (with perfect foresight) performed in the power system simulation can mean the impact of dunkelflauten is not seen directly in ENS results, as the model optimised storage levels across a year to minimise ENS as much as possible. 
In terms of thermal plants, we show in the Supplementary Material (Figure~\ref{fig:Otero_EVA}) that in a scenario with more dispatchable capacity available, the F-Score of all three dunkelflaute detection methods declines due to a lower dependence on weather-driven supply, and higher impact of non-weather dependent ENS causes, such as unplanned outages. 
Lastly, the ENS events resulting from power system simulations may be to some extent arbitrary in time and location. 
For example, in a system with significant storage capacity and uniform (i.e. time and location invariant) value of lost load, the model can assign ENS in different ways at the same overall cost, leading to multiple potential solutions (e.g. gathering ENS at a single time step or spanning ENS over multiple time steps). 
Moreover, during periods of simultaneous shortage in several countries, which is not uncommon due to high spatial correlation of RES generation in nearby regions, the model has to decide in which country or countries to allocate the ENS. 
Depending on the formulation of the model this may be an arbitrary decision, unless specific rules are implemented in the model to determine how ENS is shared between multiple connected countries. 
In ERAA2023, the so-called 'local matching' and 'curtailment sharing' rules are applied which essentially (i) prevent countries with sufficient domestic capacity from 'exporting' so much that they  experience ENS, and (ii) countries with simultaneous ENS  ‘fairly’ share this load curtailment between them.

While beyond the scope of this study, further work could be performed to better understand the reasons for the low F-scores, and try to improve them. For example, more in-depth analyses could be performed with additional information about the outage scenarios, as the knowledge of which thermal plants or transmission line is unavailable could also help discriminate the types of ENS and focus the analysis only on 'weather-induced' ENS events. Including run-of-river and hydro storage in the analysis would also likely improve the results, but this would not longer be purely on 'dunkelflaute' type events. Lastly,  the arbitrariness of the ENS allocation could potentially be circumvented by using hours (days) with high electricity prices instead of the ENS for the validation. However, using electricity prices may reduce the F-score if it leads to a higher increase in false negatives than reduction in false positives.

The recently released PECD4.1\footnote{PECD version 4.1 is openly available on the Copernicus Climate Data Store \href{https://doi.org/10.24381/cds.f323c5ec}{https://cds.climate.copernicus.eu}.}, which currently includes historical (1980-2021) and projection data (2015-2065) with 3 climate models, will be used in the 2025 ERAA study.
This new dataset will allow to study the impact of climate change on dunkelflaute events in the future EU power system.
The methods validated in this paper could be used to assess these changes \emph{ex ante}, and identify the most stressful situations to use as inputs in computationally-intensive power system models.


\section{Conclusion}
\label{sec:conc}

Detecting the most challenging weather conditions for the power system is of utmost importance to ensure the adequacy of the future power systems.
To balance the risk of low-probability events, and account for future changing climate conditions, a large amount of climate data must be considered.
However, current power system simulations are very computationally intensive, limiting the quantity of data they can handle.
Being able to robustly identify challenging adequacy situations from climate (and load) data alone would thus be of significant practical use for TSOs and other resource adequacy analysts.

To this end, we have reviewed several methods that define challenging weather-induced events for the power system, referred to in this paper under the generic term of "dunkelflaute". 
We selected three dunkelflaute detection methods with different approaches and data requirements, and validated their ability to detect the timing and severity of energy shortages using energy-not-served (ENS) results from the ERAA 2023 simulations.
We show that while these methods have some ability to identify potential energy shortages, their absolute skill is limited and other factors such as the impact of storage (battery capacities or hydro storage levels), forced outages, and model arbitrariness limit the ability of these methods to robustly identify all ENS events. 

Each method has its strengths and weaknesses.
Li'21 has the merit of simplicity, since it does not require assumption on RES installed capacities or demand, but has the least skill in identifying ENS events.
Otero'22 is based on daily data of demand and RES production, thus requiring socioeconomic assumptions on the future demand and RES installed capacities.
It can investigate the correlation between duration and total ENS, and define a normalized severity index $S$ (Eq.~\eqref{eq:OteroSeverity}), that potentially allows different energy variables or countries to be compared.
Stoop'23 is a context-informed method, which defines a severity metric (CREDI) informed by the knowledge of the power system and that can be interpretable from an energy modeler's perspective (e.g. in TWh of required storage).
This method is more complex to implement, as it requires data with hourly time resolution and defines a climatology that correctly represents all the relevant timescales (daily, weekly, and seasonal).
It allows the study of a specific class of events with a fixed duration.
Overall, we find that Otero'22 is the method that yields the best results while being straightforward to implement and requiring only data with daily resolution. 

Importantly, residual load proves to be the best energy variable to detect shortages, for all considered methods and regions. 
This stresses the need to account for both variable renewable generation and weather-induced demand to identify the challenging events that are the most likely to lead to shortages.

We have also shown that dunkelflaute detection methods are more efficient to identify the most challenging years for the power system than a simple approach based on total cumulative residual load.
TSOs could adopt such methods in adequacy studies, either for stress testing or to run the computationally expensive power system simulations on a reduced set of climate years, to mitigate against resource constraints.
In particular, new climate and energy datasets from the latest version of the PECD consider climate projections in addition to historical data. These include several climate models and greenhouse gases emission scenarios, making the entire dataset much bigger than previous versions of the PECD. The methods assessed here will help analyzing the input data and selecting a sub-ensemble of the full dataset to be run in power system models.

This work validates methods based on climate data with outage data based on power system adequacy simulations. These two types of datasets are generally produced and analyzed by different scientific communities. 
This work also intends to help bridge the gap between the energy system and climate modeling communities~\autocite{craig2022disconnect}.


\section*{CRediT Author Statement}
Conceptualization: all.
Formal analysis: BB and BC.
Methodology: all.
Investigation: BB and BC.
Visualization: BB and BC.
Writing - Original Draft: BB, BC and LS.
Writing - Review \& Editing:  all.
Funding Acquisition: LD and WZ.

\section*{Acknowledgments}
The authors would like to thank the ENTSO-E ERAA Working Group for interesting discussions and feedback.
The authors are, or have been, members of ENTSO-E's Expert Team Climate and acknowledge the support of the ENTSO-E secretariat  during the early development of this work.
The content of this paper and the views expressed in it are solely the authors' responsibility, and do not necessarily reflect the views of TenneT and/or RTE.

\section*{Open research} 
All the code used to generate the figures in the main article and the supporting information are available on GitHub via \url{https://github.com/BastienCozian/Evaluation_Dunkelflaute_Methods}

\printbibliography

\clearpage
\appendix
\beginsupplement

\begin{flushleft}
    \begin{adjustwidth}{0in}{.5in} 
        \begin{Huge} 
            \begin{spacing}{.9} 
            \textbf{Supporting information for: \newtitle}
            \end{spacing}
        \end{Huge}
    \end{adjustwidth}
    \bigskip
    \linespread{1}
    \newauthor 
    \bigskip
    \newdate
\end{flushleft}


The main text shows results for the validation of dunkelflaute detection methods.
In this Supporting Information, some additional details are provided.
Results for other European regions are discussed in Section~\ref{sec:Otero_other_countries} for the method Otero'22 and in Section~\ref{sec:Stoop_other_countries} for the method Stoop'23.
Section~\ref{sec:SI_Stoop_T} shows the result for longer CREDI event ($T=3$ and $5$~days) with Stoop'23.
Section~\ref{sec:Stoop_HWRW} provide more details on the hourly and weekly rolling window (HWRW) used in the definition of the CREDI index.
The sensitivity of the F-score results to the target year (Section~\ref{sec:SI_TY}) and the Economic Viability Assessment (EVA) scenario (Section~\ref{sec:SI_EVA}) are also discussed. 
Section~\ref{sec:SI_filter} show small improvements in F-score value when filtering out ENS events \emph{a priori} not related to weather.
Section~\ref{sec:SI_year_selection} provides additional figures on year selection.
Section~\ref{sec:SI_Fbeta} discusses the relative weight of 'precision' and 'recall' in the definition of the F-score.

\section{Otero'22: F-score and correlation for other countries}
\label{sec:Otero_other_countries}

\begin{figure}[htbp]
    \centering
    \includegraphics[width=0.3\textwidth]{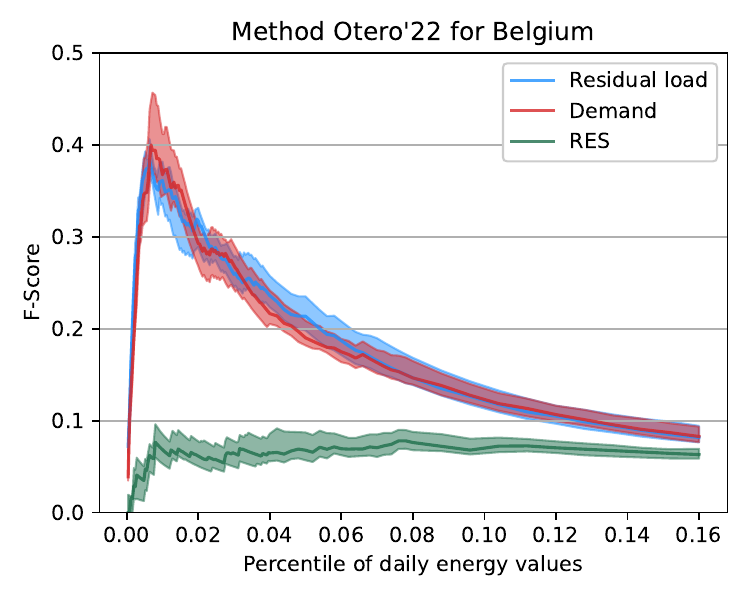}
    \includegraphics[width=0.3\textwidth]{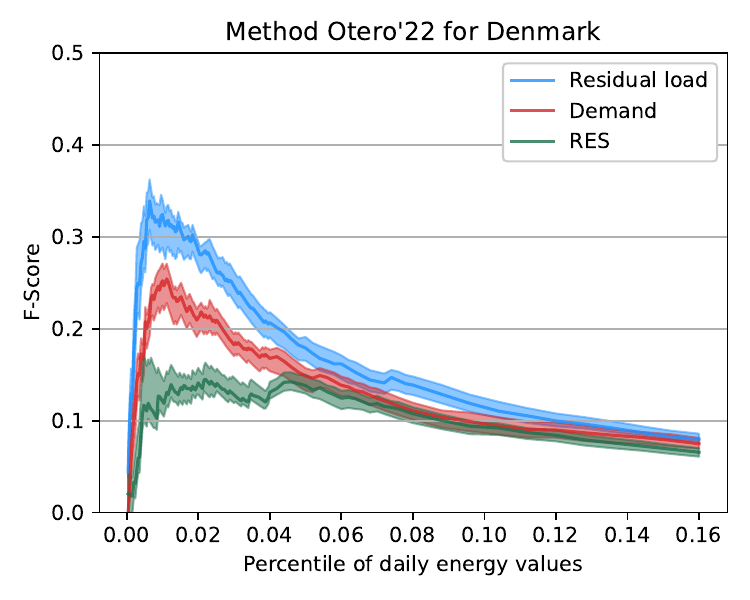}
    \includegraphics[width=0.3\textwidth]{Validation_Otero22_ENS_scenarioB_FR00.pdf}
    \includegraphics[width=0.3\textwidth]{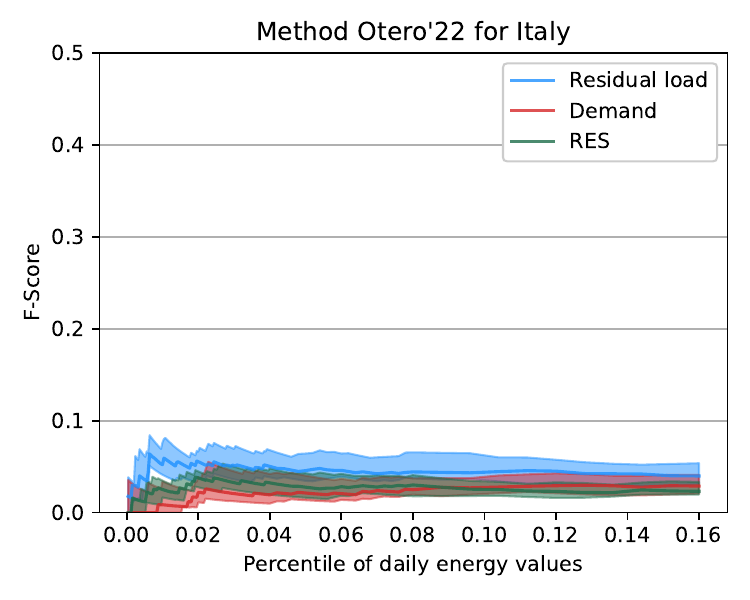}
    \includegraphics[width=0.3\textwidth]{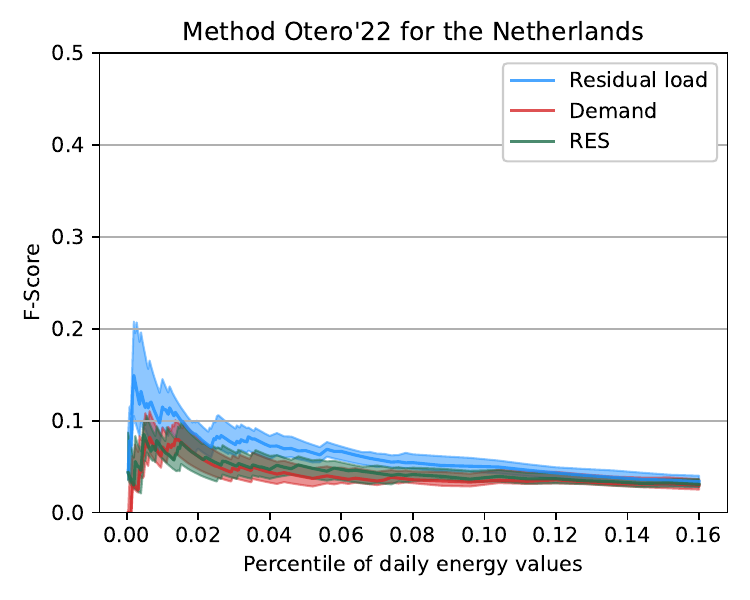}
    \includegraphics[width=0.3\textwidth]{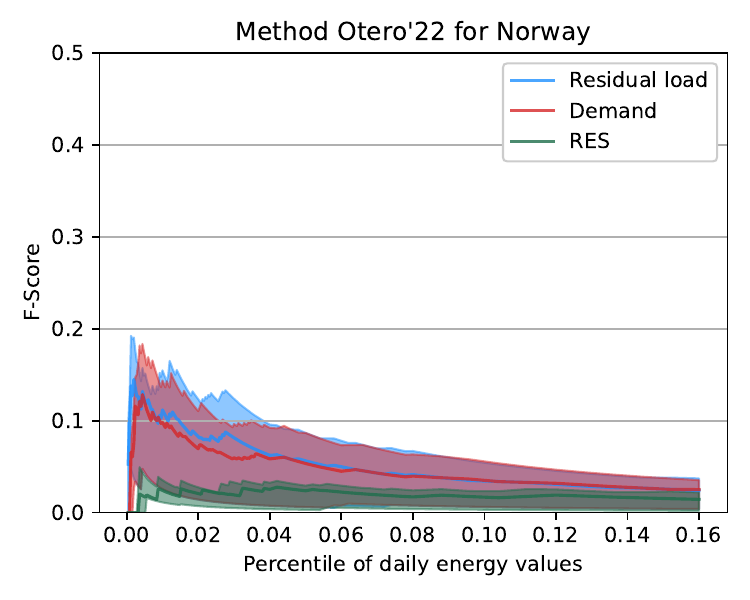}
    \includegraphics[width=0.3\textwidth]{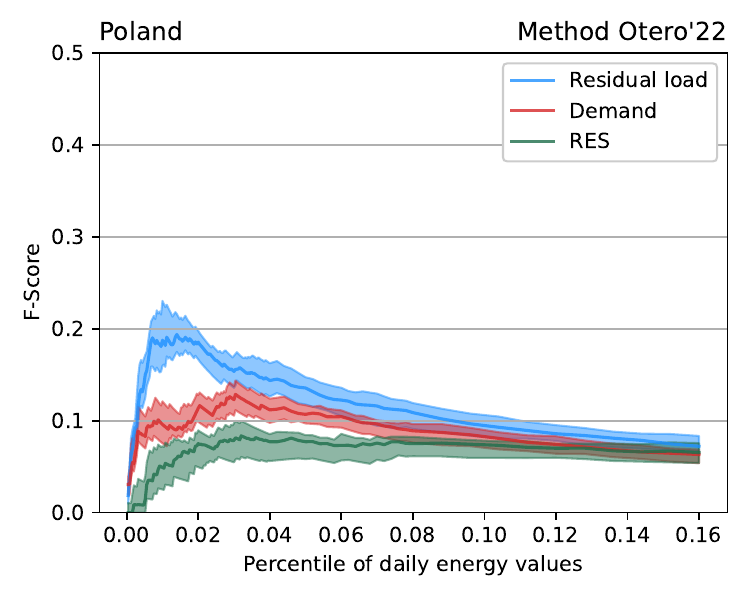}
    \includegraphics[width=0.3\textwidth]{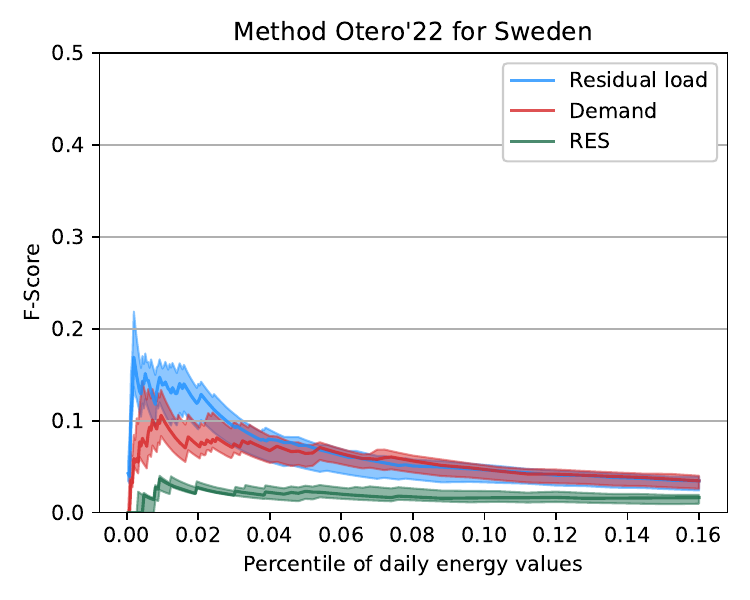}
    \includegraphics[width=0.3\textwidth]{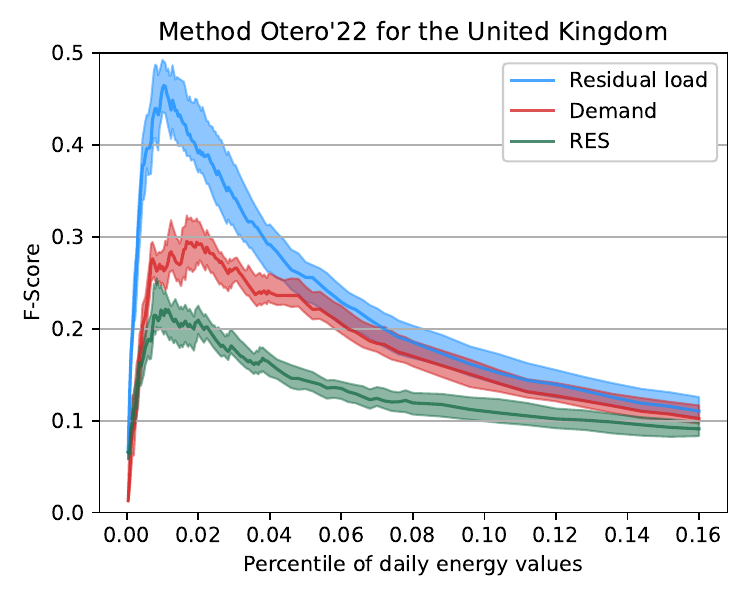}
    \includegraphics[width=0.3\textwidth]{Validation_Otero22_ENS_scenarioB_CWE.pdf}
    \includegraphics[width=0.3\textwidth]{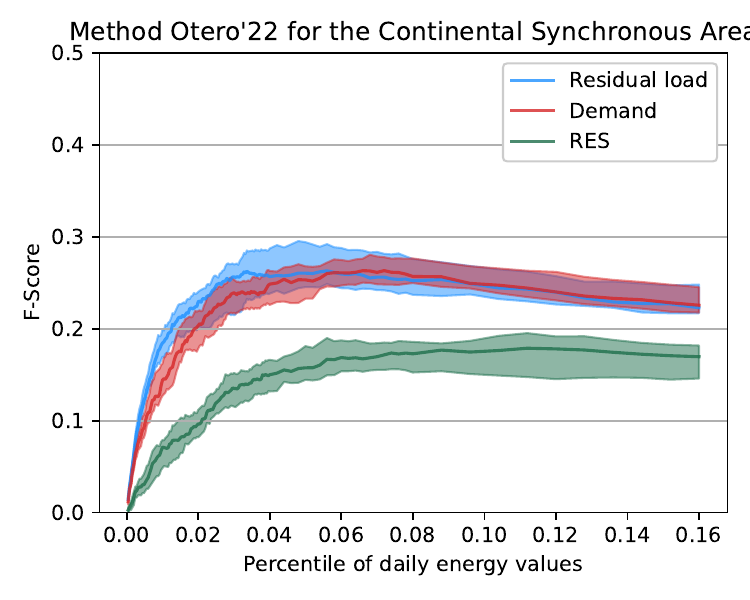}
    \caption{
    F-score for dunkelflaute detection with Otero'22.
    Same as Fig.~\ref{fig:Otero_DE00}a, for various countries and large European regions.
    }
    \label{fig:Otero_other_countries}
\end{figure}

\begin{figure}[htbp]
    \centering
    \includegraphics[width=0.3\textwidth]{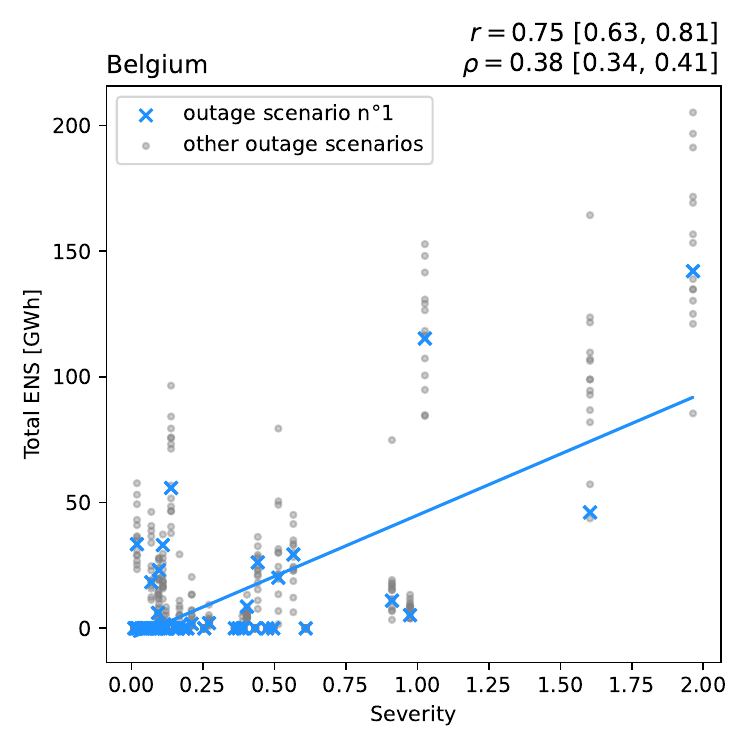}
    \includegraphics[width=0.3\textwidth]{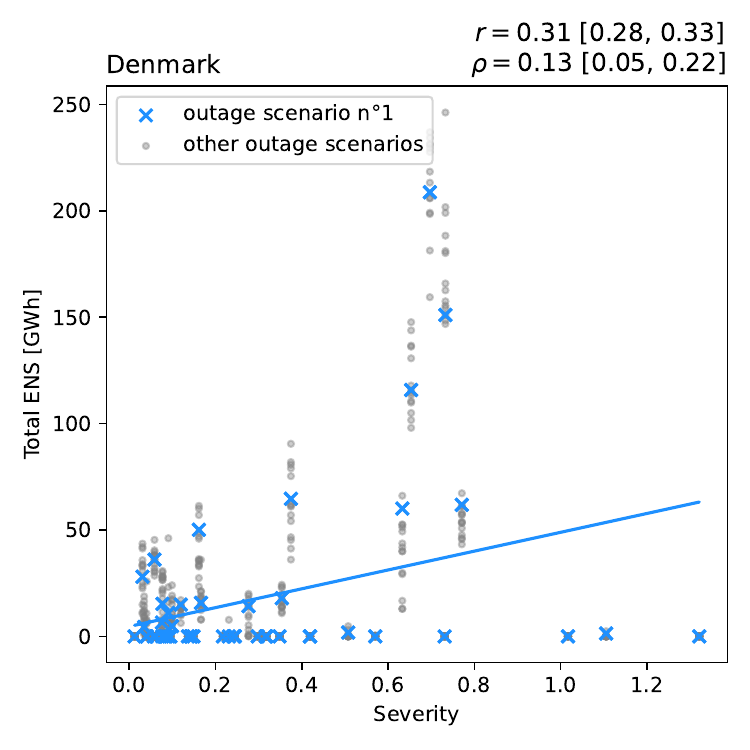}
    \includegraphics[width=0.3\textwidth]{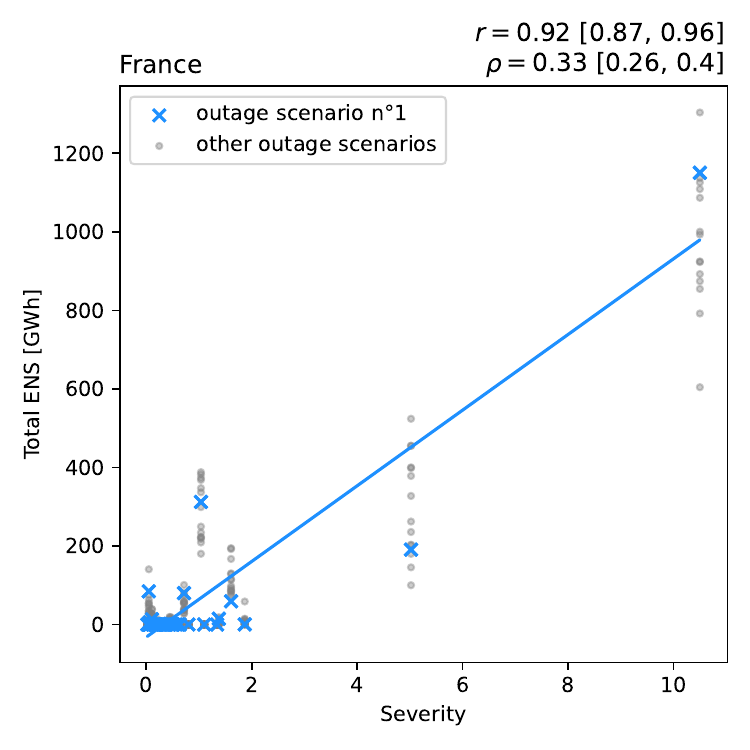}
    \includegraphics[width=0.3\textwidth]{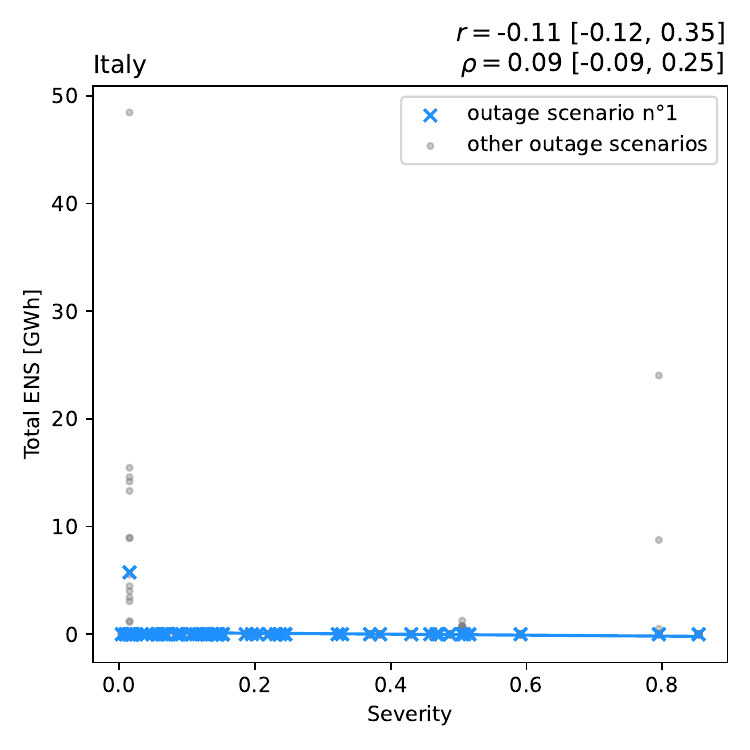}
    \includegraphics[width=0.3\textwidth]{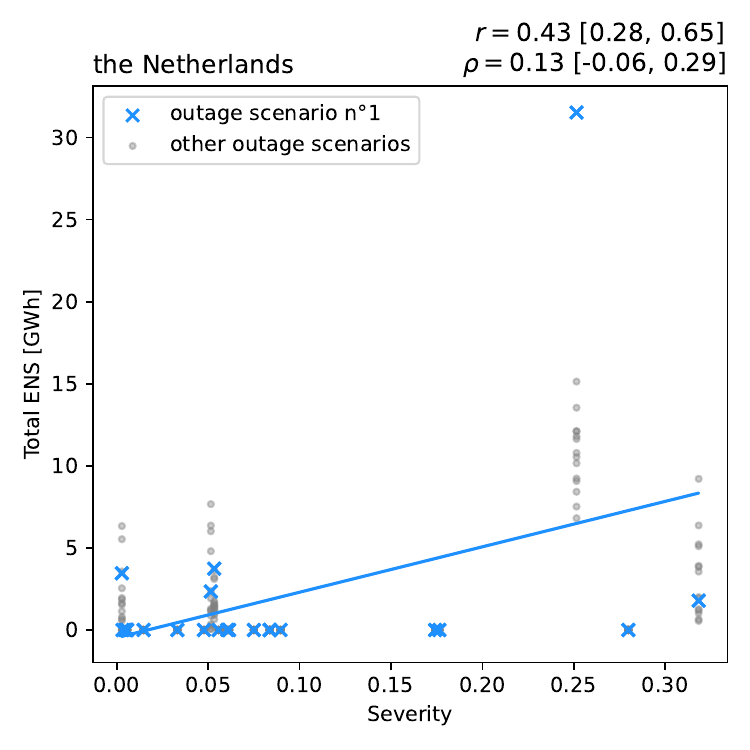}
    \includegraphics[width=0.3\textwidth]{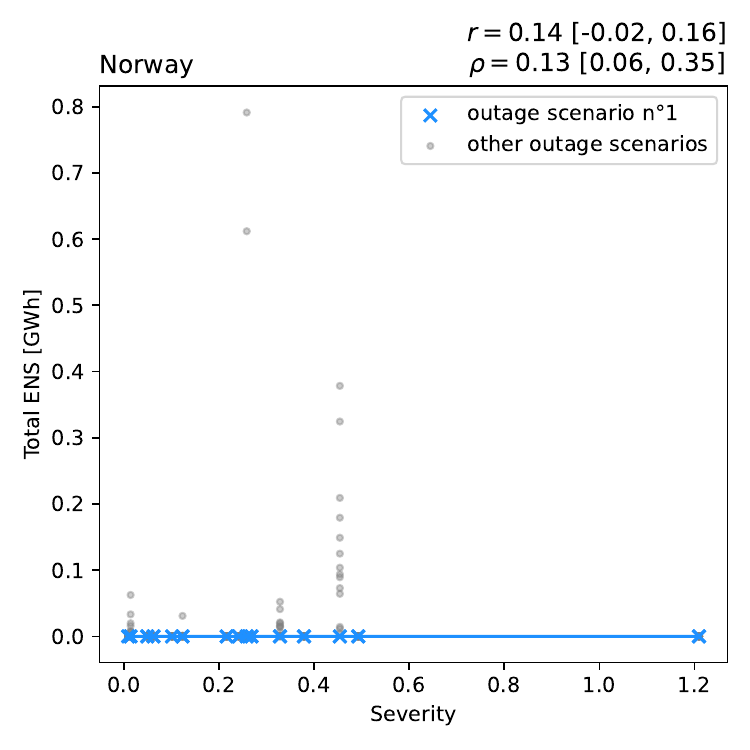}
    \includegraphics[width=0.3\textwidth]{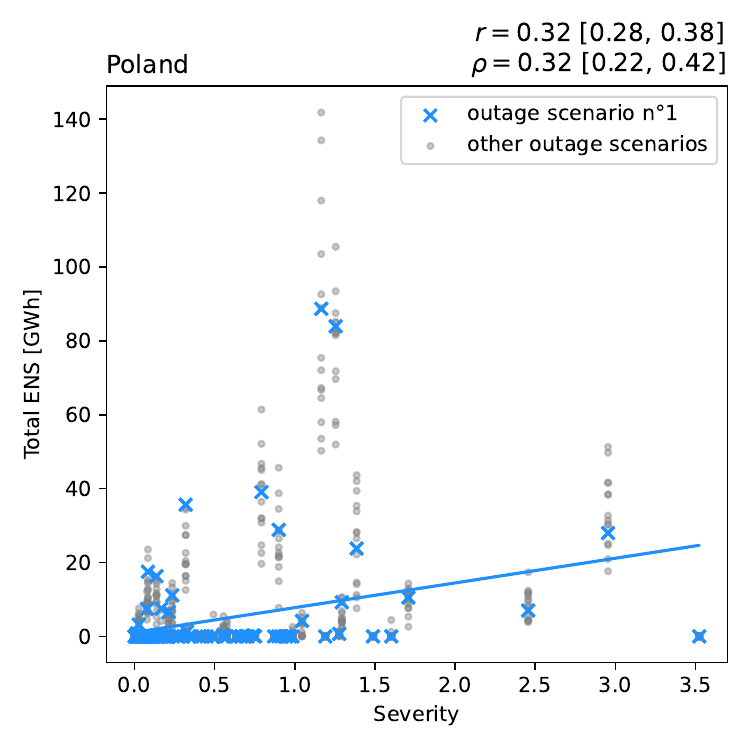}
    \includegraphics[width=0.3\textwidth]{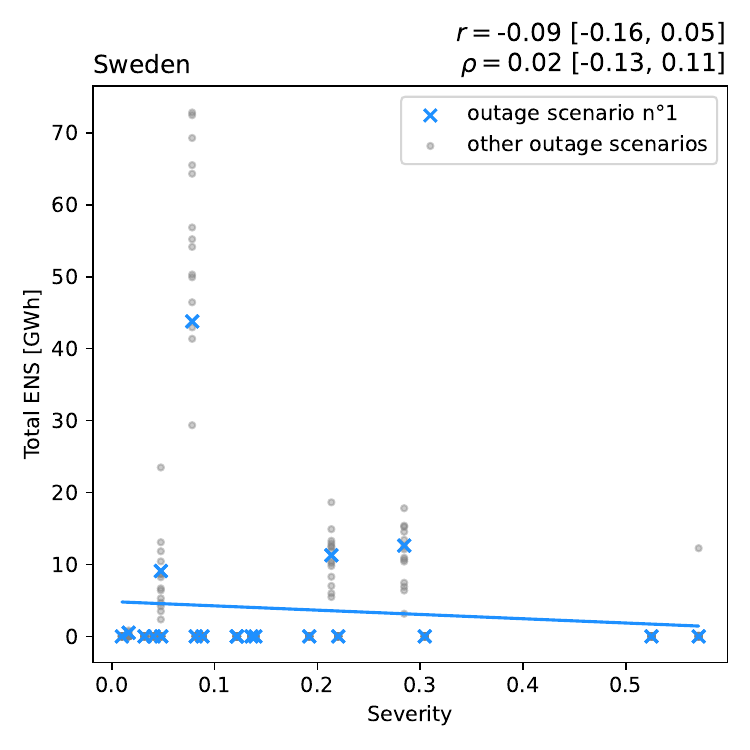}
    \includegraphics[width=0.3\textwidth]{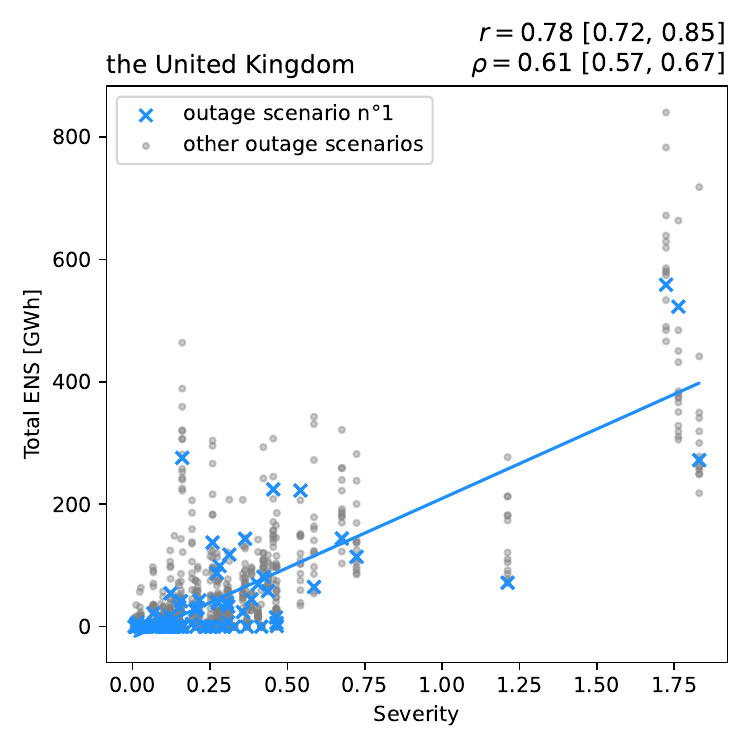}
    \includegraphics[width=0.3\textwidth]{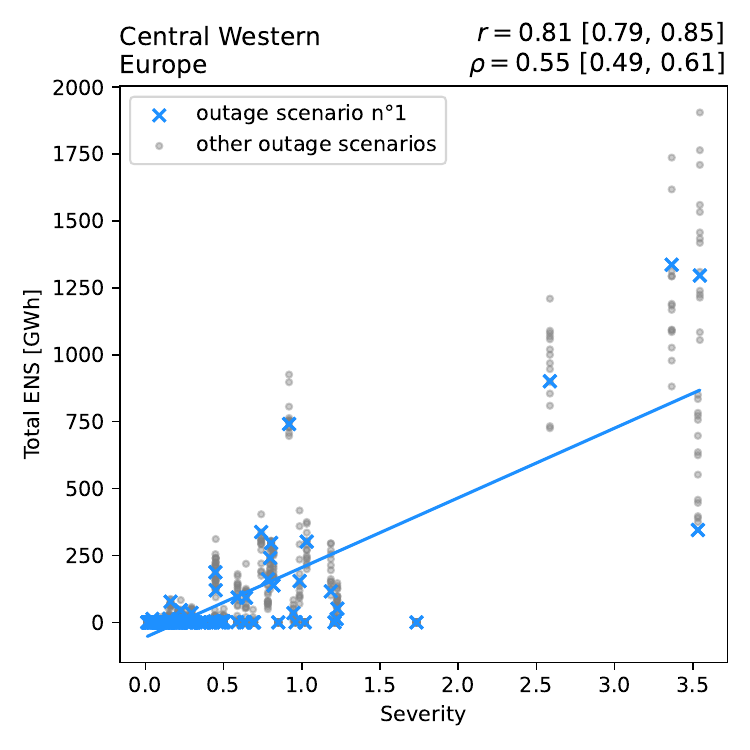}
    \includegraphics[width=0.3\textwidth]{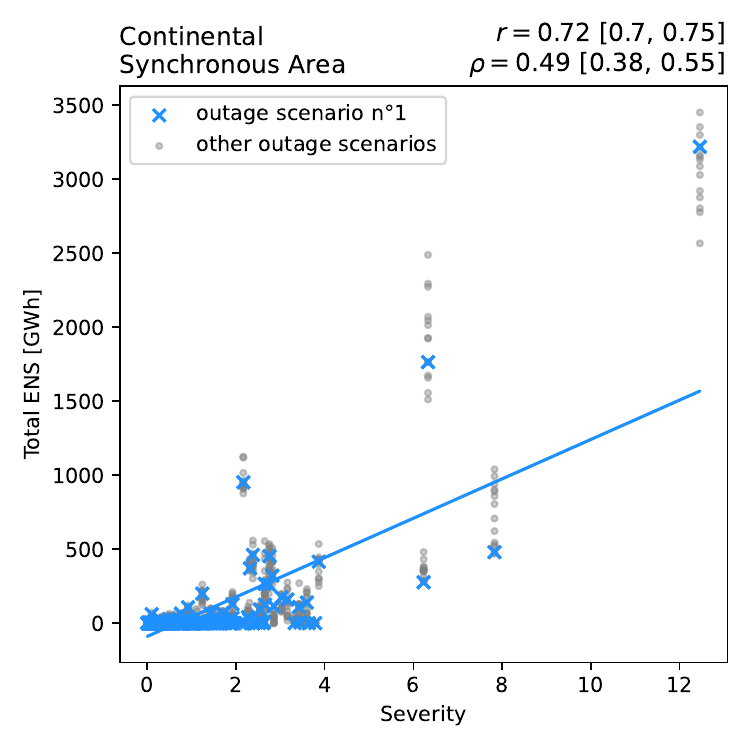}
    \caption{
    Correlation between the total ENS values of detected dunkelflaute events and their severity $S$ (Eq.~\eqref{eq:OteroSeverity}).
    Same as Fig.~\ref{fig:Otero_DE00}b, for various countries and large European regions.
    }
    \label{fig:Otero_other_countries_correlation}
\end{figure}

Figure~\ref{fig:Otero_DE00} of the main text present the results (F-score and correlation) for Germany with the method Otero'22.
This section provide some additional figures (Fig.~\ref{fig:Otero_other_countries} and Fig.~\ref{fig:Otero_other_countries_correlation})) for other countries and larger European regions. 

Figure~\ref{fig:Otero_other_countries} shows the F-score for various countries and regions.
Only a subset of country among the 34 available countries is shown.
For countries with multiple bidding zones (Denmark, Italy, Norway, Sweden, the United Kingdom) or for larger regions 
(Central Western Europe, or Continental Synchronous Area\footnote{
Continental Synchronous Area (CSA) aggregates the following bidding zones: 'PT00', 'ES00', 'BE00', 'FR00', 'NL00', 'DE00', 'DKW1', 'CH00', 'ITN1', 'ITCN', 'ITCS', 'ITS1', 'ITCA', 'ITSI', 'ITSA', 'PL00', 'CZ00', 'AT00', 'SI00', 'SK00', 'HU00', 'HR00', 'RO00', 'BA00', 'RS00', 'ME00', 'MK00', 'GR00', 'BG00'.
Note that 'DKE1' is part of the Nordic Zone, not CSA. 
The following bidding zones are not included because at least one of the required dataset (capacity factor time series, installed capacities, or demand time series) is missing: 'FR15', 'MD00', 'UA01', 'UA02', 'CR00', 'TR00'.
}), 
the respective bidding zones are aggregated.

Otero'22 have its highest accuracy skill of shortage detection in in Belgium, France, and the United Kingdom ($F \approx 0.4-0.5$), while it shows little or no skill in Italy, the Netherlands, or Norway. 
At a large scale (Central Western Europe) or very large scale (Continental Synchronous Area), Otero'22 retains significant detection skill.
In all cases, residual load stands out as the best energy variable to identify shortages.
When comparing the validation results between different countries, it is important to highlight that the robustness of the F-score is also affected by the number of ENS events in the ERAA2023 dataset. 
For example, in a country with very few days with expected energy shortage, there are fewer ENS events which can be detected. 
In the case where there are no true positive (TP) events to detect, the F-score would be 0. 
This applies to all three dunkelflaute detection methods analysed in this study.

Figure~\ref{fig:Otero_other_countries_correlation} completes Figure~\ref{fig:Otero_DE00}b by showing the correlation between the total ENS during a dunkelflaute with its severity $S$ (Eq.~\eqref{eq:OteroSeverity}).
Countries with higher F-score values show better correlation between total ENS and severity.

\section{Stoop'23: F-score and correlation for other countries}
\label{sec:Stoop_other_countries}

\begin{figure}[htbp]
    \centering
    \includegraphics[width=0.3\textwidth]{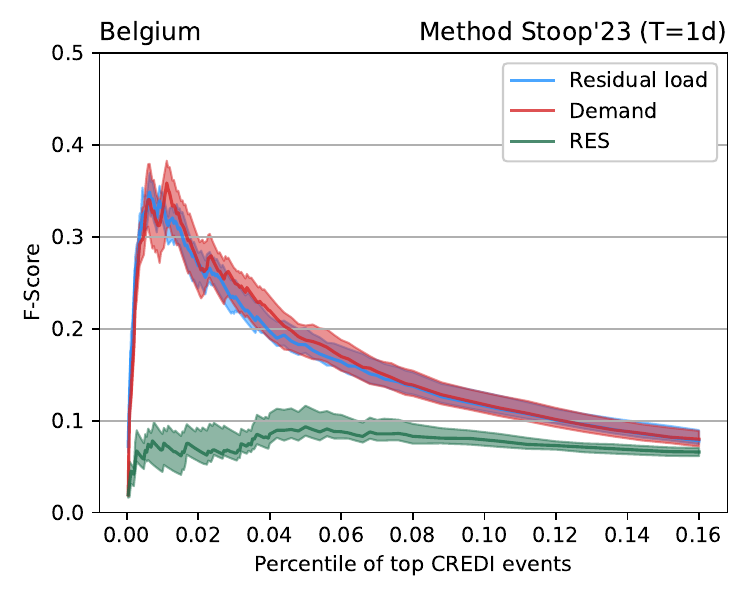}
    \includegraphics[width=0.3\textwidth]{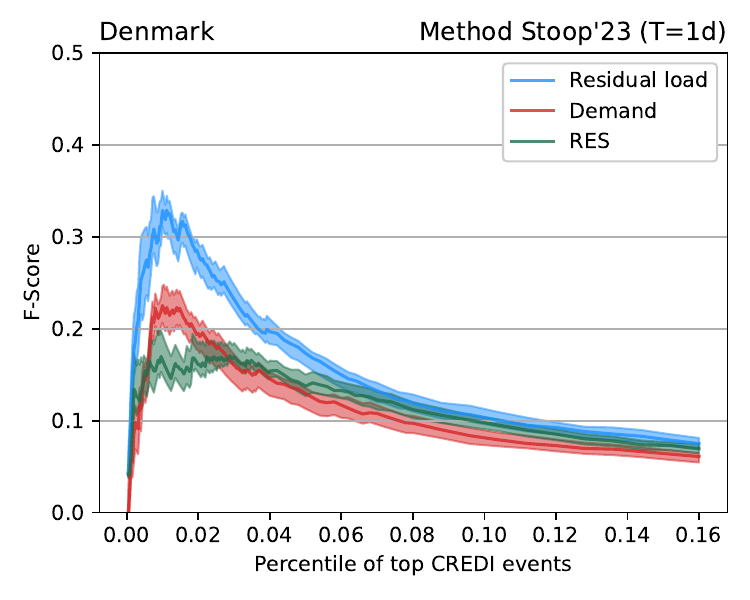}
    \includegraphics[width=0.3\textwidth]{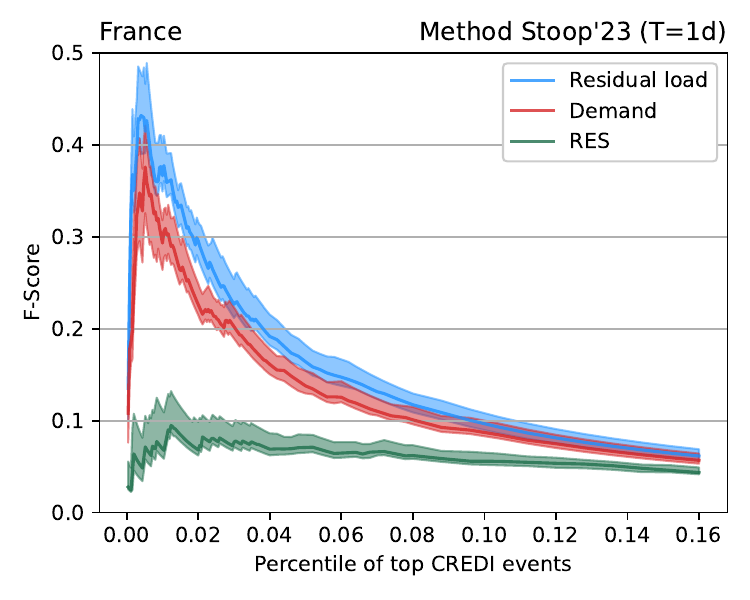}
    \includegraphics[width=0.3\textwidth]{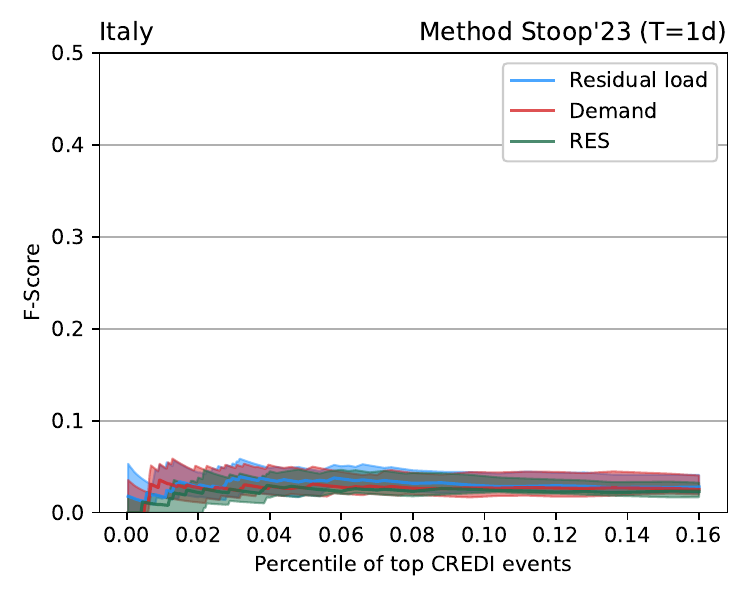}
    \includegraphics[width=0.3\textwidth]{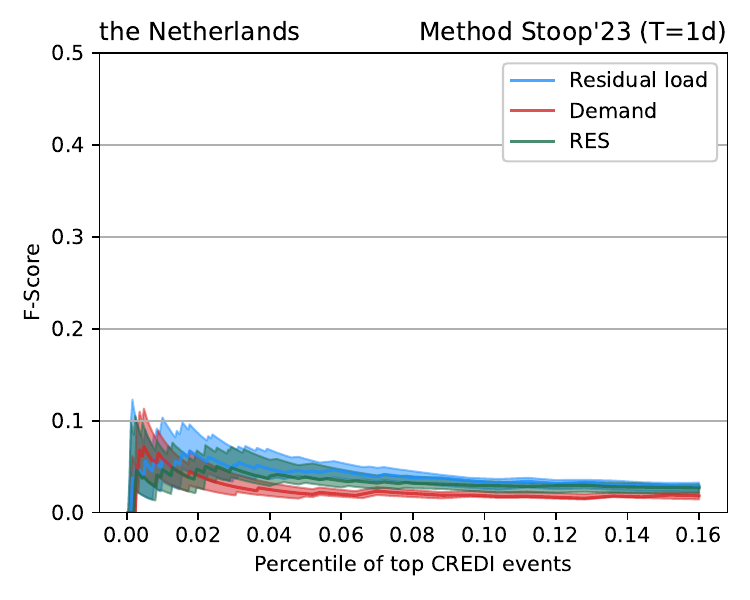}
    \includegraphics[width=0.3\textwidth]{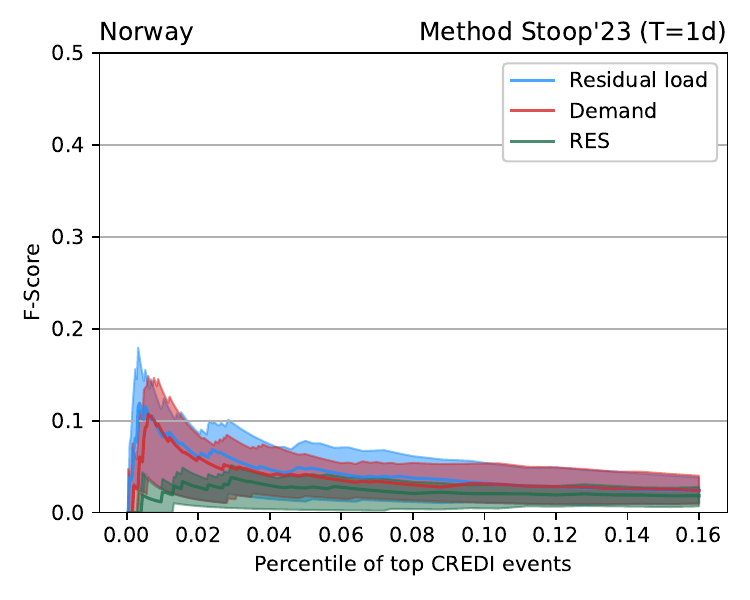}
    \includegraphics[width=0.3\textwidth]{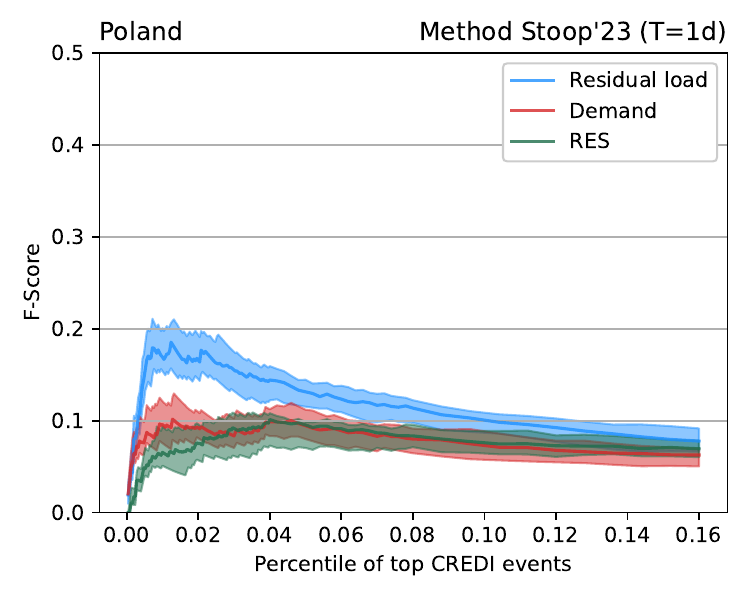}
    \includegraphics[width=0.3\textwidth]{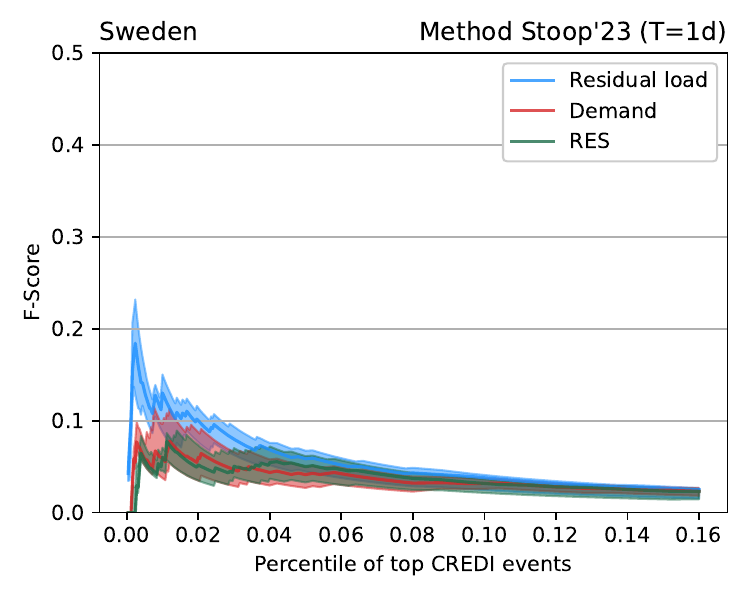}
    \includegraphics[width=0.3\textwidth]{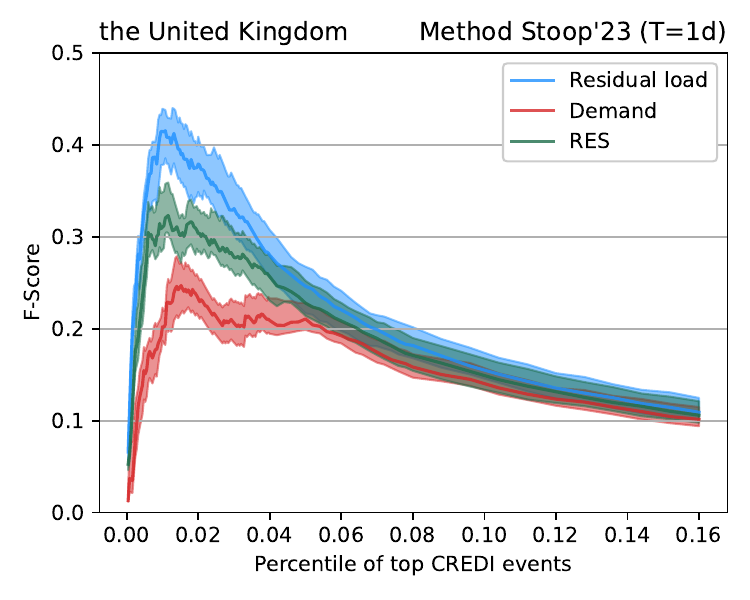}
    \includegraphics[width=0.3\textwidth]{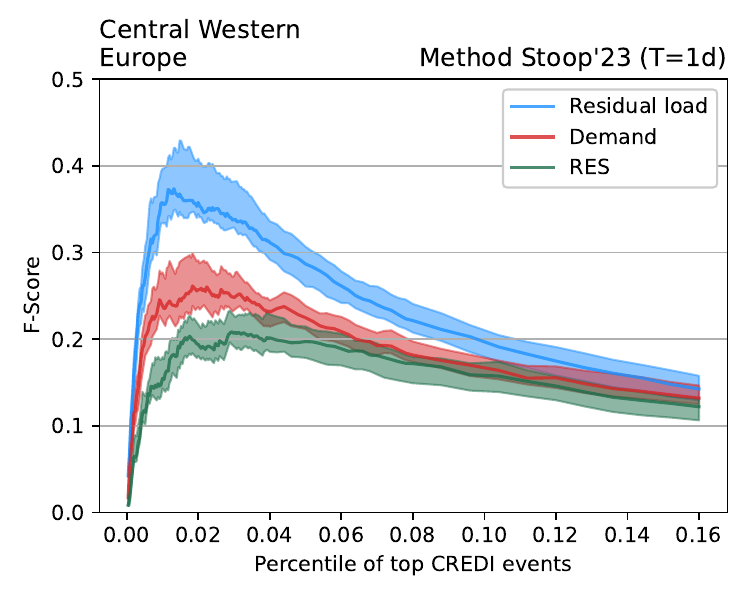}
    \includegraphics[width=0.3\textwidth]{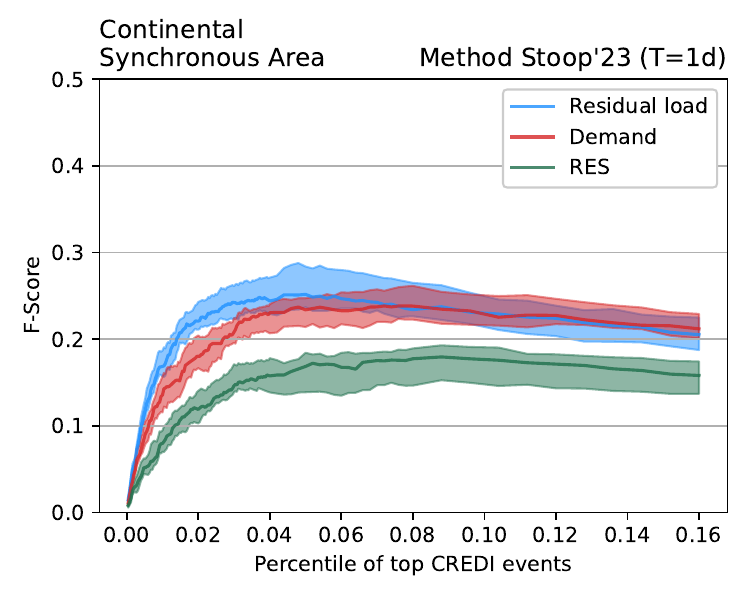}
    \caption{
    F-score for dunkelflaute detection with the method Stoop'23, for $T=1$-day CREDI events.
    Same as Fig.~\ref{fig:Stoop_DE00_T1}a, for various countries and large European regions.
    }
    \label{fig:Stoop_other_countries}
\end{figure}

\begin{figure}[htbp]
    \centering
    \includegraphics[width=0.3\textwidth]{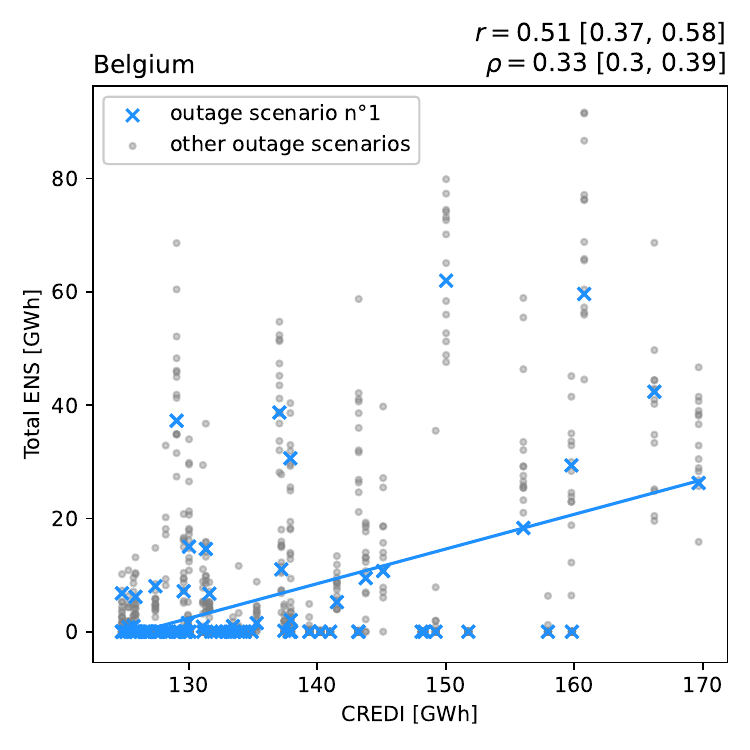}
    \includegraphics[width=0.3\textwidth]{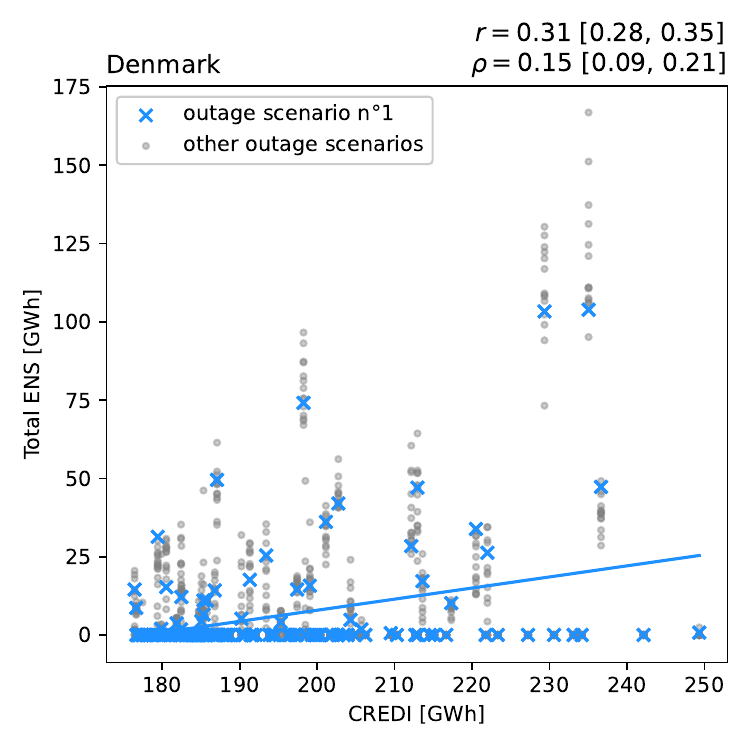}
    \includegraphics[width=0.3\textwidth]{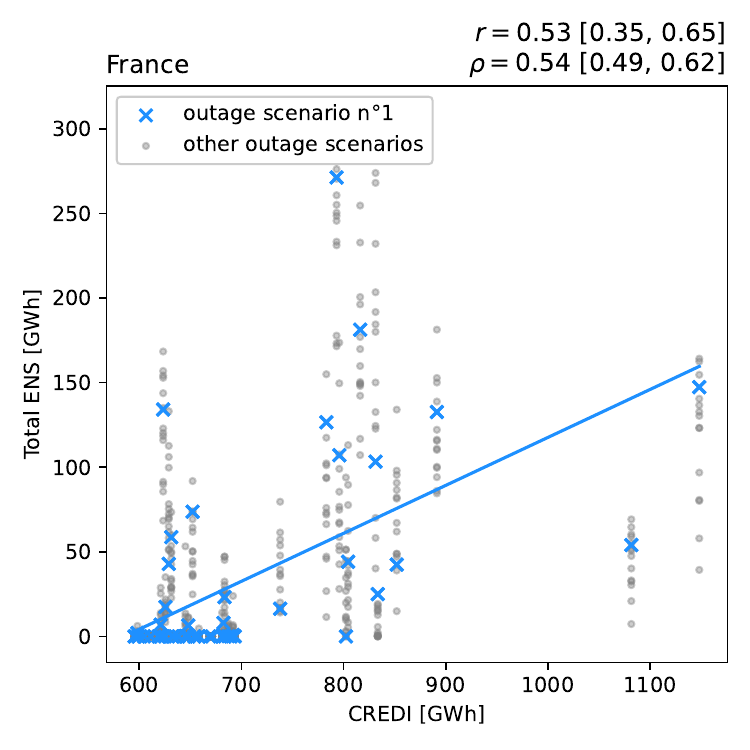}
    \includegraphics[width=0.3\textwidth]{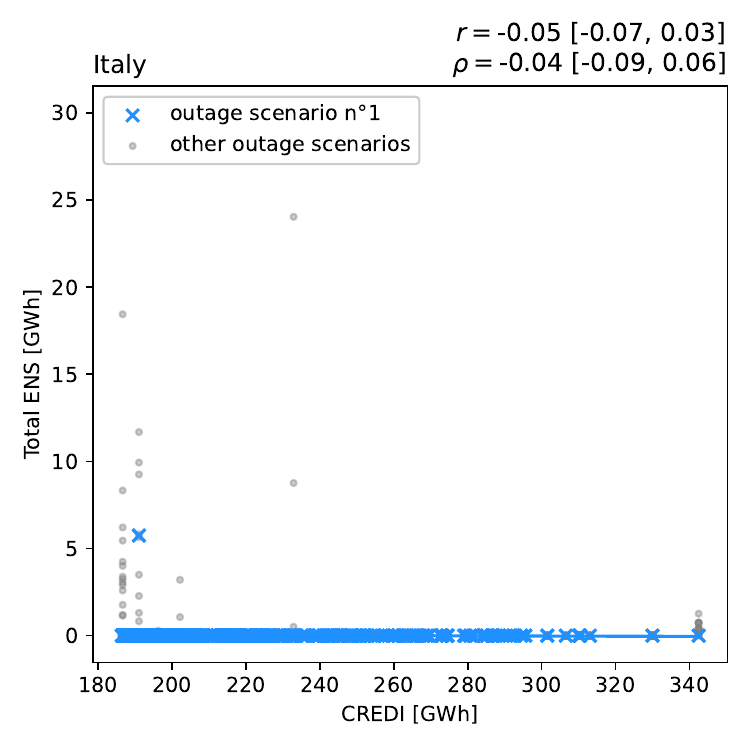}
    \includegraphics[width=0.3\textwidth]{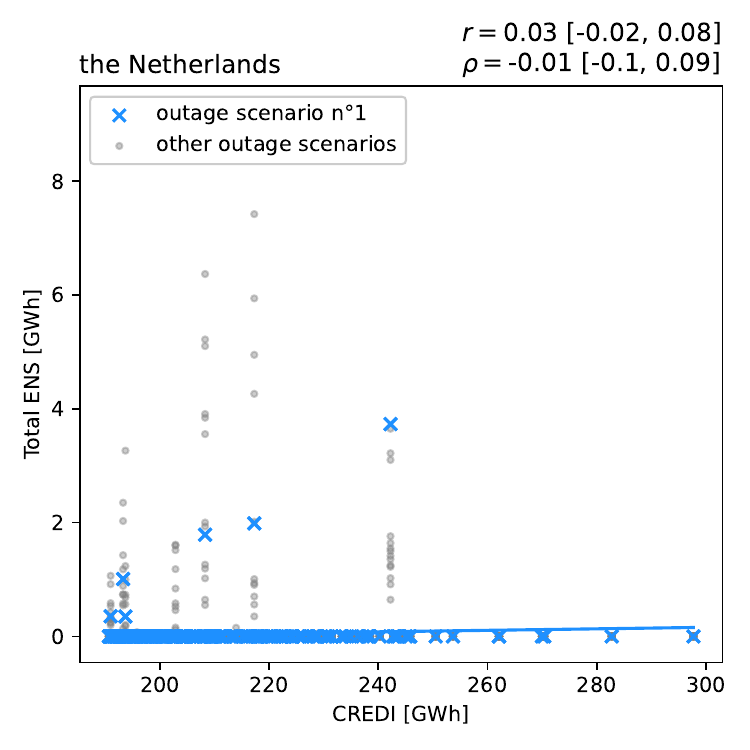}
    \includegraphics[width=0.3\textwidth]{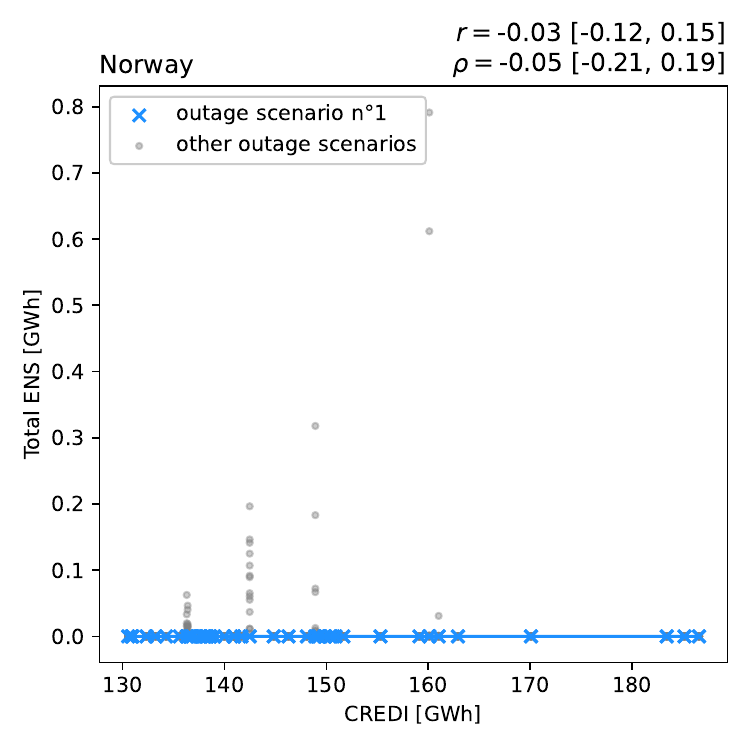}
    \includegraphics[width=0.3\textwidth]{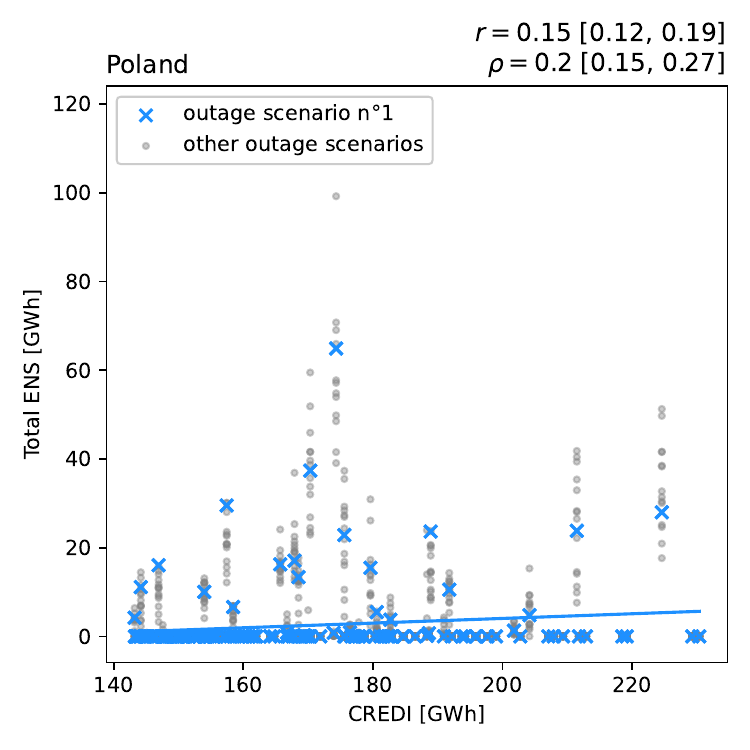}
    \includegraphics[width=0.3\textwidth]{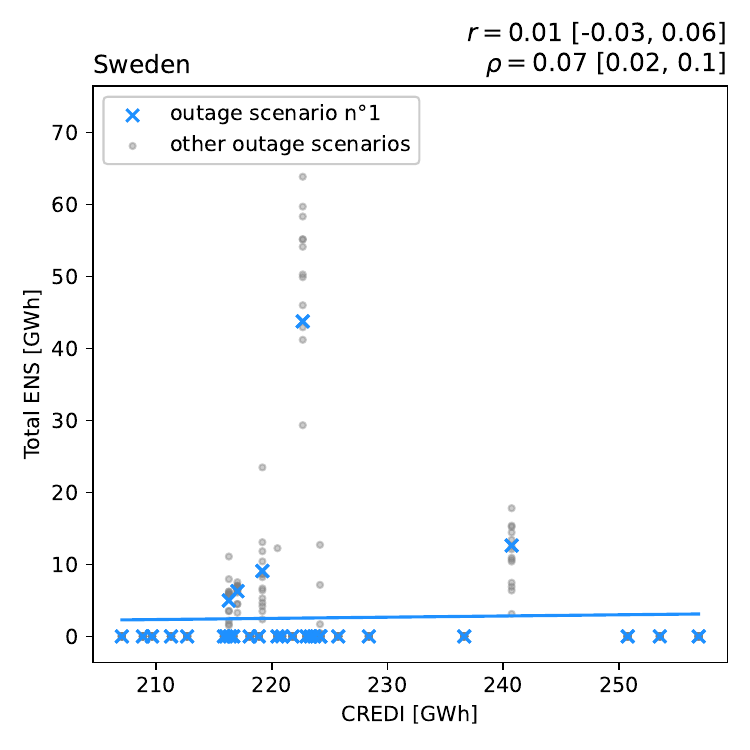}
    \includegraphics[width=0.3\textwidth]{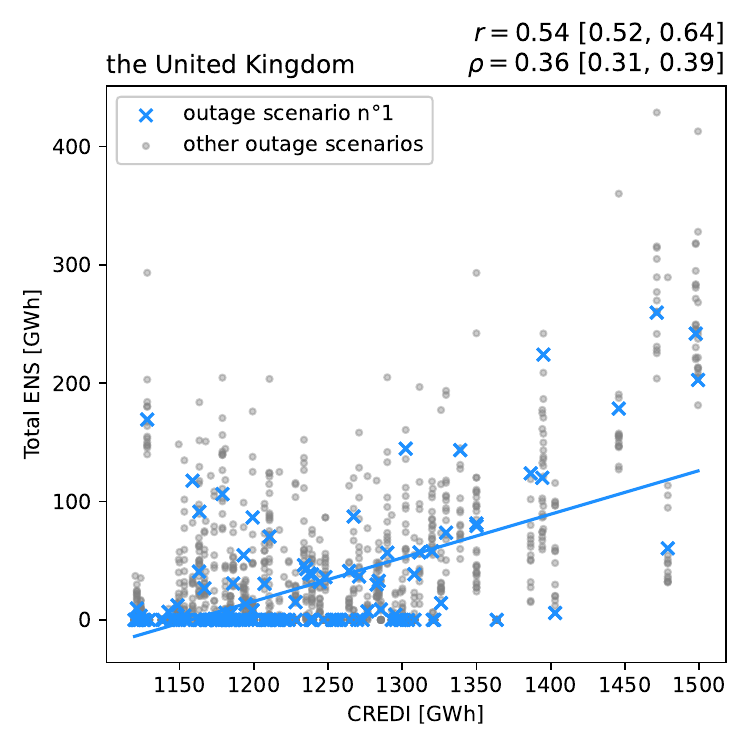}
    \includegraphics[width=0.3\textwidth]{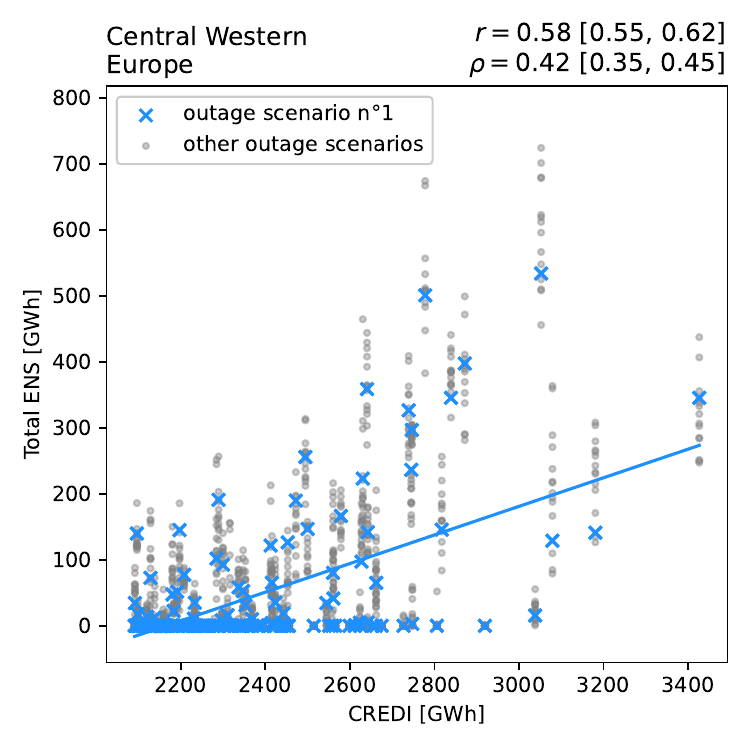}
    \includegraphics[width=0.3\textwidth]{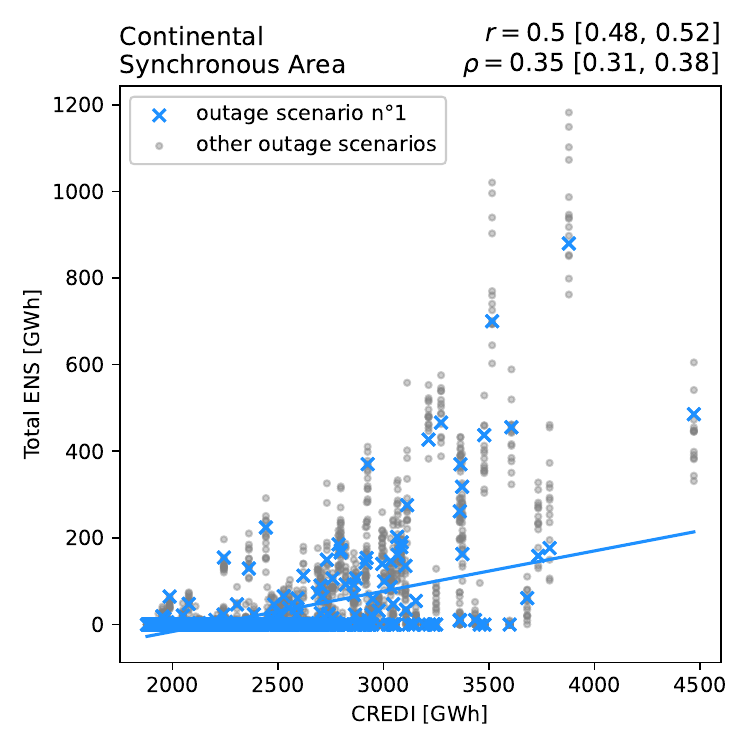}
    \caption{
    Correlation between the total ENS values of detected dunkelflaute events and their $T=1$-day CREDI values.
    Same as Fig.~\ref{fig:Stoop_DE00_T1}b, for various countries and large European regions.
    }
    \label{fig:Stoop_other_countries_correlation}
\end{figure}

Figure~\ref{fig:Stoop_DE00_T1} of the main text present the results (F-score and correlation) for Germany with the method Stoop'23.
This section provide some additional figures (Fig.~\ref{fig:Stoop_other_countries} and Fig.~\ref{fig:Stoop_other_countries_correlation})) for other countries and larger European regions. 

Figure~\ref{fig:Otero_other_countries} shows the F-score for various countries and regions.
The same conclusions as for Otero'22 are found: while Stoop'23 shows relatively high skill in some countries (Belgium, France, the United Kingdom), in others it shows little or no skill (Italy, the Netherlands, Norway).
Again, the residual load is the best energy variable to identify shortages.

Figure~\ref{fig:Stoop_other_countries_correlation} completes Figure~\ref{fig:Stoop_DE00_T1}b by showing the correlation between the total ENS during a dunkelflaute with its $T=1$-day CREDI values.
Countries with higher F-score values show better correlation between total ENS and CREDI.

\section{Stoop'23: Correlation for varying duration $T$}
\label{sec:SI_Stoop_T}

\begin{figure}[htbp]
    \centering
    \includegraphics[width=0.3\textwidth]{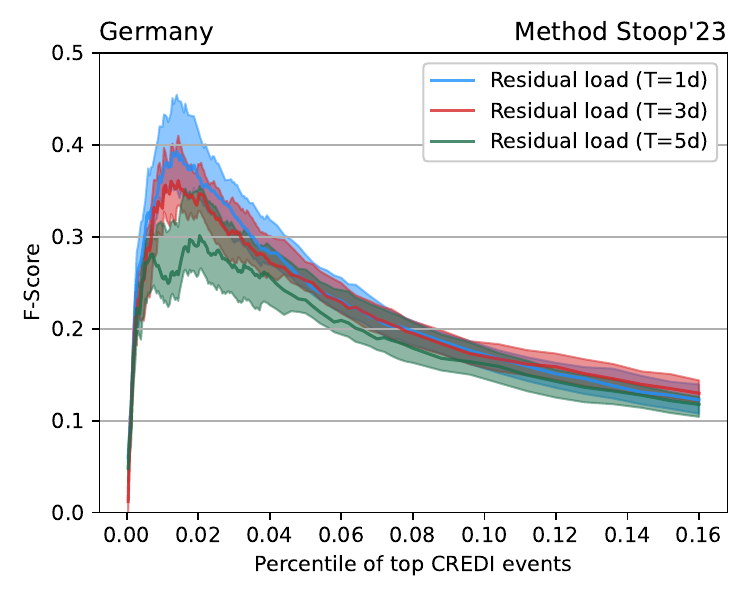}
    \includegraphics[width=0.3\textwidth]{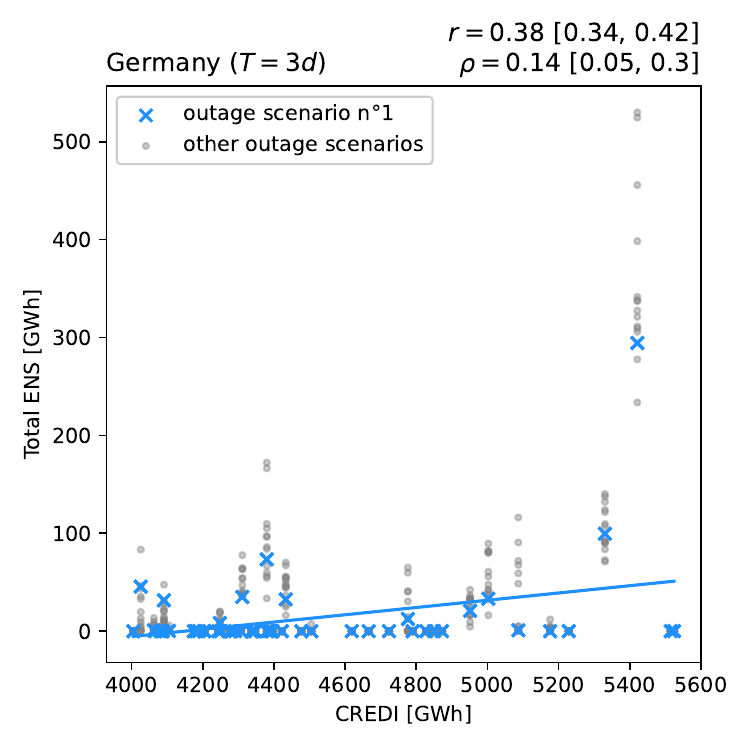}
    \includegraphics[width=0.3\textwidth]{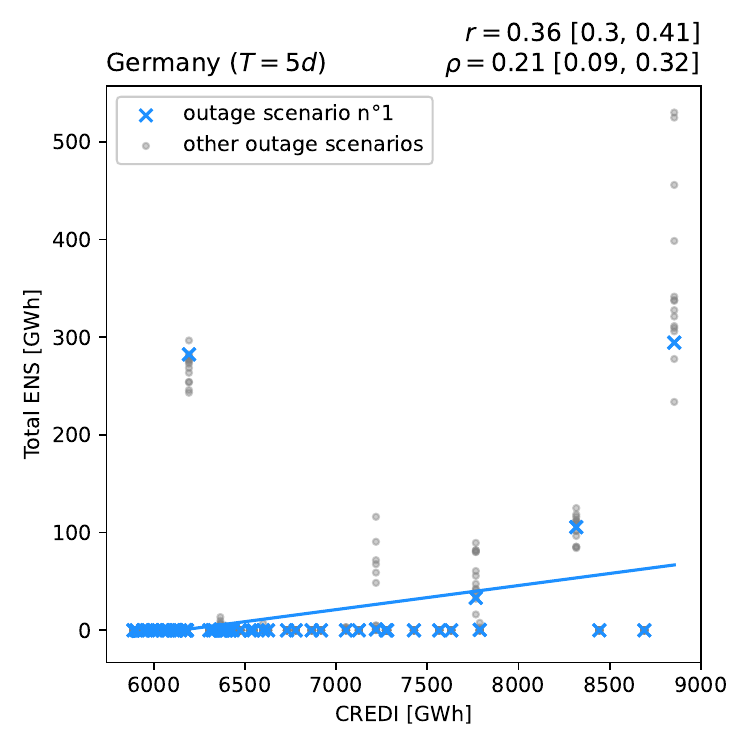}

    \includegraphics[width=0.3\textwidth]{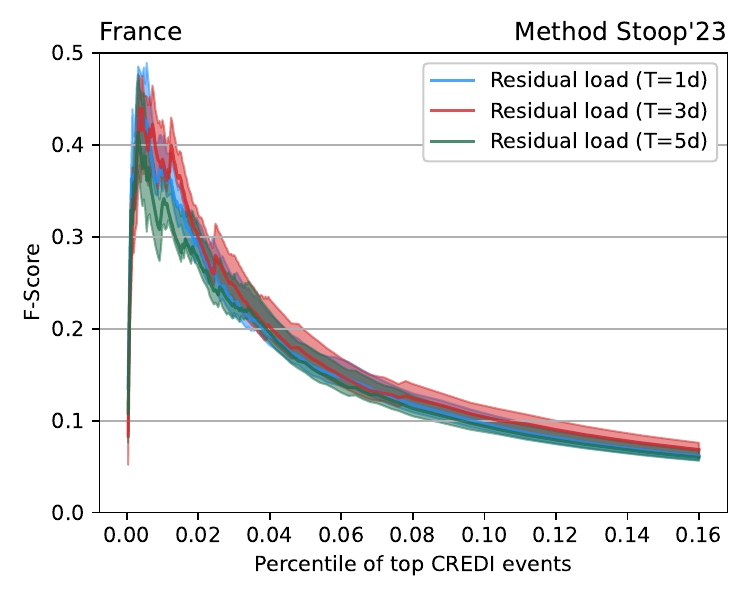}
    \includegraphics[width=0.3\textwidth]{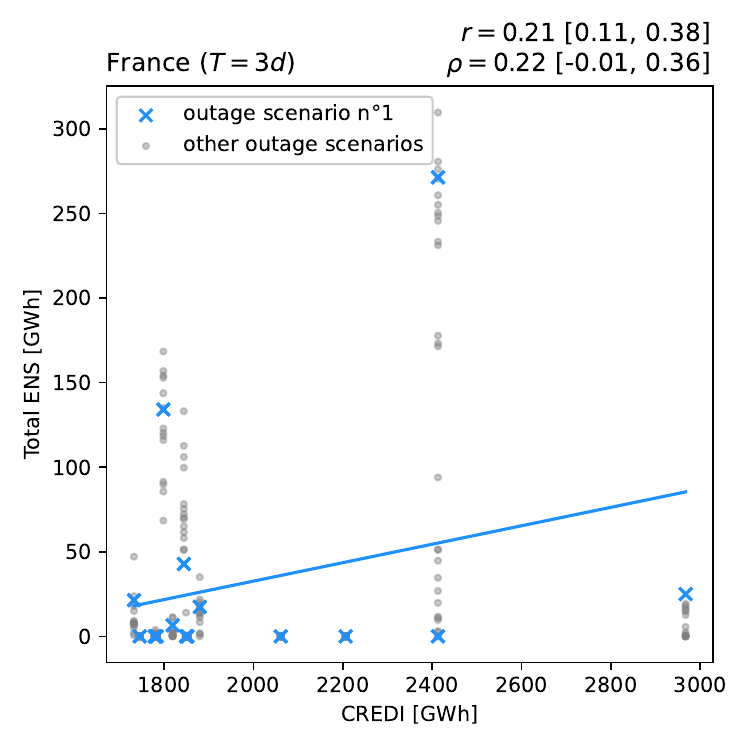}
    \includegraphics[width=0.3\textwidth]{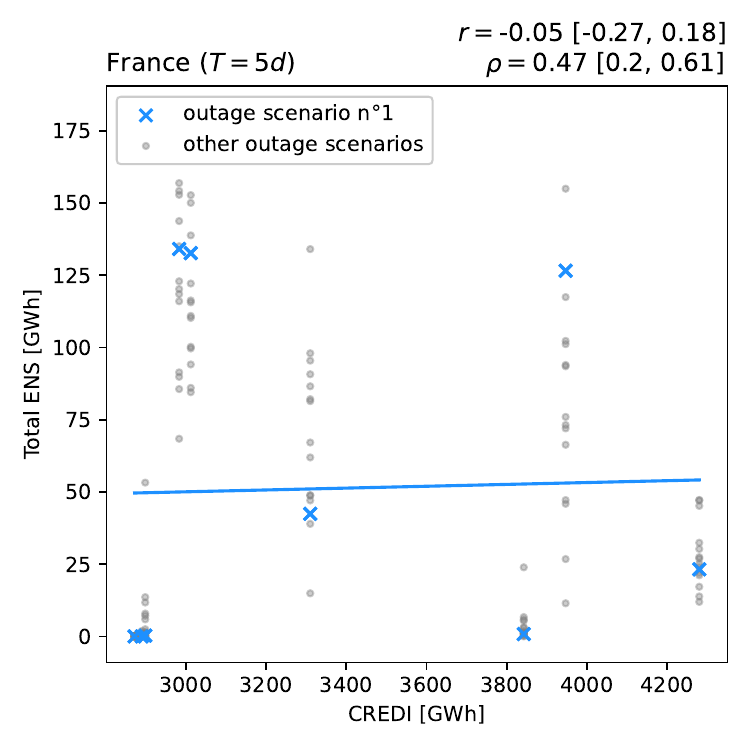}

    \includegraphics[width=0.3\textwidth]{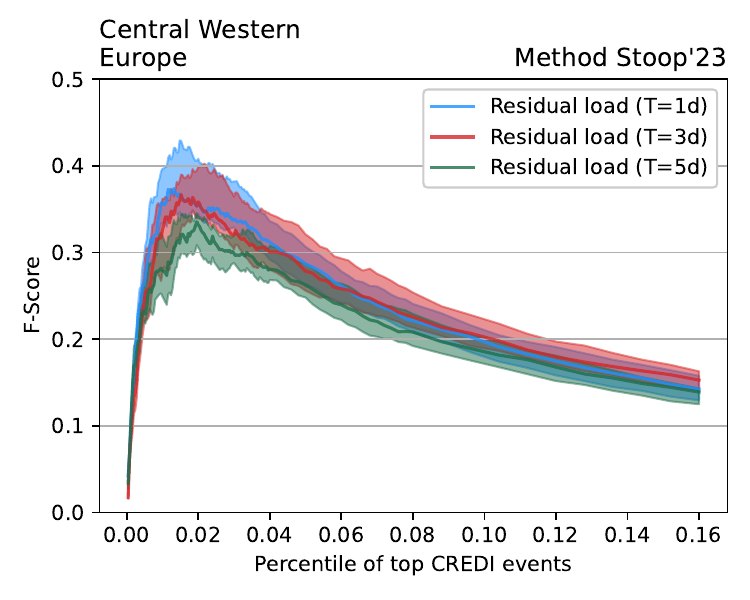}
    \includegraphics[width=0.3\textwidth]{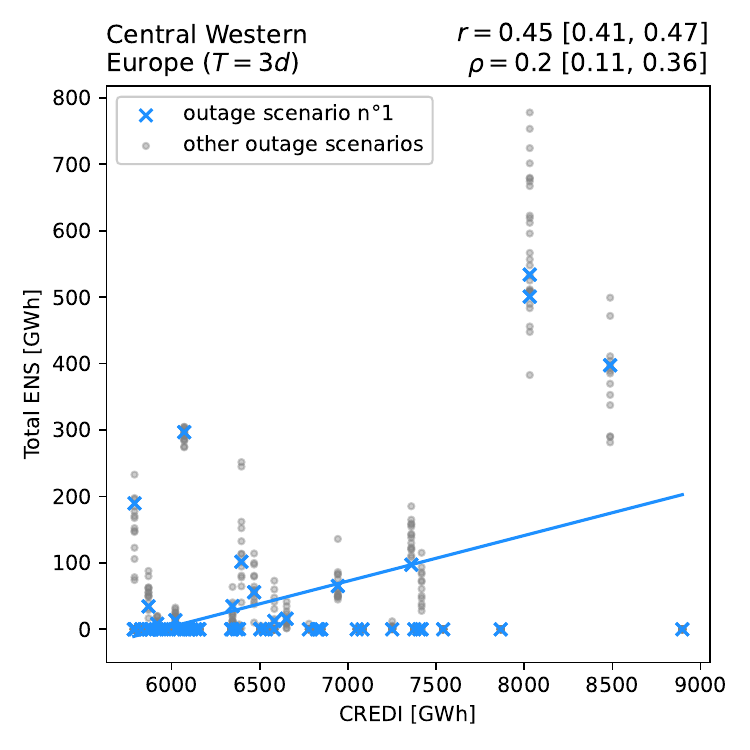}
    \includegraphics[width=0.3\textwidth]{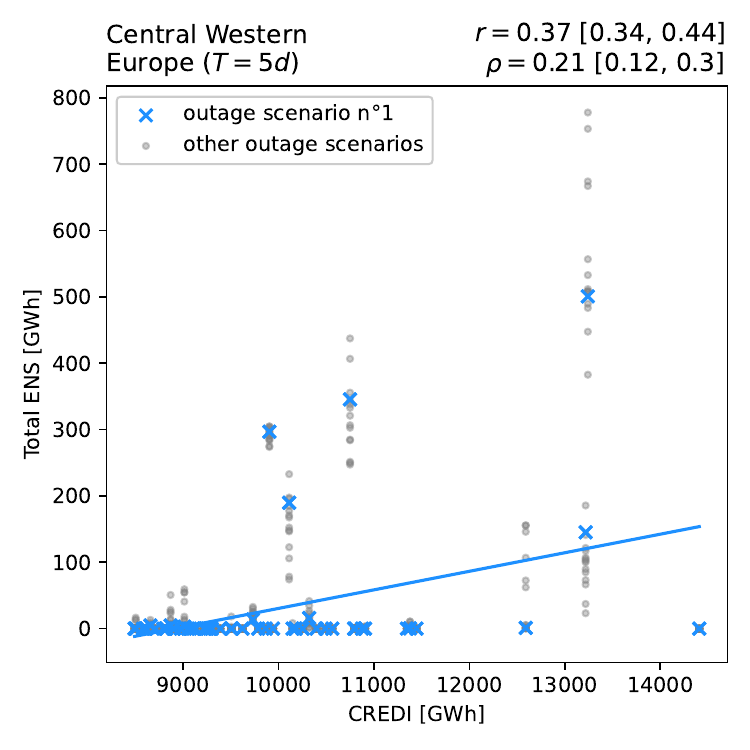}

    \caption{
    F-score (left column) and correlation between the total ENS values of detected dunkelflaute events and their CREDI values for $T=3$~days events (middle column) and $T=5$~days (right column) in Germany, France, and Central Western Europe.
    }
    \label{fig:Stoop_corr_T}
\end{figure}

The Climatological renewable energy deviation index (CREDI) defined in Eq.~\eqref{eq:CREDI} can be defined for different time period $T$.
In the main paper, only $T=1$-day events are discussed.
Figure~\ref{fig:Stoop_corr_T} shows the results (F-score and correlation) for longer events ($T=3$~days and $T=5$~days), for various countries and larger European regions.
Only the results for the 3 regions discussed in the main paper (France, Germany and Central Western Europe) are shown.

For longer events, the peak F-score decreases slightly in all regions (left column).
The correlation between the total ENS values and the CREDI values decreases as well when looking at longer event (lower Pearson coefficient $r$ and Spearman coefficient $\rho$).
In all cases, the dunkelflaute events used in the correlation plots are detected based on the threshold value that maximizes the F-score.

\section{Stoop'23: tuning of the hourly and weekly rolling window}
\label{sec:Stoop_HWRW}

This section focuses on the tuning of the \emph{hourly and weekly rolling window} (HWRW) used in the definition of the climatology $C_P(h)$ (Eq.~\eqref{eq:HWRW}).

The climatology defines the typical weather condition averaged over a long period (generally about 30 years). 
A straightforward and usually sufficient approach (hereafter called 'initial climatology') is to define the climatology at ordinal hour $h$ as the average of the same ordinal hour $h$ for all 35 years:
\begin{align*}
    C_P^{\mathrm{init}}(h) &= \frac{1}{n} \sum_{y=1}^{n} P(y,h),  \\
    h &= 1,\ldots,8760 \nonumber
\end{align*}

As explained in the main text (Sec.~\ref{sec:CREDI}), contrarily to the initial climatology, the climatology based on HWRW is smooth as it is based on a mean rolling window, therefore reducing random fluctuations, while it accounts for daily, weekly and seasonal timescales, which are all relevant for energy time series.

The climatology based on HWRW is a modification of the \emph{hourly rolling window} (HRW) of~\textcite{Stoop2024}, which only accounts for daily and seasonal timescales, not weekly timescale.
Specifically, the climatology $C_P^{\mathrm{HRW}}(h)$ based on a HRW of size $2d+1$ is the average of an energy value $P(y,h)$ at ordinal hour $h$ and of the values at the same hour up to $d$ days before and $d$ days after (i.e. average of the timesteps $h - 24 \times d, \ldots, h, \ldots, h + 24 \times d$) and all 35 years $y$, thus preserving the daily cycle:
\begin{align*}
    C_P^{\mathrm{HRW}}(h) &= \frac{1}{n} \sum_{y=1}^{n} \sum_{h'\in\{h+24d'\}_{d'=-d}^{d'=+d}} \frac{P(y,h')}{2d+1},  \\
    h &= 1,\ldots,8760 \nonumber
\end{align*}

Similarly, the climatology $C_P(h)$ based on the HWRW of size $2\Delta +1$ is the average of an energy value $P(y,h)$ at ordinal hour $h$ and of the values at the same hour and at the same day of the week up to $\Delta$ weeks before and $\Delta$ weeks after 
(i.e. average of the timesteps $h - 7 \times 24 \times \Delta, \ldots, h, \ldots, h + 7 \times 24 \times \Delta$ for all 35 years, thus also preserving the weekly cycle (see Eq.~\ref{eq:HWRW}).

\begin{figure}[htbp]
    \centering
    \includegraphics[width=0.3\textwidth]{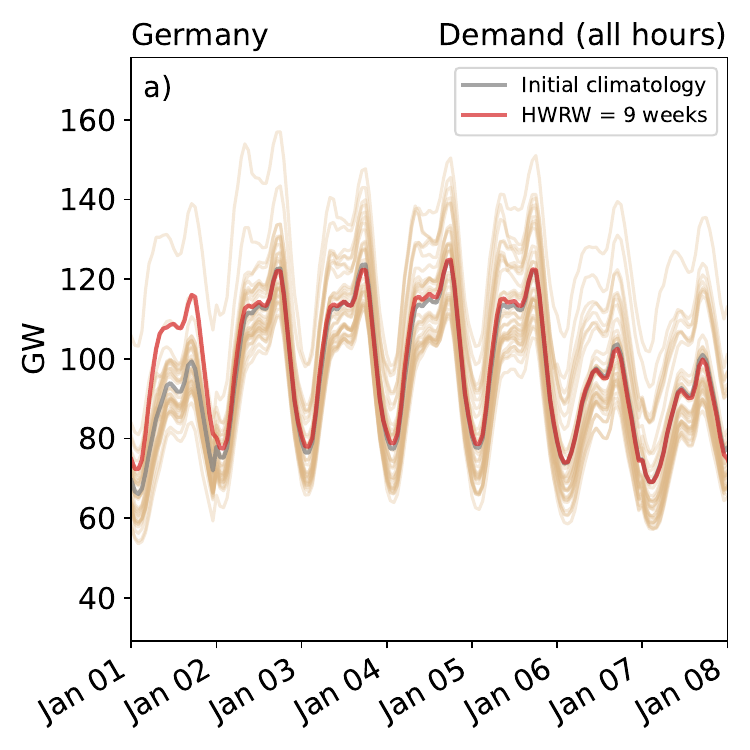}
    \includegraphics[width=0.3\textwidth]{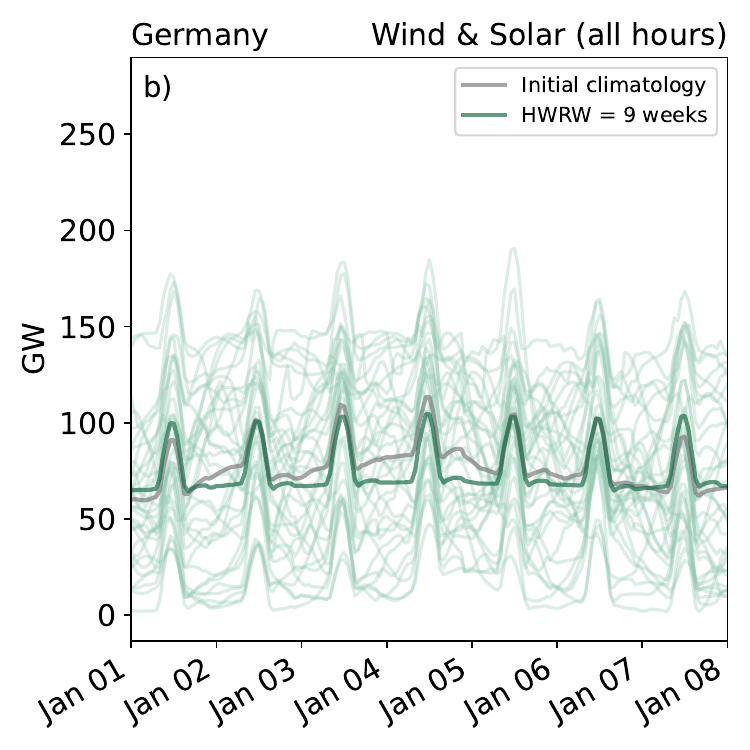}
    \includegraphics[width=0.3\textwidth]{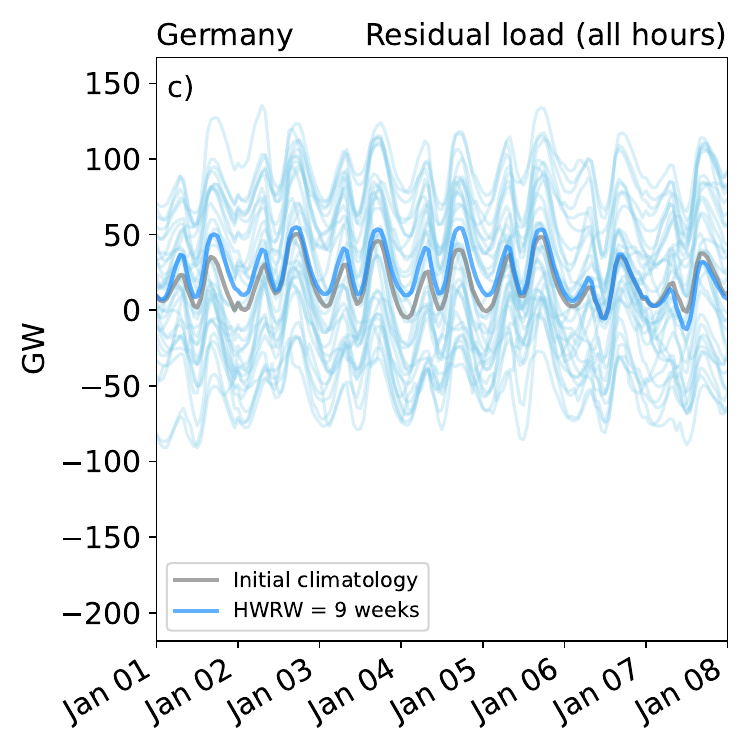}
    \includegraphics[width=0.3\textwidth]{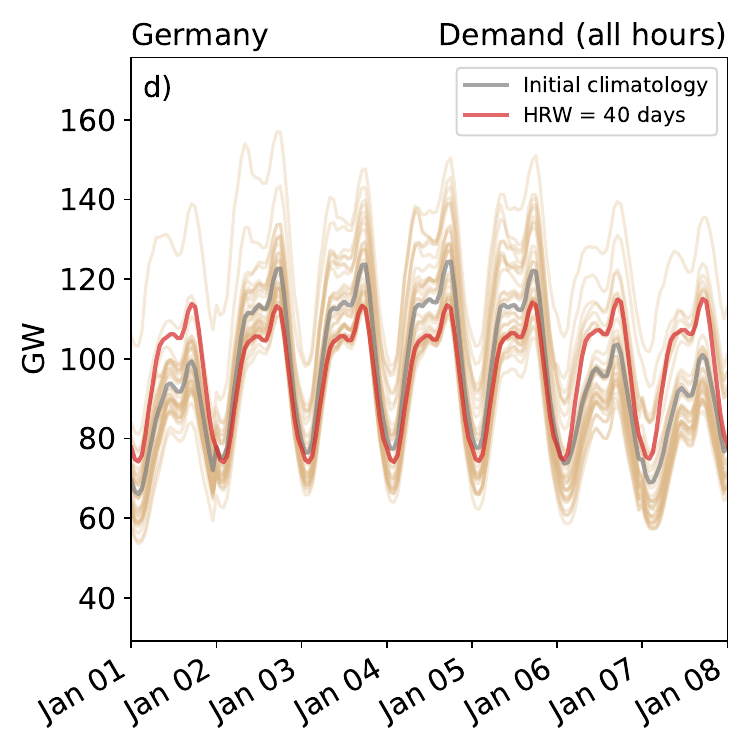}
    \includegraphics[width=0.3\textwidth]{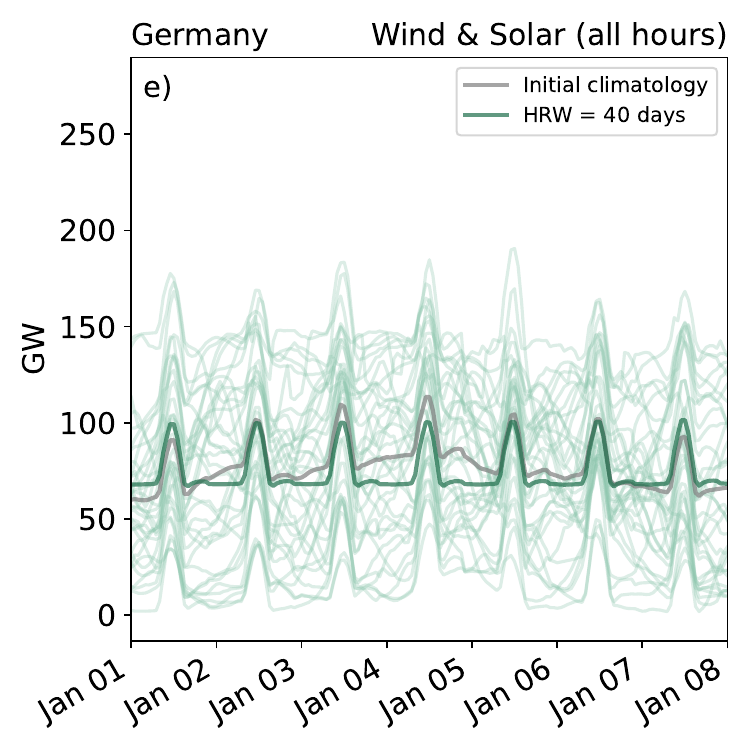}
    \includegraphics[width=0.3\textwidth]{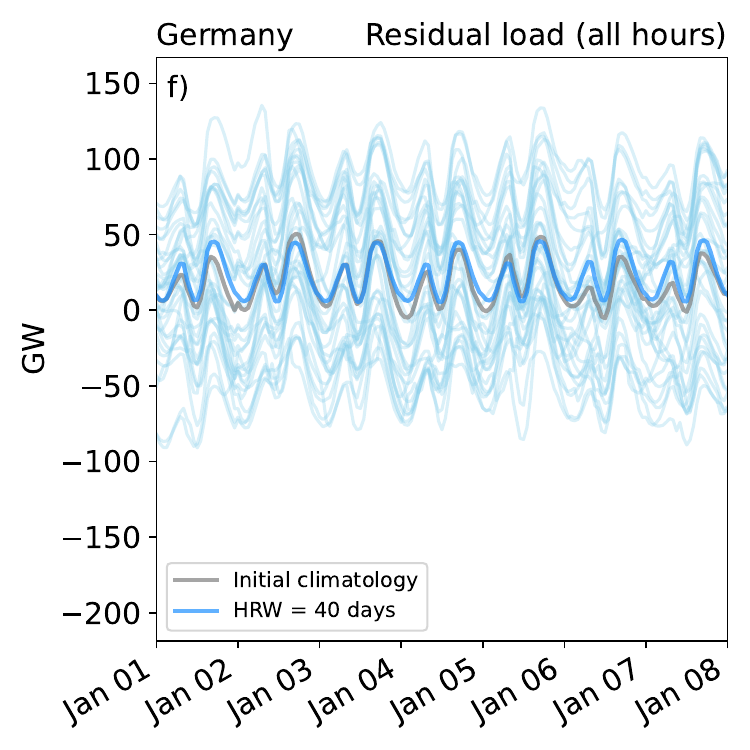}
    \caption{
    Comparison of climatology computed based on the hourly and weekly rolling window (HWRW, upper row) and hourly rolling window (HRW, bottom row) with initial climatology (grey line) for demand, RES generation, and residual load. 
    In light colours, the times series for individual years from 1982 to 2016 are shown. 
    Only the first week of January is shown.
    }
    \label{fig:SI_Comparison_HRW_HWRW}
\end{figure}

Figure~\ref{fig:SI_Comparison_HRW_HWRW} show the difference between the different approach (initial climatology, based on HRW and based on HWRW) for demand, RES generation and residual load.
As illustrated for demand, the HRW is computed in the same way in workweek and weekend (Fig.~\ref{fig:SI_Comparison_HRW_HWRW}d), thus largely underestimating the initial climatology in workweek\footnote{The HRW approach estimates a mid-day climatology in the tail of the distribution for the period from January 2nd to 5th.} and overestimating it in weekends.
On the opposite, HWRW correctly estimate the (Fig.~\ref{fig:SI_Comparison_HRW_HWRW}a) initial climatology, both in workweek and weekend.
However, these approaches overestimate the average demand in bank holidays (e.g. January 1st).

Figure~\ref{fig:SI_Comparison_HRW_HWRW}b shows peak RES production during the days due to solar PV production, and a relatively constant (over the illustrative weekly time window) mean production at night due to wind production.
The initial climatology exhibits random fluctuations (see January 3rd and 4th) which are not present with the HWRW approach.
By averaging more data than the initial climatology (35 $\times$ 9 per hourly timestep with HWRW$=9$ weeks, compared to simply 35 per hourly timestep), HWRW results in a smoother estimate of the climatology that aligns with the expected mean value.

It is important to note that the Demand Forecasting Tool (DFT), which generates demand time series for each market zone based on weather variables and societal hypotheses, utilizes the 2018 calendar for all computations, with January 1st falling on a Monday.
Therefore, care should be taken when the hourly and weekly rolling windows overlap two years.

The choice of the HWRW is a balance between having a sufficiently long time window to accurately smooth the climatology, and ensuring it is small enough to preserve the seasonal cycle.

\begin{figure}[htbp]
    \centering
    \includegraphics[width=0.33\textwidth]{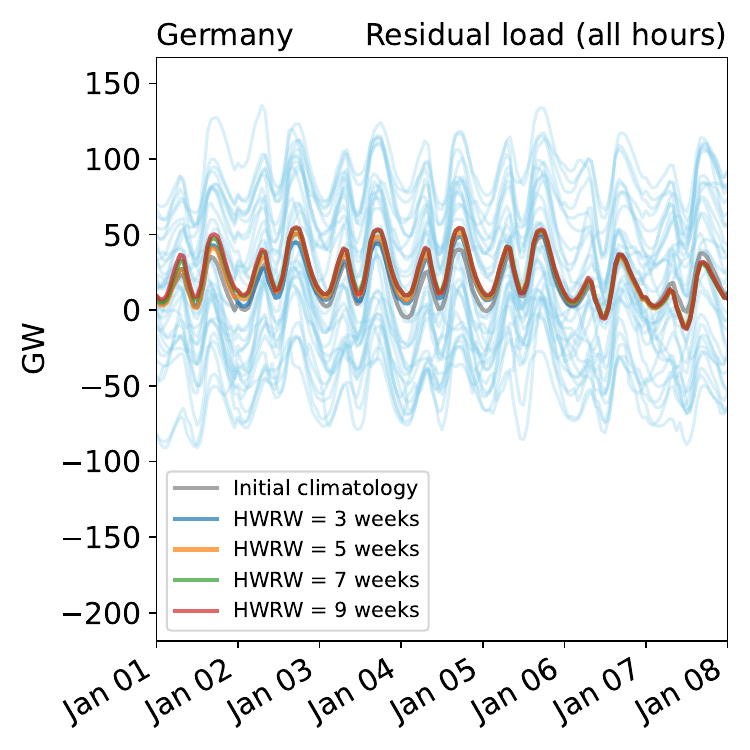}
    \includegraphics[width=0.65\textwidth]{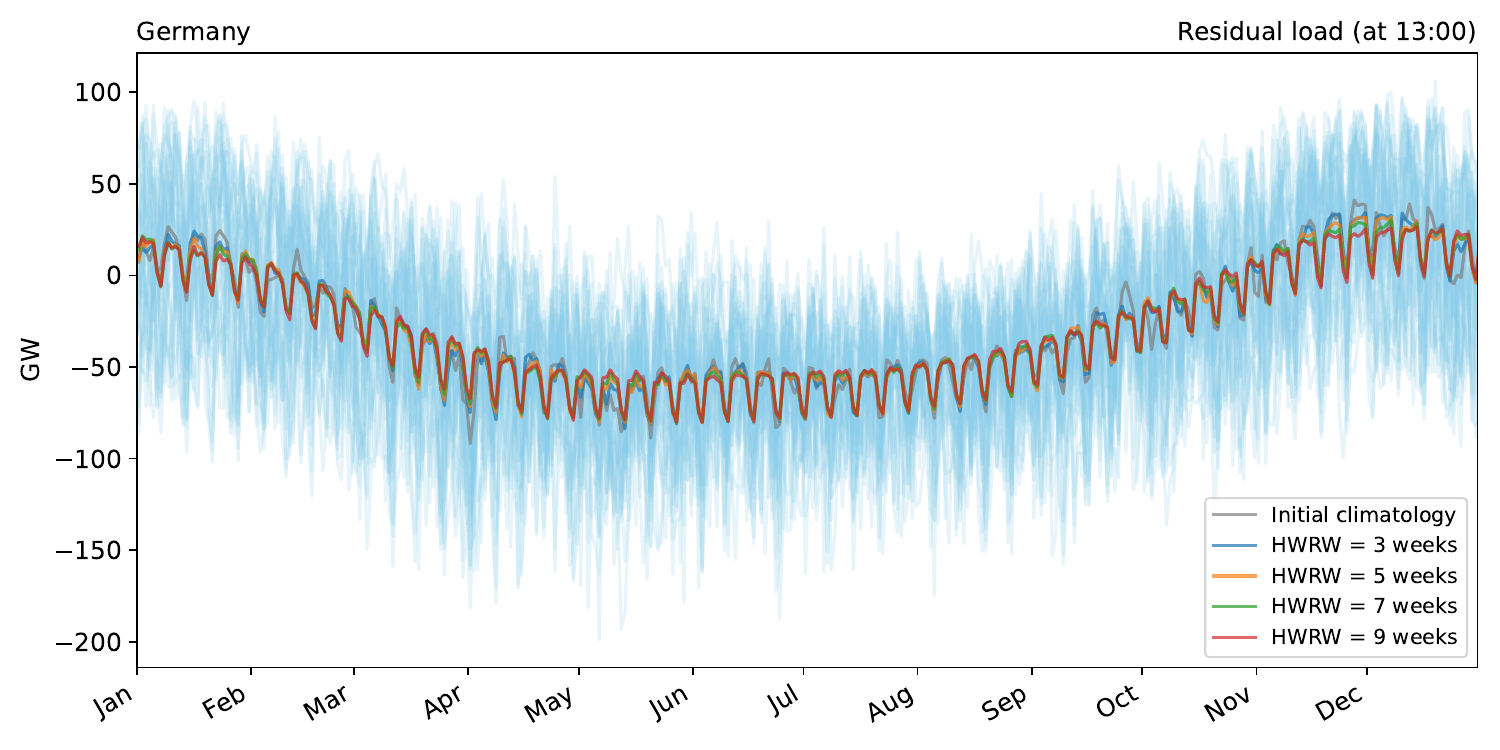}
    \includegraphics[width=0.33\textwidth]{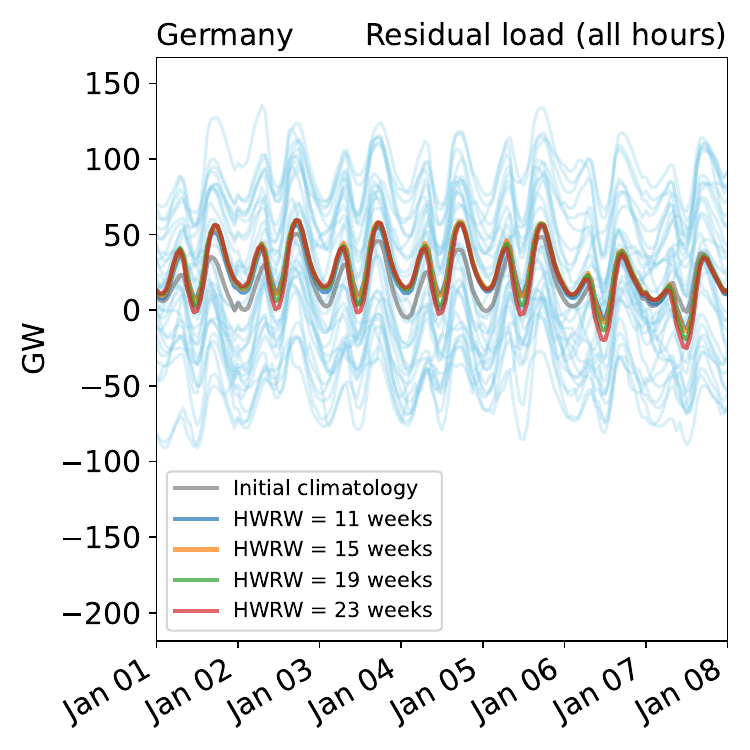}
    \includegraphics[width=0.65\textwidth]{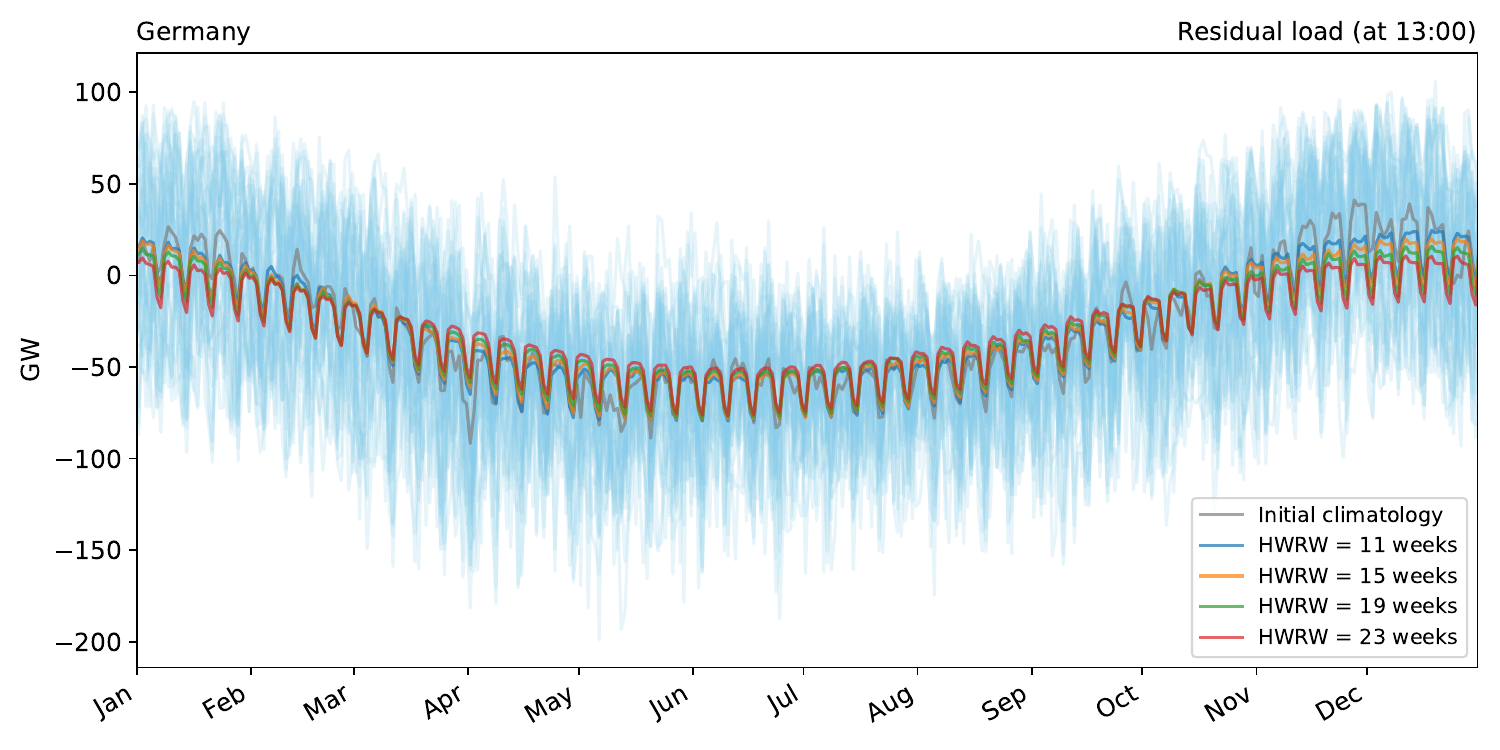}
    \caption{
    Comparison of window size for the hourly and weekly rolling window (HWRW = 3, 5, 7, 9 weeks in top panel, HWRW = 11, 15, 19, 23 weeks in bottom panel) for residual load. 
    In light blue, the times series for individual years from 1982 to 2016 are shown. 
    In the right panels, only 13:00 is shown to exhibit the variability at the weekly and seasonal scales, eliminating the large daily variability.
    }
    \label{fig:SI_tuning_HWRW}
\end{figure}

Figure~\ref{fig:SI_tuning_HWRW} shows climatology for different values of the HWRW.
Long window are found to smooth out seasonality, leading to overestimation of the mean residual load in summer (the slight increase in residual load in June disappears for HWRW$=23$ weeks) and a significant underestimation of the mean residual load in winter. 
Therefore, we have selected a HWRW of 9 weeks, which provide a suitable balance between smoothing and preserving seasonality.

\section{Comparison of target years 2025 and 2033}
\label{sec:SI_TY}

\begin{figure}[htbp]
    \centering
    \includegraphics[width=0.3\textwidth]{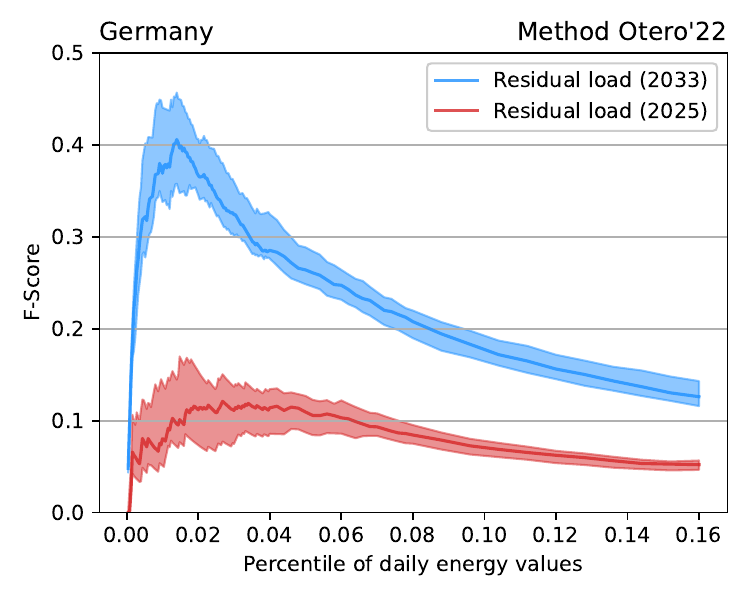}
    \includegraphics[width=0.3\textwidth]{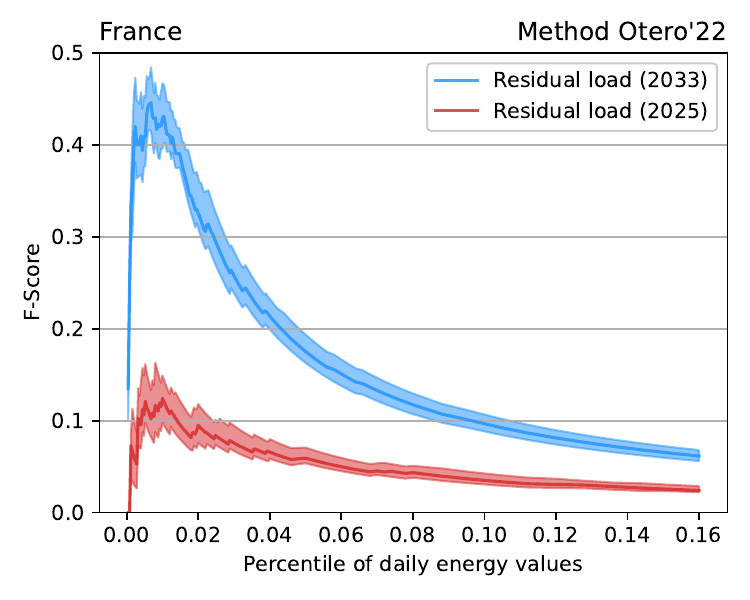}
    \includegraphics[width=0.3\textwidth]{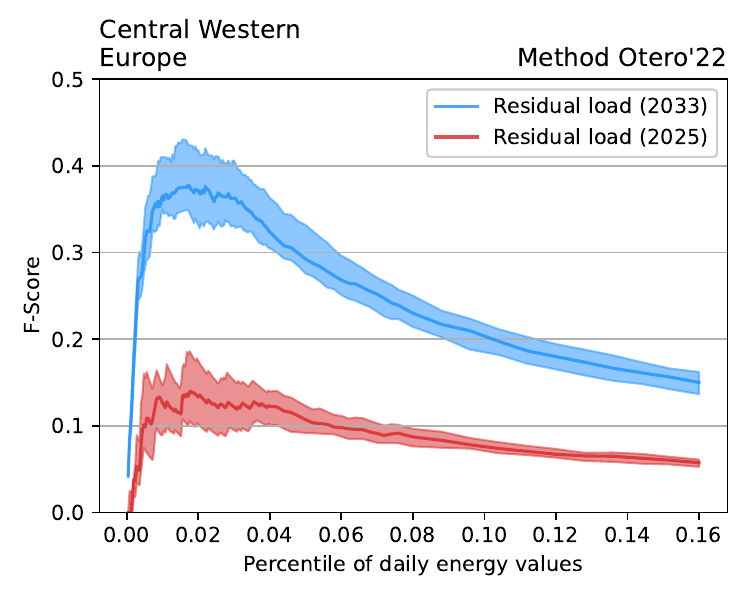}
    \caption{
    F-score for dunkelflaute detection with Otero'22 for target years 2025 and 2033, based on the residual dual for Germany, France, and Central Western Europe.
    }
    \label{fig:Otero_ty2025-2033}
\end{figure}

Figure~\ref{fig:Otero_ty2025-2033} shows F-score for target year 2025 and 2033.
In both target years, the RES installed capacities, demand time series, and ENS datasets resulting from power system simulations are different.

This figure illustrates that the accuracy of shortage identification through dunkelflaute detection is significantly higher for the target year 2033 considered in the main paper compared to 2025, due to a higher share of RES in the electricity mix. 
The fact that the simulations for target year 2033 also show significantly higher ENS for most countries than in target year 2025 may also contribute to the higher F-Score in 2033.

\section{Sensitivity to EVA scenarios}
\label{sec:SI_EVA}
The Economic Viability Assessment (EVA) step of the ERAA 2023 study~\autocite{ERAA2023} simulates potential economically-driven investment and retirement decisions in different capacity resources, by performing a minimisation of total system costs. In particular, the model endogenously considers potential decommissioning and new investments in coal, gas, and oil-fired capacities, as well as demand-side response and batteries (investments only), on top of the baseline data provided by TSOs in the National Trends scenario. Investment and retirement of RES and nuclear are not considered.
The ERAA2023 considered two post-EVA scenarios based on the same three historical climate years (1985, 1988, 2003), but with different weights applied to these climate years in the EVA investment model. In scenario B (presented in the main paper), a lower weighting is assigned to the most challenging climate year (1985), leading to more retirement and less investment in new capacity than in scenario A. Thus, scenario B showed more adequacy concerns (i.e. higher ENS) than scenario A, which was the main reason scenario B was used for the validation performed in the main paper. However, by comparing the results of the F-Scores of scenario B with scenario A, one can assess the impact of having different levels of dispatchable capacity in the scenarios, and its impact on the ability to detect dunkelflauten.

\begin{figure}[htbp]
    \centering
    \includegraphics[width=0.3\textwidth]{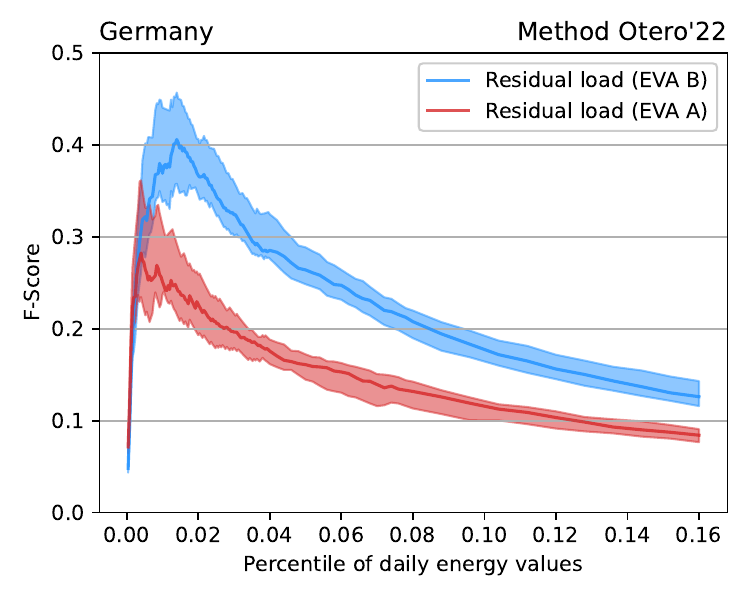}
    \includegraphics[width=0.3\textwidth]{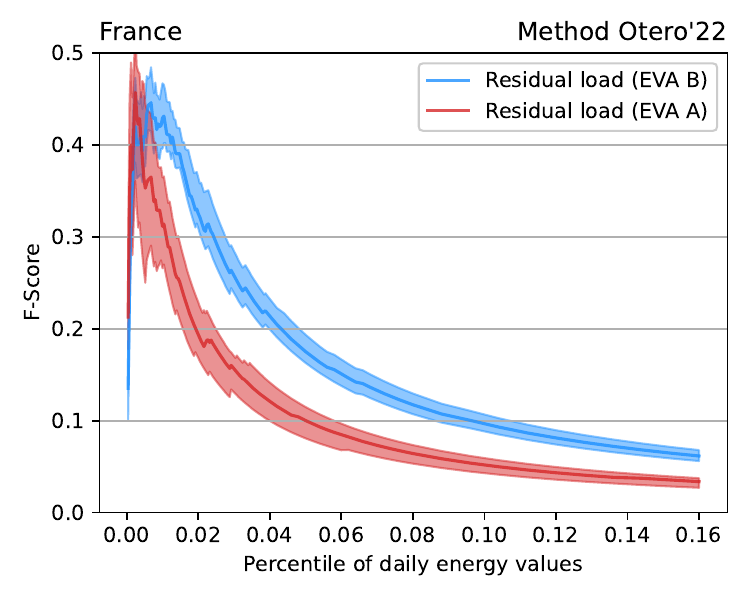}
    \includegraphics[width=0.3\textwidth]{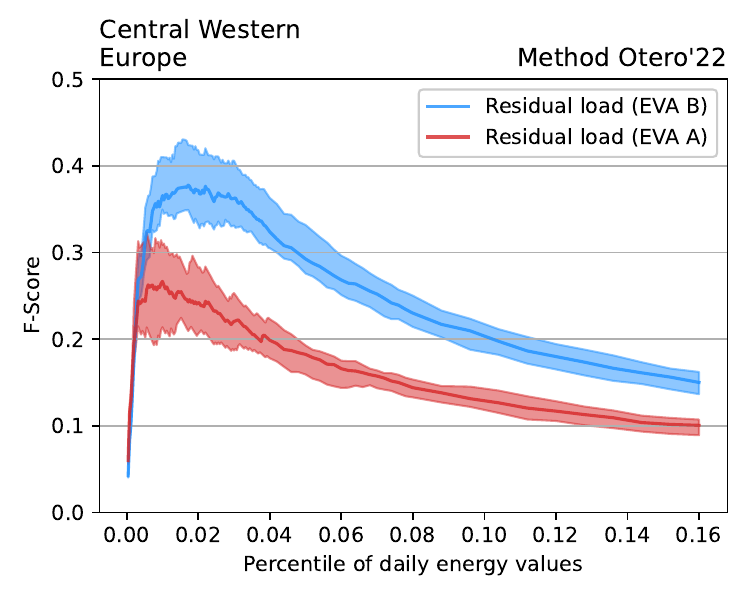}
    \caption{
    F-score for dunkelflaute detection with Otero'22 for Economic Viability Assessment (EVA) scenarios A and B, based on the residual dual for Germany, France, and Central Western Europe.
    }
    \label{fig:Otero_EVA}
\end{figure}

Figure~\ref{fig:Otero_EVA} compares the F-score obtained for both EVA scenarios using residual load, in Germany, France, and Central Western Europe.
In both scenarios, the time series of RES generation and demand are identical; the only difference is in the ENS datasets, which reflect the differences in investment and decommissioning in thermal power plants.
The F-scores show that the accuracy of the dunkelflaute detection methods is better in scenario B than in scenario A. This shows that the ability to correctly detect dunkelflauten is affected by how much non-RES dispatchable capacity is in the scenario, and likely the level of ENS. In scenario A, the additional investment in dispatchable non-RES capacity reduces the dependence of the system on weather-driven RES generation, reducing the overall volume of ENS. This means other non-weather related factors which can cause ENS, such as unplanned outages, likely cause a higher proportion of ENS events in scenario A than in scenario B. 

\section{Filtering out non-weather ENS events}
\label{sec:SI_filter}

One question that arises is whether it is possible to filter \emph{a priori} weather-related ENS (Energy Not Supplied) days. 
By excluding ENS occurrences that are not weather-related, the accuracy of the detection method could be significantly enhanced. 
To achieve this, we combine the ENS dataset encompassing all 15 outage scenarios and select only those days where outages occur in a minimum number of these scenarios.

\begin{figure}[htbp]
    \centering
    \includegraphics[width=0.3\textwidth]{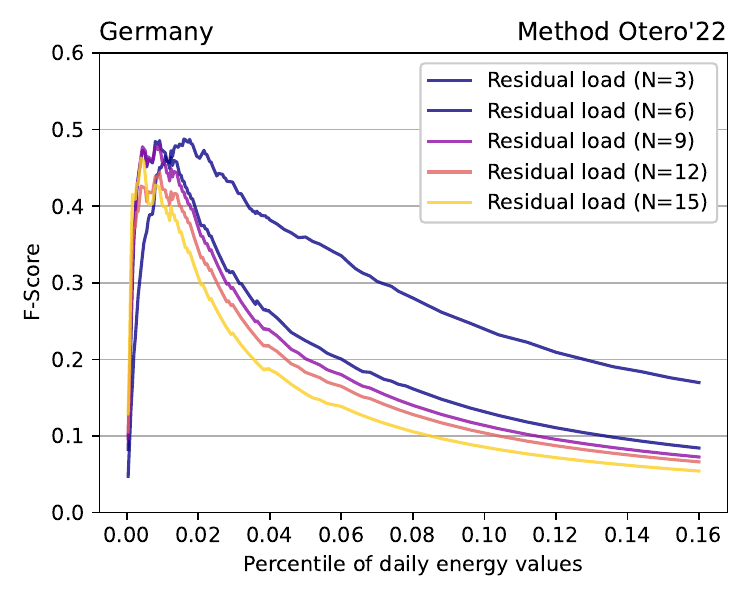}
    \includegraphics[width=0.3\textwidth]{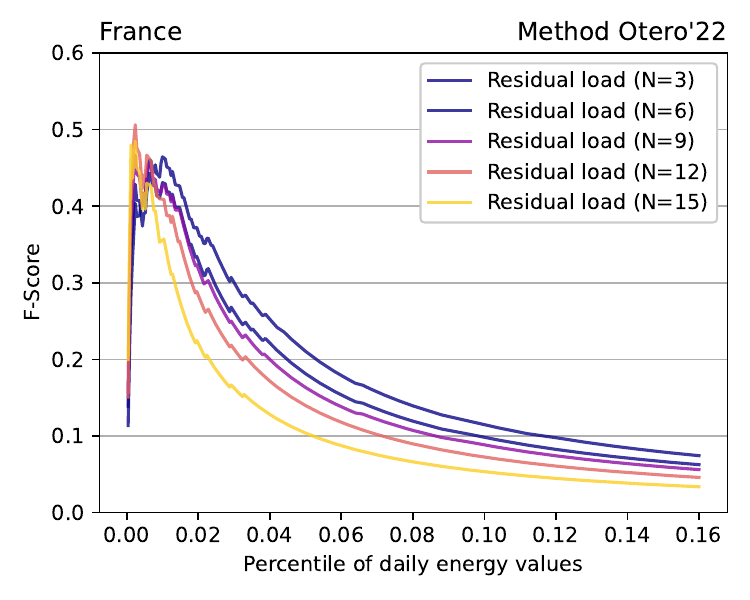}
    \includegraphics[width=0.3\textwidth]{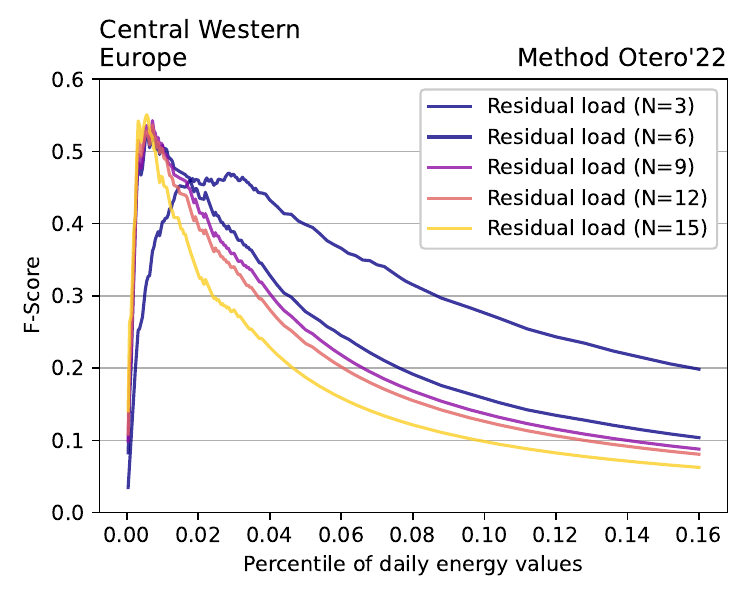}
    \caption{
    F-score for dunkelflaute detection with Otero'22, taking only ENS events that occur on a specific calendar day in at least $N$ outage scenarios (out of 15).
    }
    \label{fig:Otero_filterFOS}
\end{figure}

Figure~\ref{fig:Otero_filterFOS} only consider the days that are classified as ENS in at least $N$ outage scenarios ($N=3, 6, 9, 12, 15$).
Interestingly, while the peak F-score value does increase when compared to considering all ENS events (compare Figure~\ref{fig:Otero_filterFOS} with Figure~\ref{fig:Otero_DE00} and Figure~\ref{fig:Otero_FR_CWE}), the improvement is marginal. 
Specifically, the F-score rises to $F=0.49$ for $N=3$, compared to $0.41$ in Germany; $F=0.51$ for $N=12$ vs. $0.45$ in France; $F=0.51$ for $N=12$ vs. $0.38$ in Central Western Europe.

Filtering out \emph{a priori} non-weather-related ENS events thus marginally improve shortages identification based on dunkelflaute detection.

\section{Year selection}
\label{sec:SI_year_selection}

\begin{figure}[htbp]
    \centering
    \includegraphics[width=\textwidth]{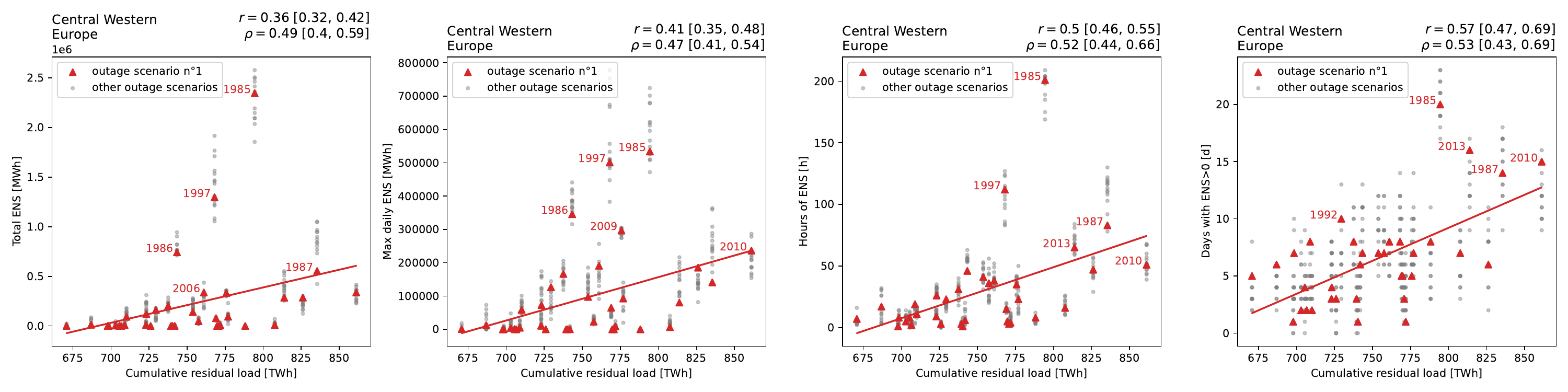}
    \caption{
    Correlation between the cumulative residual load of each year and four annual shortage metrics: total cumulative ENS of each year, maximum daily ENS value of each year, number of hours of ENS for each year (same as Fig.~\ref{fig:yeark_ranking}a), and number of days with at least one hour of ENS. 
    The five years with the highest annual shortage metric are labelled in each figure. 
    }
    \label{fig:year_rank_RL}
\end{figure}

Figure~\ref{fig:year_rank_RL} completes Figure~\ref{fig:yeark_ranking}a by showing the correlation between the cumulative residual load of each year and other annual metrics of shortages: the total cumulative ENS of each year, the maximum daily ENS value of each year, the number of hours of ENS for each year (same as Fig.~\ref{fig:yeark_ranking}a), and the number of days with at least one hour of ENS.

This rather simple approach, which takes the cumulative residual load of each year to identify the most challenging years, yields rather low correlation values ($ r\approx 0.3-0.6$) and struggles to rank the climate years from the most extreme to the least extreme ($\rho \approx 0.5$), for all four annual shortage metrics.

\begin{figure}[htbp]
    \centering
    \includegraphics[width=\textwidth]{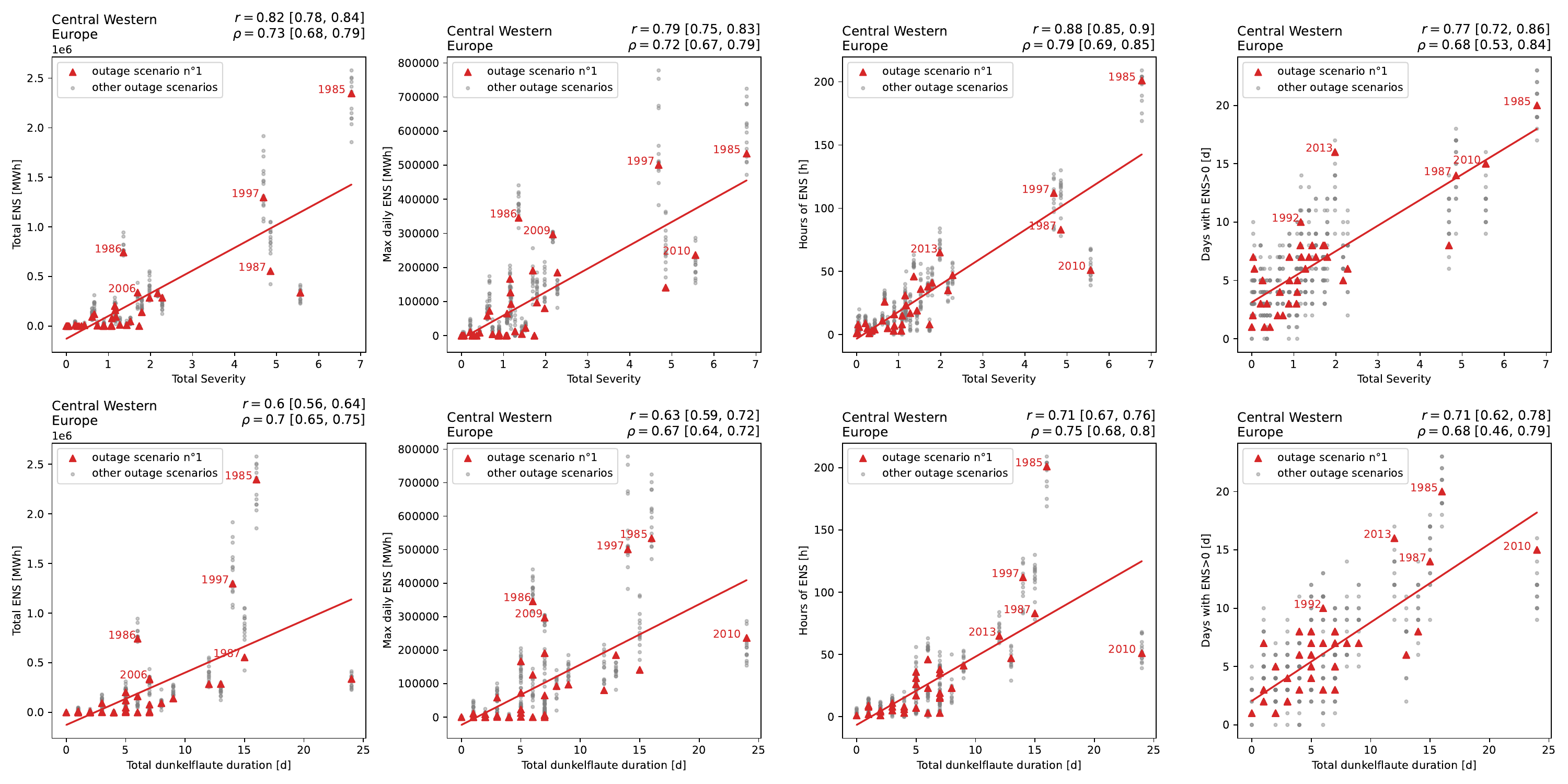}
    \caption{
    Correlation between the total severity determined with the Otero'22 method (sum of the severity $S$ (see Eq.~\eqref{eq:OteroSeverity}) of identified dunkelflaute events, upper row) or the total duration of identified dunkelflaute events (bottom row), with four annual shortage metrics: total cumulative ENS of each year, maximum daily ENS value of each year, number of hours of ENS for each year, and number of days with at least one hour of ENS. 
    The five years with the highest annual shortage metric are labelled in each figure. 
    }
    \label{fig:year_rank_Otero}
\end{figure}

Similarly, Figure~\ref{fig:year_rank_Otero} completes Figure~\ref{fig:yeark_ranking}b.
It shows the correlation between the four previously mentioned shortage annual metrics and the total severity, defined as the sum of the severity $S$ (see Eq.~\eqref{eq:OteroSeverity}) of identified dunkelflaute events, as determined by the Otero'22 method.
It also shows the same correlation plots with the total duration, i.e. the sum of the durations of all identified dunkelflaute events.

For all annual metrics, the dunkelflaute detection methods Otero'22 improves the identification of the most challenging years (i.e. years with the highest annual shortage metrics) compared to the simpler approach that calculates the cumulative residual load of each year.
This is shown by higher correlation coefficients $r$ and Spearman coefficients $\rho$.

\begin{figure}[htbp]
    \centering
    \includegraphics[width=\textwidth]{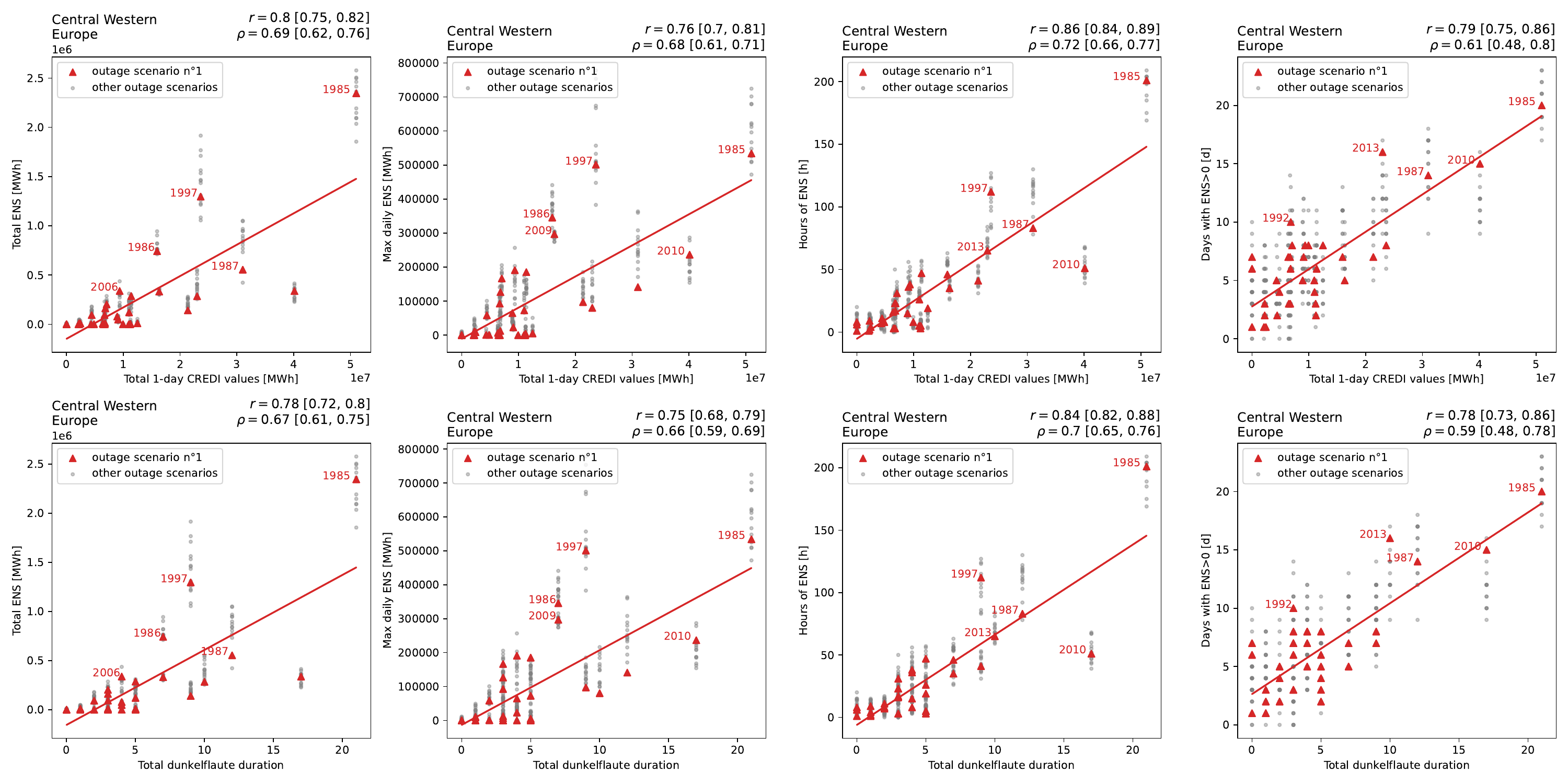}
    \caption{
    Correlation between the $T=1$-day CREDI values of dunkelflaute events as determined with the Stoop'23 method (upper row) or the total duration of identified dunkelflaute events (bottom row), with four annual shortage metrics: total cumulative ENS of each year, maximum daily ENS value of each year, number of hours of ENS for each year, and number of days with at least one hour of ENS. 
     The five years with the highest annual shortage metric are labelled in each figure. 
    }
    \label{fig:year_rank_Stoop}
\end{figure}

The same analysis is performed for the Stoop'23 approach.
Figure~\ref{fig:year_rank_Stoop} shows the correlation between the four annual shortage metrics and the $T=1$-day CREDI values\footnote{
The accuracy of ENS detection is notably higher when using $T=1$-day events compared to longer duration events (see Sec~\ref{sec:SI_Stoop_T}).
}
and total duration of CREDI events of each year.
The same conclusions as with the approach of Otero'22 can be drawn.

\section{$F_{\beta}$ score: favouring precision or recall}
\label{sec:SI_Fbeta}

The F-score measures the accuracy of the shortage detection. 
However, this metric arbitrarily assigns the same importance to $\mathrm{precision} = \frac{\mathrm{TP}}{\mathrm{TP}+\mathrm{FP}}$ (i.e. how many detected dunkelflaute are actually ENS events) and $\mathrm{recall} = \frac{\mathrm{TP}}{\mathrm{TP}+\mathrm{FN}}$ (how many ENS events were picked up as dunkelflaute).

A more generic $F_{\beta}$ score can be used, where $\beta$ expresses the relative weight of precision and recall:
\begin{equation}
    F_\beta = \frac{(1+\beta^2) \cdot \mathrm{precision} \cdot \mathrm{recall}}{\beta^2 \cdot \mathrm{precision} + \mathrm{recall}}  
            = \frac{(1+\beta^2)\cdot \text{TP}}{(1+\beta^2)\cdot\text{TP} + \beta^2 \cdot \text{FN} + \text{FP}}.
    \label{eq:F-score_SI}
\end{equation}
Note that the F-score is retrieved for $\beta=1$.
Depending on the user's primary interest, either precision or recall can be prioritized by tuning $\beta$:
\begin{itemize}
    \item Favouring precision ($\beta < 1$) would result in identifying most of the ENS events, but at the expense of detecting numerous dunkelflaute events that are not ENS.
    \item Favouring recall ($\beta > 1$) would increase the proportion of ENS events among the detected dunkelflaute events, potentially at the cost of missing many ENS events.
\end{itemize}

\begin{figure}[htbp]
    \centering
    \includegraphics[width=0.3\textwidth]{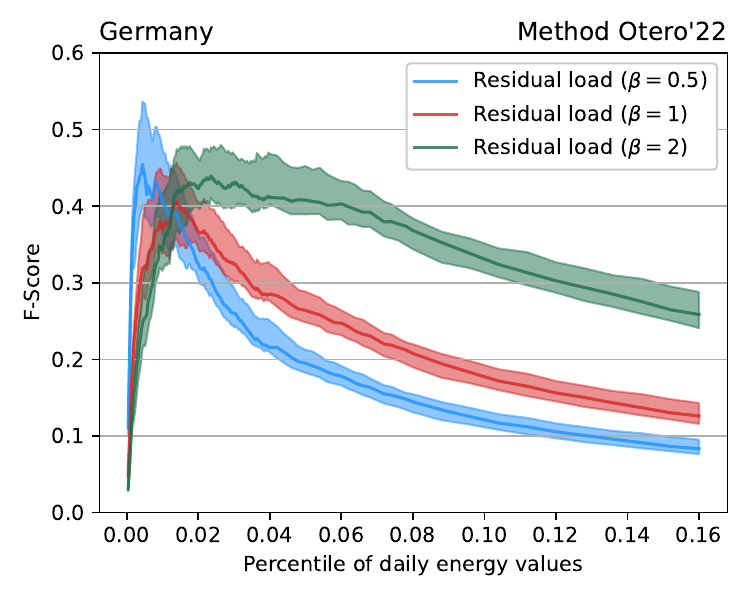}
    \includegraphics[width=0.3\textwidth]{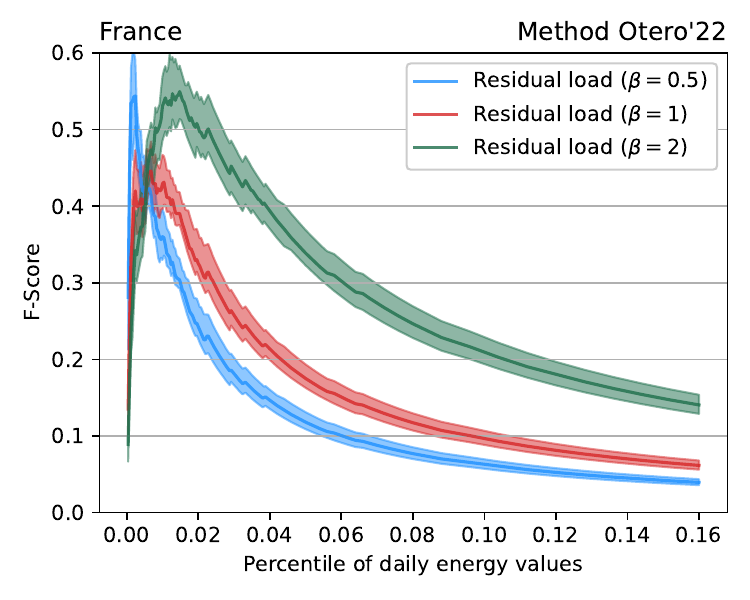}
    \includegraphics[width=0.3\textwidth]{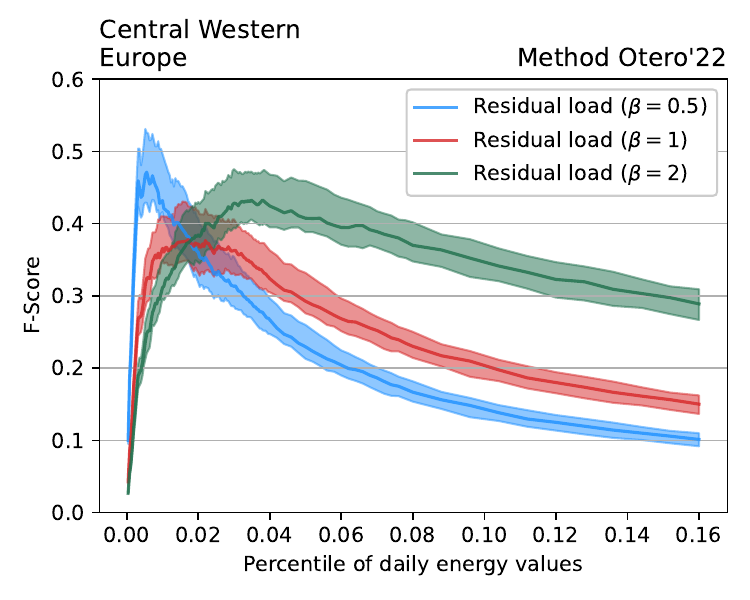}
    \includegraphics[width=0.3\textwidth]{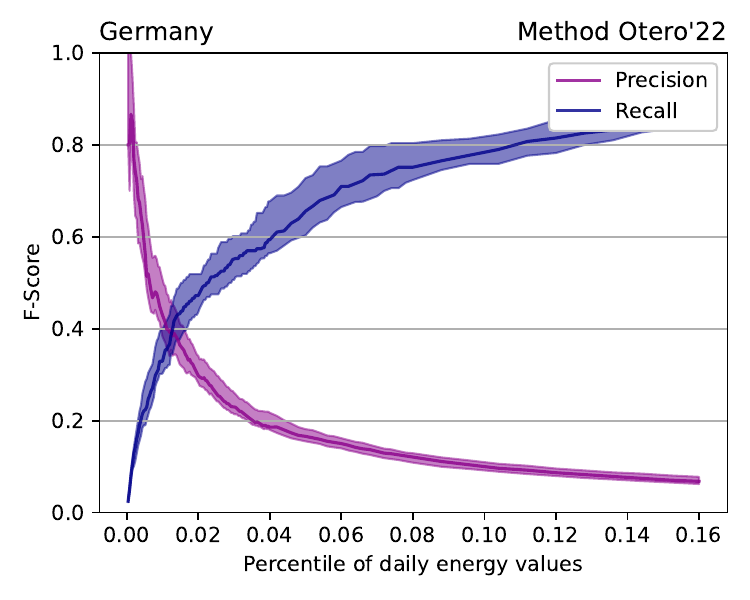}
    \includegraphics[width=0.3\textwidth]{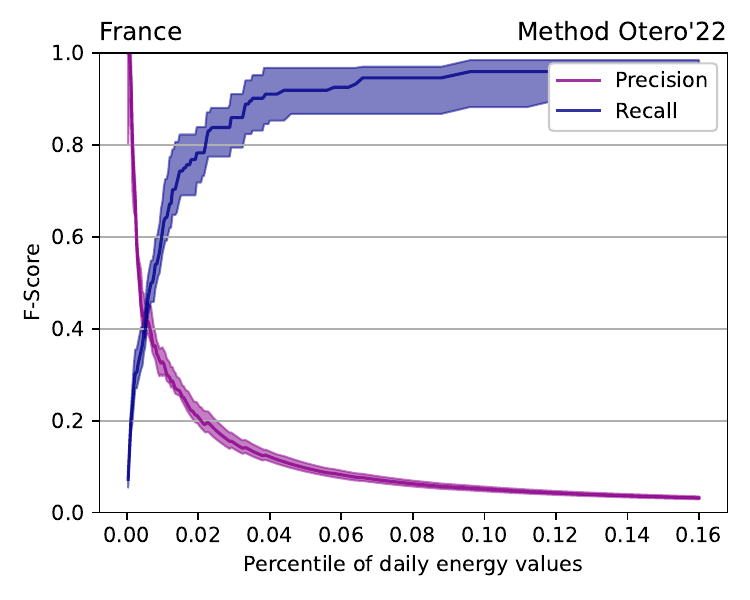}
    \includegraphics[width=0.3\textwidth]{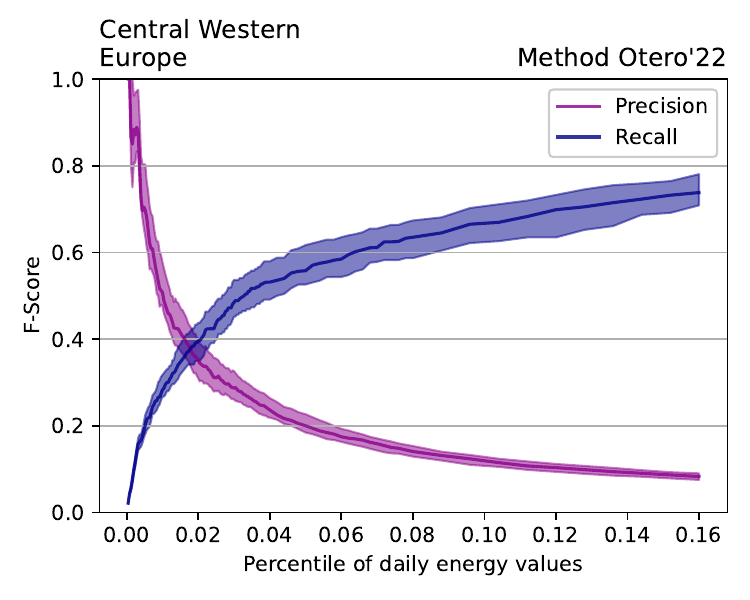}
    \caption{
    Top panels: $F_{\beta}$ score for $\beta = 0.5, 1, 2$ (relative weight between precision and recall), based on residual load in Germany, France and Central Western Europe.
    The associated precision and recall are shown in the bottom panels.}
    \label{fig:Otero_Fbeta}
\end{figure}

Figure~\ref{fig:Otero_Fbeta} (upper row) shows the $F_{\beta}$ score for $\beta=$ 0.5, 1, and 2, in Germany, France and Central Western Europe.
In all region, the score is improved when deviating from $\beta=1$, i.e. favouring either precision or recall.

The increase of the peak $F_{\beta}$ value is explained by looking at the precisions and recall for each region (Fig.~\ref{fig:Otero_Fbeta}, bottom row).
When applying a stringent threshold (using a small percentile of the daily residual load distribution), precision significantly increases, meaning that most of the detected dunkelflaute events are indeed ENS events. 
However, this comes at the cost of a substantial decrease in recall, resulting in the detection of very few ENS events.
When the threshold is less extreme, we find the opposite: precision drops substantially while recall strongly increases.
Thus, adjusting $\beta$ in either direction would increases the $F_{\beta}$ score relative to the F-score either for low or high threshold, leading to an increase of the peak $F_{\beta}$ score. 
Given that this study aims to assess the overall detection performance of these methods in terms of both recall and precision, we have set $\beta=1$.

\end{document}